\documentclass[aps,onecolumn,prd,showpacs,showkeys,preprintnumbers,superscriptaddress,nobibnotes,floatfix,longbibliography,notitlepage,nofootinbib]{revtex4-1}

\pdfoutput=1
\usepackage{amsmath}
\usepackage{amsfonts}
\usepackage{amssymb}
\usepackage{mathrsfs}
\usepackage{graphicx}
\usepackage{cases}
\usepackage{comment}
\usepackage{fontawesome}
\usepackage[dvipsnames]{xcolor}

\usepackage{hyperref}
\usepackage{multirow}
\usepackage{upgreek}
\usepackage[capitalise]{cleveref}

\usepackage{siunitx}
\DeclareSIUnit \s {\second}
\DeclareSIUnit \ns {\nano\second}
\DeclareSIUnit \mus {\micro\second}
\DeclareSIUnit \ms {\milli\second}
\DeclareSIUnit \MB {\mega\byte}
\DeclareSIUnit \GB {\giga\byte}
\DeclareSIUnit \TB {\tera\byte}
\DeclareSIUnit \PB {\peta\byte}
\DeclareSIUnit \Mbps {\mega\bit/\s}
\DeclareSIUnit \Gbps {\giga\bit/\s}
\DeclareSIUnit \Tbps {\tera\bit/\s}
\DeclareSIUnit \Pbps {\peta\bit/\s}
\DeclareSIUnit \kton {\kilo\tonne} % changed  back to kton
\DeclareSIUnit \kt {\kilo\tonne}
\DeclareSIUnit \kty {\kilo\tonne-\year}
\DeclareSIUnit \Mt {\mega\tonne}
\DeclareSIUnit \eV {\electronvolt}
\DeclareSIUnit \keV {\kilo\electronvolt}
\DeclareSIUnit \MeV {\mega\electronvolt}
\DeclareSIUnit \GeV {\giga\electronvolt}
\DeclareSIUnit \TeV {\tera\electronvolt}
\DeclareSIUnit \PeV {\peta\electronvolt}
\DeclareSIUnit \EeV {\exa\electronvolt}
\DeclareSIUnit \m {\meter}
\DeclareSIUnit \cm {\centi\meter}
\DeclareSIUnit \nm {\nano\meter}
\DeclareSIUnit \in {\inchcommand}
\DeclareSIUnit \km {\kilo\meter}
\DeclareSIUnit \kV {\kilo\volt}
\DeclareSIUnit \kW {\kilo\watt}
\DeclareSIUnit \MW {\mega\watt}
\DeclareSIUnit \MHz {\mega\hertz}
\DeclareSIUnit \mrad {\milli\radian}
\DeclareSIUnit \year {years}
\DeclareSIUnit \POT {POT}
\DeclareSIUnit \sig {$\sigma$}
\DeclareSIUnit\parsec{pc}
\DeclareSIUnit\lightyear{ly}
\DeclareSIUnit\foot{ft}
\DeclareSIUnit\ft{ft}
\DeclareSIUnit \ppb{ppb}
\DeclareSIUnit \ppt{ppt}
\DeclareSIUnit \samples{S}
\DeclareSIUnit \pe{PE}
\DeclareSIUnit \GeVmwe{GeV/mwe}
\DeclareSIUnit \mwe{mwe}

\newcommand{\enu}{\E_\enu}

\begin{document}

\preprint{FERMILAB-PUB-21-214-T}

\title{Millicharged Particles From the Heavens:\\ Single- and Multiple-Scattering Signatures\\
\textcolor{BlueViolet}{\href{https://github.com/Harvard-Neutrino/HeavenlyMCP}{\faGithub}}\vspace{-0.3cm}}

\author{Carlos A. Arg{\"u}elles}
\email{carguelles@fas.harvard.edu}
\affiliation{Department of Physics \& Laboratory for Particle Physics and Cosmology, Harvard University, Cambridge, MA 02138, USA}

\author{Kevin J. Kelly}
\email{kkelly12@fnal.gov}
\affiliation{Theoretical Physics Department, Fermilab, P.O. Box 500, Batavia, IL 60510, USA}

\author{V\'ictor M. Mu\~noz}
\email{vicmual@ific.uv.es}
\affiliation{Instituto de F\'isica Corpuscular, Universidad de Valencia and CSIC, Edificio Institutos Investigac\'ion, Catedr\'atico Jos\'e Beltr\'an 2, 46980 Spain}

\date{\today}

\begin{abstract}
For nearly a century, studying cosmic-ray air showers has driven progress in our understanding of elementary particle physics.
In this work, we revisit the production of millicharged particles in these atmospheric showers and provide new constraints for XENON1T and Super-Kamiokande and new sensitivity estimates of current and future detectors, such as JUNO.
We discuss distinct search strategies, specifically studies of single-energy-deposition events, where one electron in the detector receives a relatively large energy transfer, as well as multiple-scattering events consisting of (at least) two relatively small energy depositions. We demonstrate that these atmospheric search strategies --- especially this new, multiple-scattering signature --- provide significant room for improvement in the next decade, in a way that is complementary to anthropogenic, beam-based searches for MeV-GeV millicharged particles.
Finally, we also discuss the implementation of a Monte Carlo simulation for millicharged particle detection in large-volume neutrino detectors, such as IceCube.
\end{abstract}

\maketitle
\vspace{-35pt}
\tableofcontents

\section{Introduction}
\label{sec:intro}
% Carlos

Since the earliest times, humans have stared at the sky and wondered.
By observing the low-orbit Earth skies, they discovered the presence of extraterrestrial charged particles, now called cosmic rays~\cite{Pacini:2010hv,Hess:2018twh,DeAngelis:2014doa,Grupen:2013gja}.
These experimental efforts started from attempts to understand the origin of environmental radioactivity, and through their study led to the discovery of muons in 1936~\cite{Neddermeyer:1937md} and pions in 1947~\cite{Lattes:1947mw}.
In this work, we return to contemplating cosmic-ray air showers' products in order to shed light on the nature of electric charge by looking for signatures of small electrically charged particles, often known as millicharged particles~\cite{Dobroliubov:1989mr}.

In the Standard Model (SM), all electric charges are multiples of the $d$-quark charge, a principle known as charge quantization.
However, though the SM imposes charge conservation by its gauge symmetries and anomaly cancellation, there is no firm theoretical evidence for the principle of charge quantization~\cite{Dobroliubov:1989mr}.
There are two ways in which this principle can be tested: by searching for small deviations between proton and positron charges or by looking for particles with small electric charges, $\varepsilon \equiv q/e \ll 1$. 
In this work, we will follow the second approach, but first we briefly summarize the leading results on these two directions.
\begin{figure}[!b]
\begin{center}
\includegraphics[width=0.6\columnwidth]{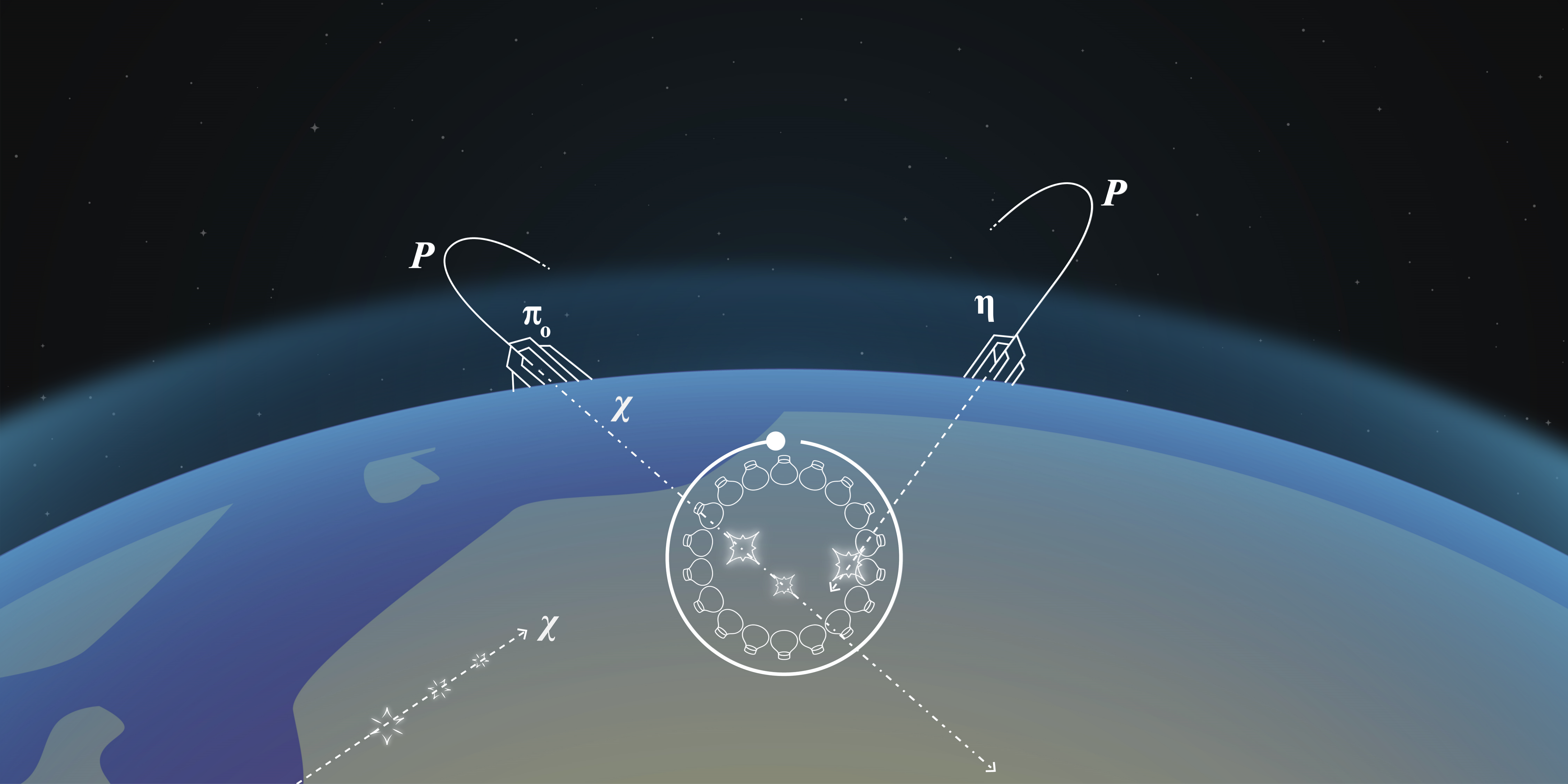} 
\caption{\textbf{\textit{Artistic rendition of this work.}} Millicharged particles, $\chi$, are produced in cosmic-ray showers from neutral meson decay.
They lose energy as they travel through the Earth, with only the highest energy ones reaching the detector preferentially from directions of small overburden.
These produce electromagnetic signatures in underground neutrino and dark matter detectors.
In this work, we study both single- and double-hit signatures and relate their sensitivities.}
\label{fig:MCP_Artisitic}
\end{center}
\end{figure}

The introduction of quarks~\cite{GellMann:1964nj} prompted the search for fractional charge particles between the 1960s and the early 1980s~\cite{Unnikrishnan_2004}, an effort diminished by the discovery of color confinement~\cite{Nishijima:1996ji}.
It was suggested in Ref.~\cite{GellMann:1964nj} that quarks, produced by cosmic-ray interactions in the Earth's surface~\cite{RevModPhys.49.717,LYONS1985225}, could be detected by examining the electrical neutrality of atoms and of bulk matter~\cite{doi:10.1146/annurev-nucl-121908-122035,Unnikrishnan_2004}.
This led to constraints on the number of free quarks per unit mass~\cite{Marinelli:1983nd} and constraints on the difference between proton and electron charges~\cite{osti_4211117}, $\eta = | q_p + q_{e^-} | / q_{e^-} $.
In addition to suggestions from fundamental particle physics, it was pointed out in Ref.~\cite{Lyttleton:1959zz} that the expansion of the universe could be accounted for by a small difference between the electron and proton charges at the level of $\eta \sim 10^{-18}$.
However, soon after this proposal, laboratory experiments using de-ionized gas constrained it to be $\eta < 2 \times 10^{-20}$~\cite{osti_4211117}.
Currently, constraints on this quantity using diverse methods --- such as gas efflux, acoustic resonators, Millikan drop style experiments, and atomic and neutron beams, among others~\cite{Unnikrishnan_2004} --- have limited $\eta$ to be less than $10^{-21}$~\cite{Zyla:2020zbs}.

Complementary to the above, direct searches for particles with small charge and sub-GeV mass have been motivated by dark matter models~\cite{Holdom:1985ag,FOOT_2004} and cosmological puzzles~\cite{doi:10.1146/annurev-nucl-121908-122035}.
In recent years, more attention has been given to these so-called millicharged particles (MCP) in the MeV-GeV mass regime.
Searches for this scenario have been proposed for beam-based neutrino experiments~\cite{Magill:2018tbb,Harnik:2019zee} and for dedicated experiments situated near accelerator complexes~\cite{Haas:2014dda,Yoo:2018lhk,Kelly:2018brz,Foroughi-Abari:2020qar,Kim:2021eix,Gorbunov:2021jog}.
Existing data have been reanalyzed in this context in Refs.~\cite{Magill:2018tbb,Marocco:2020dqu}.
Recently, the first dedicated neutrino-experiment analysis for MCP in this mass range was carried out by the ArgoNeuT collaboration~\cite{Acciarri:2019jly}, setting the strongest constraints for a range of MCP masses and demonstrating the capability of neutrino experiments for these searches for decades to come.

In tandem with accelerator-based searches, atmospheric-based production of MCP has been proposed~\cite{Plestid:2020kdm} (see also Refs.~\cite{Harnik:2020ugb,Pospelov:2020ktu} which considered even more distant production mechanisms that can contribute to these searches), with large-volume neutrino detectors like Super-Kamiokande serving as the best candidates to search for these particles.
Ref.~\cite{Plestid:2020kdm} demonstrated that atmospheric MCP searches can be as or more powerful than beam-based ones.
We build on this previous work, revisiting calculations of MCP production, discussing the uncertainty on the MCP flux, and proposing further searches that can be done with atmospheric MCP.
Motivated by Ref.~\cite{Harnik:2019zee}, we explore multiple-hit signatures in which a given MCP traversing a detector can scatter off two or more electrons, leaving a faint track.
This search is highly advantageous in detectors that can identify low-energy electrons, such as the upcoming multi-kiloton-scale unsegmented liquid-scintillator neutrino experiment JUNO~\cite{An:2015jdp,Abusleme:2021zrw}.

Fig.~\ref{fig:MCP_Artisitic} offers an artistic rendering of this work's main ideas.
High-energy SM particles, like protons, bombard the atmosphere, producing rich particle showers.
If MCP exist, then they can emerge in these showers and travel through the Earth.
Large-volume detectors provide excellent targets for these MCP, which can scatter once (right track), potentially imparting enough energy on the target electron to be a strong signal in these detectors.
If the MCP scatters multiple times (left track), the faint track it provides is difficult for background processes to mimic.
Some flux of MCP could also travel through the Earth (bottom track), if its mean free path through the Earth is long enough, and come upwards through the detector.

The remainder of this work is organized as follows: in Section~\ref{sec:production} we revisit previous simulations of atmospheric MCP production and discuss the tools we use for our simulation.
Section~\ref{sec:Propagation} discusses MCP propagation through the Earth, including energy loss and possible absorption, leading to an attenuation of the upward going MCP flux passing through a detector.
In Section~\ref{sec:ExperimentalSearches} we provide the details of our experimental simulations for both current and upcoming experiments and construct sensitivity estimates for both the single-scattering and multiple-scattering searches.
Section~\ref{sec:NeutrinoTelescopes} discusses some Monte Carlo techniques for searches in even larger detectors, such as IceCube and the upcoming IceCube Upgrade.
Finally, in Section~\ref{sec:Conclusions} we offer some concluding remarks.

\section{Millicharged Particle Production}
\label{sec:production}

A careful consideration of the production of millicharged particles in the upper atmosphere is crucial to this proposed search strategy. We provide the details of this approach with respect to its formalism in Section~\ref{subsec:ProductionFormalism}. Several sources of uncertainty are relevant as well, and these uncertainties, unfortunately, can plague any search for millicharged particles that relies on their production from the decays of neutral pseudoscalar/vector mesons. We discuss these uncertainties in Section~\ref{sec:production_uncertainties}.

We make a minimal assumption regarding the nature of this new millicharged particle --- that it has some mass and small coupling to the Standard Model photon via a small electric charge. It is possible that such an MeV/GeV-scale particle constitutes some fraction of the Dark Matter in the universe. If this is the case, a number of additional constraints apply. This has been explored as a potential solution to the EDGES anomaly~\cite{Bowman:2018yin} in, for instance, Refs.~\cite{Berlin:2018sjs,Kovetz:2018zan,Creque-Sarbinowski:2019mcm}. Various effects of millicharged particles as dark matter, leading to stringent constraints in the MeV-GeV mass range, are discussed in Refs.~\cite{Boehm:2013jpa,Vogel:2013raa,Chang:2018rso,Emken:2019tni,Harnik:2020ugb,Pospelov:2020ktu,Carney:2021irt}. If a millicharged particle is discovered via the atmospheric-production and scattering-detection approach we propose, then it is \textit{at most} a very small fraction of the relic dark matter in the universe. After such a discovery, a great deal of scrutiny must be applied to determine a consistent picture of this new particle with cosmological and astrophysical observations.

\subsection{Formalism}\label{subsec:ProductionFormalism}

Given that cosmic rays are composed mostly of protons, their collision with the upper layers of Earth's atmosphere mimics the setup encountered in a proton beam dump experiment, with nuclei in the air playing the role of the target.
An extensive cascade of radiation, ionized particles, and hadrons is generated with energies ranging from a few GeV up to dozens of EeV~\cite{Bird:1994uy}.
Among the mesons produced in the cascade, it is possible to find pseudoscalar mesons --- such as $\pi^{0}$, $\eta$ --- and vector mesons --- such as $\rho$, $\omega$, $\phi$ and $J/\Psi$.
If a millicharged particle exists with a mass below half of any given meson mass, then it can be produced in two- or three-body decays, replacing the final-state electrons in relatively common processes such as $\pi^0 \to \gamma e^+ e^-$ and $J/\Psi \to e^+ e^-$.

All of these mesons are unstable and have very short lifetimes and although none of them reaches the surface of the Earth, they can decay via a photon-mediated process to millicharged particles.
The MCP are assumed to be stable particles and can reach the surface of the Earth and propagate through it\footnote{Despite having charge, the interaction rate of MCP with the Earth is small enough that, for most of the parameter space of interest, they can travel through the bulk of the Earth without significant energy loss. We discuss this effect and how we include it in our simulation in Section~\ref{sec:Propagation}.} in such a way that they could be detected in underground detectors such as neutrino and dark matter experiments.

In this work, we will adopt a minimal model were the MCP ($\chi$) is described by a stable particle coupled to the photon with strength $\varepsilon \times e$ and mass $m_\chi$.
Taking this into account, the production profile of millicharged particles generated in air-showers from a parent meson $\mathfrak{m}$, can be described by the cascade equation~\cite{Gondolo:1995fq}
\begin{equation}
\frac{d\Phi_\chi}{dE_\chi\,d\cos\theta\,dX} = \int d E_\mathfrak{m} \,\frac{1}{\rho(X) \lambda_\mathfrak{m}(E_\mathfrak{m})} \frac{d \Phi_\mathfrak{m}}{ dE_{\mathfrak{m}} \,d\cos\theta}(E_\mathfrak{m},\cos\theta,X)\, \frac{d n}{d E_\chi}(\mathfrak{m} \rightarrow \chi;\, E_{\mathfrak{m}}, E_\chi),
\label{eq:production}
\end{equation}
where $\rho(X)$ is the atmospheric density at column depth $X$, $\lambda_{\mathfrak{m}} = \gamma_{\mathfrak{m}}\beta_{\mathfrak{m}}c\tau_{\mathfrak{m}}$ is the decay length of the parent meson $\mathfrak{m}$, and $\frac{d \Phi_\mathfrak{m}}{ dE_\mathfrak{m} \,d\cos\theta}$ is the production rate of the meson with energy $E_\mathfrak{m}$ at zenith angle $\theta$. Here, $\frac{d n}{d E_\chi}$ is the energy distribution of the millicharged particle in the decay, which can be written in terms of the branching fraction and the decay rate distribution as
\begin{equation}
\frac{dn}{d E_\chi}(\mathfrak{m} \rightarrow \chi) =  \mathrm{Br}(\mathfrak{m} \rightarrow \chi)\times \frac{1}{\Gamma(\mathfrak{m} \rightarrow \chi)} \frac{d \Gamma}{dE_\chi}(\mathfrak{m} \rightarrow \chi).
\label{eq:distribution}
\end{equation}
The branching fraction $\mathrm{Br}(\mathfrak{m} \rightarrow \chi)$ and decay width $\Gamma(\mathfrak{m} \rightarrow \chi)$ must be specified accordingly for the three-body decays of the pseudoscalar mesons and the two-body decay of the vector mesons.
The expressions for these quantities, as well as the kinematic integration limits in Eq.~\eqref{eq:production}, are specified in Appendix~\ref{app:decay}.
The production rate of the mesons in the atmospheric cascade can be solved numerically. We have used the Matrix Cascade Equation (\texttt{MCEq}) software package~\cite{Fedynitch:2015zma,Fedynitch:2012fs}, which includes several models for the cosmic ray spectrum, hadronic interactions, and atmospheric density profiles.
In this work, for our benchmark results we have used the \texttt{SYBILL-2.3} hadronic interaction model~\cite{Fedynitch:2018cbl}, the Hillas-Gaisser cosmic-ray model $H3a$~\cite{Gaisser:2011cc}, and the \texttt{NRLMSISE-00} atmospheric model~\cite{Picone:2002go} to obtain the production rate of mesons.

\begin{figure}
\begin{center}
\includegraphics[width=1.0\columnwidth]{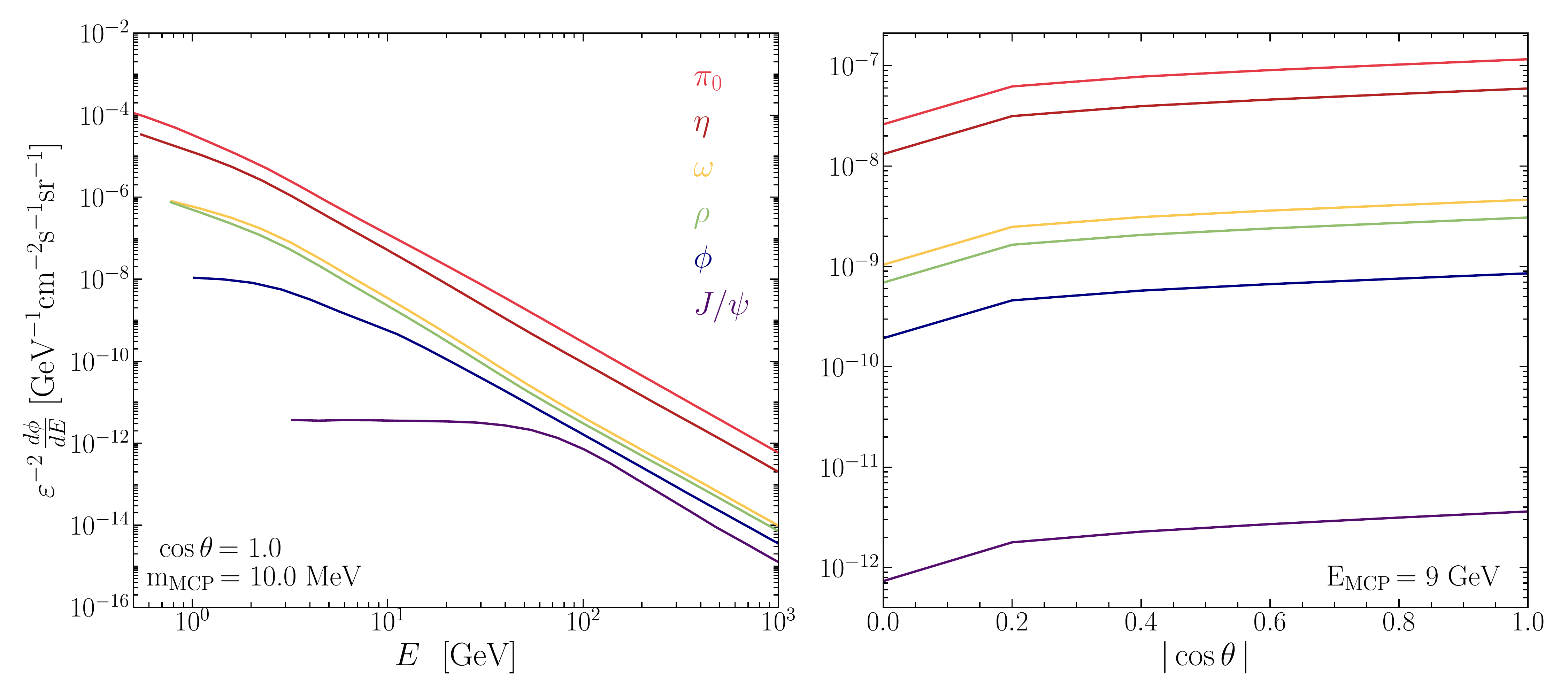} 
\caption{
\textbf{\textit{Flux of millicharged particles from different parent mesons.}}
Left: Spectra, which scales as $\varepsilon^2$, of millicharged particles arriving at the surface of the Earth with zenith angle $\ang{0}$, for each one of the different parent mesons.
Right: Angular dependence of the flux at $\SI{9}\GeV$.
In both panels the MCP mass is set to $\SI{10}\MeV$.
}
\label{fig:MCP_flux}
\end{center}
\end{figure}

Given that millicharged particles are stable, a direct integration of Eq.~\eqref{eq:production} in the column depth $X$ can be performed to find the differential energy spectra for a fixed zenith angle $\theta$.
In Fig.~\ref{fig:MCP_flux}, we show the energy and angular dependence of the MCP flux at the surface of the Earth, for each one of the parent mesons considered in this work.

\subsection{Uncertainties}
\label{sec:production_uncertainties}

The uncertainties in the production of MCPs are mostly due to the cosmic-ray model (CRM) chosen to generate the primary spectrum, as well as the hadronic interaction model (HIM) used to simulate the production rate of mesons in the atmospheric shower, the latter providing the dominant source of uncertainties, which grow as the energy increases.
Ref.~\cite{Kachelriess:2021lpm} recently explored these uncertainties in a fashion similar to how we have done here.

The composition and energy spectra of the primary cosmic radiation are characterized by a CRM that describes the primary spectrum with a power law that is ultimately fitted to air-shower data.
The power law spectrum may break or not depending on the origin of the cosmic ray and the energy range observed.
Additionally, the spectrum is typically characterized by the steepening that occurs for proton energies between $10^{6}$ and $10^{7}$ GeV, the so-called ``knee,'' and an extra feature around $10^{9}$ GeV called the ``ankle.''
The origin of these features remains unclear, and it is an active research topic (see, for instance, Refs.~\cite{Gaisser:2005tu,Gaisser:2006sf}).

\begin{figure}
\begin{center}
\includegraphics[width=1.0\columnwidth]{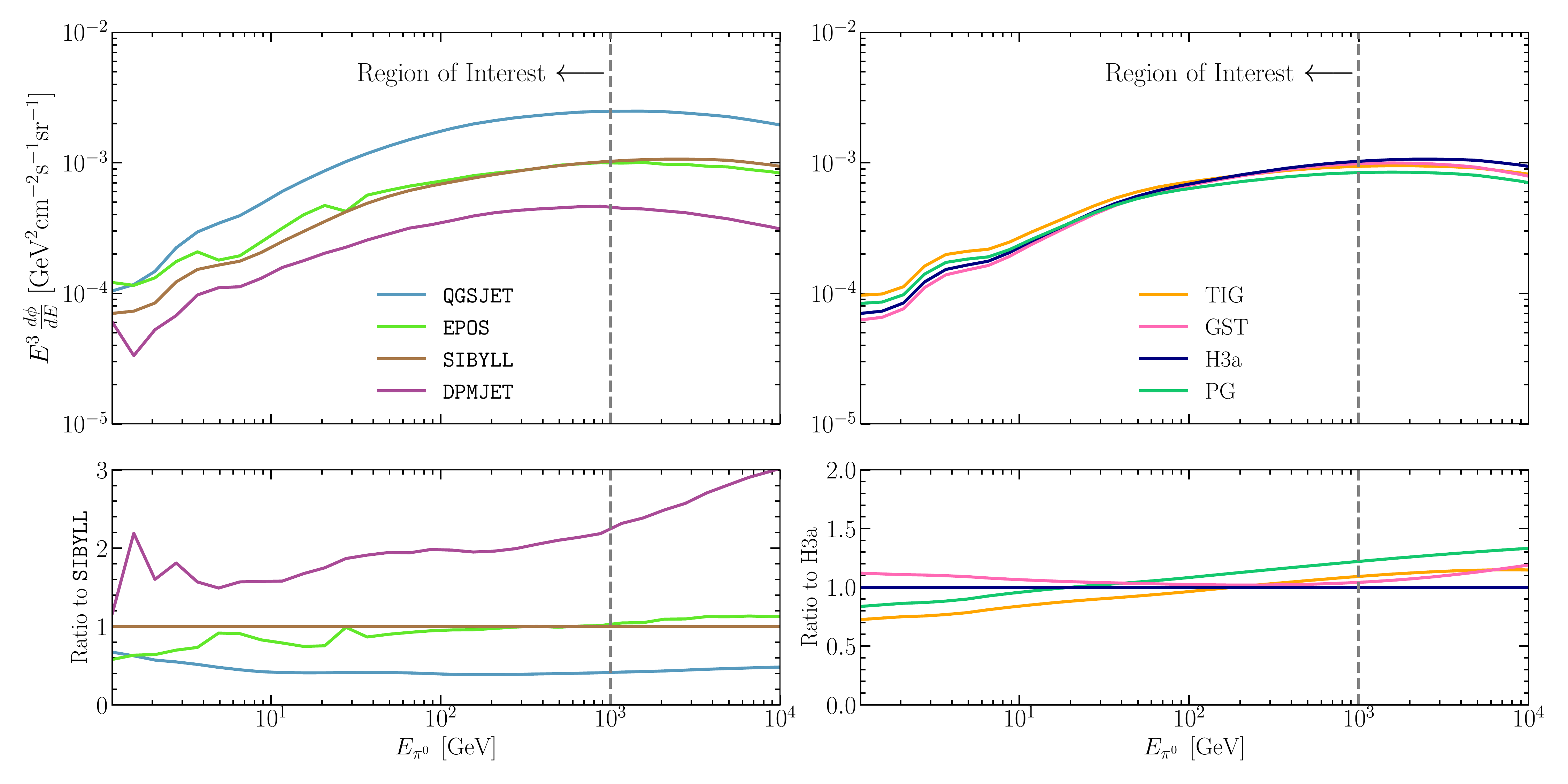} 
\caption{
\textbf{\textit{Production Rate of $\pi^{0}$ in cosmic showers and variations in different HIMs and CRMs.}}
Left: Comparison of different hadronic interaction models, including QGSJET (blue), EPOS (green), SIBYLL (brown), and DPMJET (purple). The bottom panel depicts the production ratio relative to SIBYLL.2.3.c, which we use for our production estimates.
Right: Comparison of production rates for different cosmic ray models: TIG (orange), GST (pink), H3a (blue), and PG (green). The bottom panel shows the ratio relative to H3a, which we use for our production estimates. See text for further discussion of CRMs and HIMs.
}
\label{fig:Flux_Uncert}
\end{center}
\end{figure}

As mentioned before, the benchmark CRM we use to compute the MCP flux is the $H3a$ model, which is widely used for calculations of atmospheric lepton fluxes~\cite{Fedynitch_2019,Aartsen_2016}.
However, there are other realistic CRMs which could be considered and that would yield to a more optimistic/pessimistic estimate of the MCP flux. 
To illustrate this point, we have considered the production rate of the $\pi^{0}$ meson in other CRMs, such as the Thunman-Ingelman-Gondolo model~\cite{Gondolo:1995fq}, the Gaisser-Stanev-Tilav model~\cite{gaisser2013cosmic}, and the poly–gonato model~\cite{H_randel_2003}.\footnote{This last model is however not applicable at energies above the ``knee.''} 
The comparison of the differential production rate for these other CRMs as well as the ratio with respect to our benchmark model $H3a$ as a function of the energy, can be seen in the right panel of Fig.~\ref{fig:Flux_Uncert}.

On the other hand, to model the interactions of a primary cosmic ray with the atmosphere, we need a suitable model of hadron-hadron, hadron-nucleus, and nucleus-nucleus collisions.
Yet, hadronic collisions at very high energies involve the production of particles with low transverse momenta, where present theoretical tools such as QCD are not enough to understand this feature. 
To address this problem, phenomenological models with Monte Carlo implementations are used.
Since a hadronic interaction model provides the interaction coefficients of the coupled cascade equation used to describe the production rate of mesons, we can directly prove its impact by estimating the production rate, for a fixed CRM. In this work, we choose \texttt{SYBILL-2.3c} as benchmark model, and the \texttt{QGSJET-II-02}~\cite{Ostapchenko_2011}, \texttt{DPMJET-III}~\cite{Roesler_2001}, and \texttt{EPOS-LHC}~\cite{Pierog_2009} HIMs for comparison. 
The left panel of Fig.~\ref{fig:Flux_Uncert} displays the production rate of $\pi^{0}$ for all of these models as well as the ratio to our benchmark HIM.

To estimate the impact of the benchmark models used in this work, we evaluate the ratio of total production with respect to a different CRM/HIM, defined by
\begin{equation}
\Delta_{j}(\mathfrak{m}) = \frac{\int_{E_{\rm min}}^{\Lambda} dE\, \frac{d\phi_{\textrm{BM}}}{dE}}{\int_{E_{\rm min}}^{\Lambda} dE\, \frac{d\phi_{j}}{dE}} ,
\label{eq:uncertain}
\end{equation}
where $\mathfrak{m}$ is the meson of interest, $j$ the index used to denote the model that is being compared with the benchmark model $\textrm{BM}$, $E_{\rm min}$ the minimum energy available in \texttt{MCEq}, and $\Lambda = 10^{3}\si\GeV$ is the upper energy cut that we use in order to obtain the MCP flux from a parent meson.
Table~\ref{tab:uncert_val} shows the $\Delta$ coefficients for the different CRMs and HIMs for $\pi^{0}$, $\eta$, and $\phi$.
A similar result can be found for the other mesons, except for $J/\Psi$, in which case there are not enough statistics on the meson production rate to evaluate the uncertainty properly.
As can be seen from this result, the biggest source of uncertainty comes from the hadronic interaction models which would induce a difference in the MCP flux from about $\sim16\%$ up to $\sim68\%$, depending on the parent meson.

Even though all of the event generators considered here are updated with LHC data, the cross-section measurements for hadron production have rather large uncertainties~\cite{AguilarBenitez:1991yy}.
On the other hand, we must also take into account the differences encountered in the phenomenological models that arise from the treatment of inelastic hadronic collisions within the framework of Reggeon Field Theory~\cite{Gribov:1968fc}.
We stress that HIM uncertainties are present in most if not all modern MCP searches, as any beam-based\footnote{A notable exception is the SLAC mQ experiment~\cite{Prinz:2001qz}, which considered production of MCP in an electron beam dump. Such production mechanisms would be subject to far smaller uncertainties.} search needs to simulate hadronic interactions at some level.    

\begin{table}[t]
\centering
\begin{tabular}{c||c|c| l} 
 \hline
  \hline
 Model  &  $\Delta(\pi^{0})$&  $\Delta(\eta)$&  $\Delta(\phi)$  \\ 
\hline
 TIG  & 0.754& 0.766& 0.827 \\ 
 PG & 0.868& 0.88& 0.944 \\
 GST & 1.106& 1.101& 1.072 \\
 \hline
 \texttt{QGSJET} & 0.570& 1.160& ------ \\
 \texttt{DPMJET} & 1.586& 1.677& 1.541 \\
 \texttt{EPOS} & 0.664& 0.707& 0.750 \\
 \hline
  \hline
\end{tabular}
\caption{\textbf{\textit{Comparison of meson production rate using different hadronic interaction models.}} Numerical values of $\Delta$, the relative integrated production rate of a given meson, for the various cosmic ray models (top three rows) and hadronic interaction models (bottom three rows) considered in this work.}
\label{tab:uncert_val}
\end{table}

\section{Propagation through Earth and Energy Loss}
\label{sec:Propagation}

As we will show in Section~\ref{sec:ExperimentalSearches}, underground neutrino detectors provide an excellent opportunity to search for MCPs produced in cosmic-ray air showers.
However, the detection of MCPs requires detailed knowledge about charged particle propagation in the medium, since these experiments uses the Earth's crust to shield from atmospheric muons, which usually constitutes the main source of background.
Because of this, it is expected that the MCPs that arrive at the detector lose part of the energy they had when reaching the Earth's surface.
Just as with any other charged particle, the MCP would lose energy by ionizing the medium through which they propagate and by interacting with nuclei.
The average energy loss along the MCP trajectory can be parametrized by
\begin{equation}
-\frac{dE}{dX} = \varepsilon^{2} \left( a_{\rm ion.} + b_{\rm el.-brem.} \varepsilon^{2} E + b_{\rm inel.-brem.} E + b_{\rm pair} E + b_{\rm photo-had.} E \right ) \approx  \varepsilon^{2} \left( a + b E \right ),
\label{eq:Eloss}
\end{equation}
where $dX$ is the column density traversed, $\varepsilon$ is the MCP coupling, and $a_{\rm x}$ and $b_{\rm x}$ are various categories of energy loss with (potentially) different scaling with the MCP energy $E$ and $\varepsilon$. We simplify this expression, adopting the right-hand side of Eq.~\eqref{eq:Eloss} in our simulations, where $a$ and $b$ are the energy loss parameters given in units of [GeV/mwe] and [mwe] respectively (``mwe'' being ``meters of water equivalent'').
On the other hand, we can estimate the overburden length traversed by an MCP approaching a detector located at depth $d$ and along a trajectory with angle $\theta$ by
\begin{equation}
D = \sqrt{(R_{\oplus}-d)^{2}\cos^{2}{\theta} + d(2R_{\oplus}-d)} - (R_{\oplus}-d)\cos{\theta},
\label{eq:overburden}
\end{equation}
with $R_{\oplus}$ being the Earth's radius.
 
The probability that a millicharged particle with energy at the surface $E_{i}$ arrives at the detector with an energy $E_{f}$ will depend on the coupling $\alpha_{em}\varepsilon$ and the incident direction $\cos\theta$.
This probability is given by:
\begin{equation}
\mathcal{P} = \exp{\left(-\frac{D(\cos{\theta})}{R(E_{i},\varepsilon,E_{f})}\right)},
\label{eq:Parrive}
\end{equation}
where the average distance $R(E_{i},\varepsilon,E_{f})$ can be obtained from Eq.~\eqref{eq:Eloss}, and is given by
\begin{eqnarray}
R =\frac{1}{\varepsilon^{2} b} \, \ln\left( \frac{1+\frac{a}{b}E_{i}}{1+\frac{a}{b}E_{f}} \right).
\label{eq:AvgDist}
\end{eqnarray} 

\begin{figure}
\begin{center}
\includegraphics[width=1.0\columnwidth]{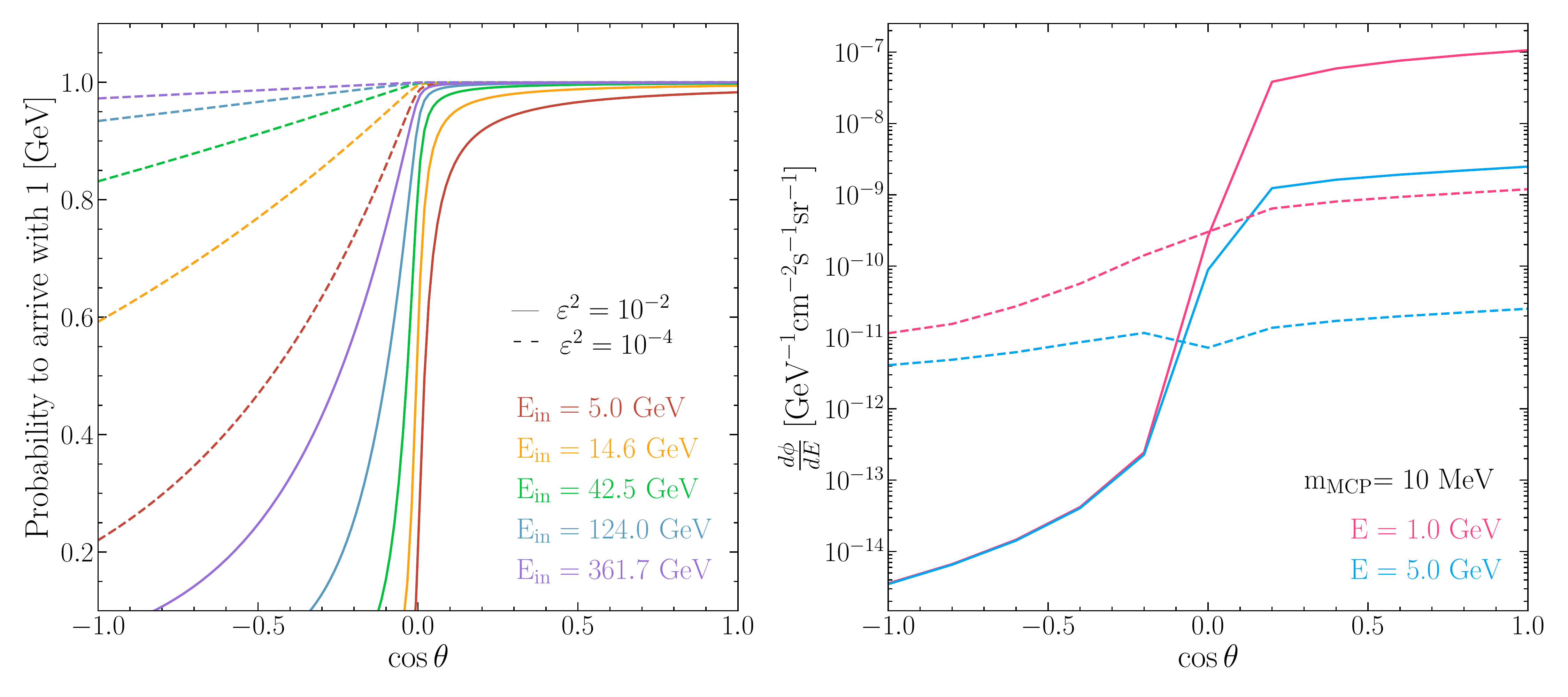} 
\caption{
\textbf{\textit{Survival probability of MCPs and expected angular distribution.}} 
Left: Probability that an MCP with initial energies as indicated by the labels arrives at the detector with a total energy of $\SI{1}\GeV$ as a function of the cosine of the zenith angle.
Solid lines are shown for $ \varepsilon = 10^{-2}$ and dashed lines for $ \varepsilon = 10^{-4}$.
Right: Flux at detector as a function of cosine of the zenith angle for a $\SI{10}\MeV$ mass MCP, arriving with an energy of $\SI{1}\GeV$ (pink) and $\SI{5}\GeV$ (blue).
Solid and dashed lines corresponds to the same values of $\varepsilon$ as in the left panel.
}
\label{fig:P_arrival}
\end{center}
\end{figure}

We have taken the energy loss parameters for a standard rock shielding with $a = \SI{0.223}\GeVmwe$ and $b= 4.64\times 10^{-4} \si\mwe$ as reported in Ref.~\cite{Koehne:2013gpa}.
The left panel of Fig.~\ref{fig:P_arrival} shows the probability that an MCP arrives at a detector at a depth of $\SI{1.5}\km$ with $E_\chi > \SI{1}\GeV$ for two different values of the coupling $\varepsilon$, with a variety of different initial energies as depicted in the label.
The flux at the detector can be obtained by convolving the flux at the surface with the survival probability $\mathcal{P}$.
The right panel of Fig.~\ref{fig:P_arrival} shows the angular distribution of the flux at detector for an MCP with a mass of $\SI{10}\MeV$ arriving at the detector with energies of 1.0 and $\SI{5.0}\GeV$. 
 
We note here that for relatively large $\varepsilon^2 = 10^{-2}$, the upward-going flux, $-1 \leq \cos\theta \leq 0$ is highly attenuated.
On the other hand, for smaller $\varepsilon^2 = 10^{-4}$, because energy losses are suppressed by two orders of magnitude, the flux is mostly independent of $\cos\theta$.
For experiments that can be sensitive to such small millicharges, the effective area of the sky that the detector can search will nearly double.
Moreover, searches for upward-going events can assist in reduction of background events, for instance from atmospheric muons entering the detector from above.

\section{Current and Future Experimental Searches}
\label{sec:ExperimentalSearches}

To date, the most stringent searches for millicharged particles in the ${\sim}$MeV--GeV mass range have used antropogenic beams, where the MCP are produced in either the collision of two beams (collider searches) or when a beam impacts a target (beam-dump searches).
Colliders place the strongest constraint on MCPs with masses above ${\sim}\SI{1}\GeV$, primarily through a combination of direct searches at LEP and measurements of the invisible width of the $Z$ boson~\cite{Davidson:2000hf}.
The LHC can probe even larger, ${\gtrsim}\SI{30}\GeV$, masses~\cite{Jaeckel:2012yz}.
Below $\SI{100}\MeV$, the strongest constraint is from the SLAC mQ electron fixed-target experiment~\cite{Prinz:2001qz}.
Future dedicated experiments have been proposed in the context of both collider~\cite{Haas:2014dda,Foroughi-Abari:2020qar}\footnote{Including prototype results found in Refs.~\cite{Ball:2020dnx,Ball:2021qrn}.} and fixed-target~\cite{Kelly:2018brz,Kim:2021eix,Gorbunov:2021jog} environments, reaching sensitivity to $\varepsilon^2 \gtrsim 10^{-6}$ and a wide range of masses, up to nearly $\SI{5}\GeV$.

Recently, searches for MCP in neutrino experiments have garnered attention.
Sensitivities for accelerator production have been estimated in Refs.~\cite{Magill:2018tbb,Harnik:2019zee} and carried out by the ArgoNeuT collaboration in Ref.~\cite{Acciarri:2019jly}.
The latter has set some of the strongest constraints over a wide MCP range.
Atmospheric production has been considered in a similar fashion to this work for the Super-Kamiokande detector and its future successor Hyper-Kamiokande~\cite{Abe:2018uyc} in Ref.~\cite{Plestid:2020kdm}, as have limits from particles produced in the interstellar medium and from Earth relics recently discussed in Refs.~\cite{Harnik:2020ugb} and~\cite{Pospelov:2020ktu}, respectively.

In the remainder of this section, we will discuss and derive constraints for MCP masses between $\SI{10}\MeV$ and $\SI{1.5}\GeV$.
Our results for current experiments, particularly Super-Kamiokande, demonstrate an improvement on current constraints for some masses, in agreement with the results of Ref.~\cite{Plestid:2020kdm}. We divide the search strategy into two different types of analyses, those relying on single-hit signals (Section~\ref{subsec:SingleHit}) and those that rely on multiple energy depositions from a single MCP particle traversing the detector (Section~\ref{subsec:MultiHit}).
The former is advantageous for large-volume, high-energy-threshold experiments (\textit{e.g.}, water Cherenkov detectors like Super-Kamiokande), whereas the latter is strongest in scintillator experiments with precision timing and low-energy thresholds (for instance, JUNO).
We find that these multiple-scattering searches offer great potential for discovering millicharged particles in the $\SI{10}\MeV$ to $\SI{1.5}\GeV$ mass range and warrant additional focus in the coming decade.

\subsection{Single-Scattering Searches}
\label{subsec:SingleHit}

Via the same coupling $\varepsilon$ that allows for the production of MCP in the upper atmosphere, the particles that reach a detector are capable of scattering (via a $t$-channel photon exchange) off detector materials.
We will consider scattering off electrons for the remainder of this work.
This massless-mediator scattering yields a differential cross section that peaks for small electron recoil energy, and so detectors with capabilities of identifying and reconstructing low-energy electrons will be advantageous in this endeavor.
However, as we consider electrons of lower and lower energy, more and more backgrounds become relevant.
In the following subsections, we will discuss the characteristics of the signal events, the various backgrounds, and the experimental limits we are able to derive in this single-hit analysis including the statistical techniques employed.

\subsubsection{Signal}

Once the flux of MCPs arrives at the detector, the millicharged particles can interact with the detector medium by scattering off electrons.
We will be interested in low-energy recoils, as those are more frequent due to the shape of the differential cross section.
The differential cross section between MCPs and electrons is given by~\cite{Magill:2018tbb}
\begin{equation}
\frac{d\sigma}{dE_{r}} =  \varepsilon^{2}\alpha_{EM}^{2}\pi \frac{m_{e}(E_{r}^{2} + 2E_{\chi}^{2}) - E_{r}( m_{e}(2E_{\chi} + m_{e})+m_{\chi}^2 ) )}{E_{r}^{2}m_{e}^2(E_{\chi}^{2} - m_{\chi}^{2}) },
\end{equation}
where $E_{\chi}$ is the energy of the incoming MCP, $E_{r}$ is the recoil energy (assuming an initial stationary electron in the laboratory frame), $\alpha_{EM}$ is the electromagnetic coupling constant, and $m_{\chi}$ and $m_e$ are masses of the MCP and the electron, respectively.
From the equation above, it is easy to verify that in the ultrarelativistic limit $E_{\chi} \gg E_{r}, m_{\chi},m_{e}$, the differential cross section scales like $E_{r}^{-2}$, and therefore as the threshold of a given experiment becomes lower, the signal becomes much stronger.

\begin{figure}
\begin{center}
\includegraphics[width=1.0\columnwidth]{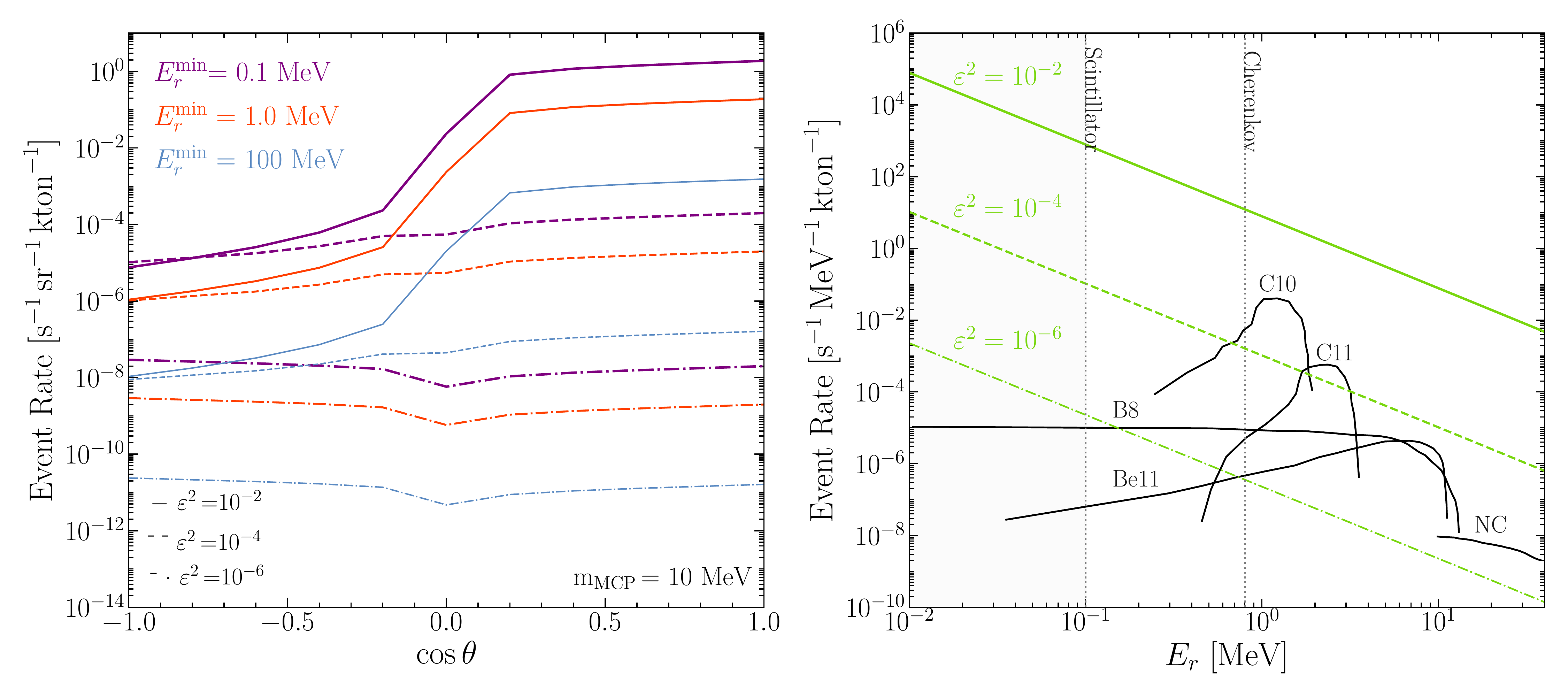} 
\caption{
\textbf{\textit{MCP event rate angular and energy distributions as well as relevant backgrounds for neutrino detectors.}}
Left: Expected event rate as a function of the incoming direction of the MCPs.
Solid, dashed, and dot-dashed lines correspond to different values of $\varepsilon^2$ as depicted in the legend. 
The colors indicate the three different minimum electron recoil energies with values shown in the legend.
Right: Expected event rate as a function of the electron recoil energy for MCPs coming in all directions.
The vertical dotted lines indicate the different thresholds for scintillator and Cherenkov detectors.
The solid black lines indicate the background expected in JUNO from different components (see background section for details).
Both figures are shown for an incident millicharged particle with a mass of $\SI{10}\MeV$.
}
\label{fig:event_rate}
\end{center}
\end{figure}

The event rate from MCPs can be obtained by a convolution of the millicharged particle flux with the differential cross section multiplied by the number of targets and the detection efficiency.
This is given by 
\begin{equation}
\frac{dN}{dE_{r}\, d\cos\theta} = n_e \epsilon(E_r)\int dE_{\chi} \, \frac{d\phi}{dE_{\chi}\, d\cos\theta} \frac{d\sigma}{dE_{r}},
\label{eq:diff_rate}
\end{equation}
where $n_e$ is the total number of electrons in the detector and $\epsilon(E_r)$ is the (detector-dependent) reconstruction efficiency of these single-electron events. 
The event rate distribution for a $\SI{1}\kton$ effective mass detector is shown in Fig.~\ref{fig:event_rate}, as a function of the incoming direction from zenith angle and the recoil energy, for different values of $\varepsilon^{2}$ and an MCP mass of $\SI{10}\MeV$.
Notice that for the angular distribution an integration in the recoil energy is needed, and the values in the label indicate different detection thresholds.
The MCP event rate shown in this figure is a general result, which can be applied to any underground particle detector located at a depth of $\sim \SI{1.5}\km$, while the mass dependence scales in a similar way as the MCP flux, with higher event rates at lower masses, where the millicharged particle receives contributions from most of the mesons.

%%%%%%%%%%%%%%%%%%%%%%%%%%%%%%%%%%%%%%%%%%%%%%%%%%%%%%%%%%%%%%%%%%%%%%%%%%%%%%%%%%%%%%%%%%%%%%%%
%%%%%%%%%%%%%%%%%%%%%%%%%%%%%%%%%%%%%%%%%%%%%%%%%%%%%%%%%%%%%%%%%%%%%%%%%%%%%%%%%%%%%%%%%%%%%%%%

\begin{figure}
    \centering
    \includegraphics[width=0.45\linewidth]{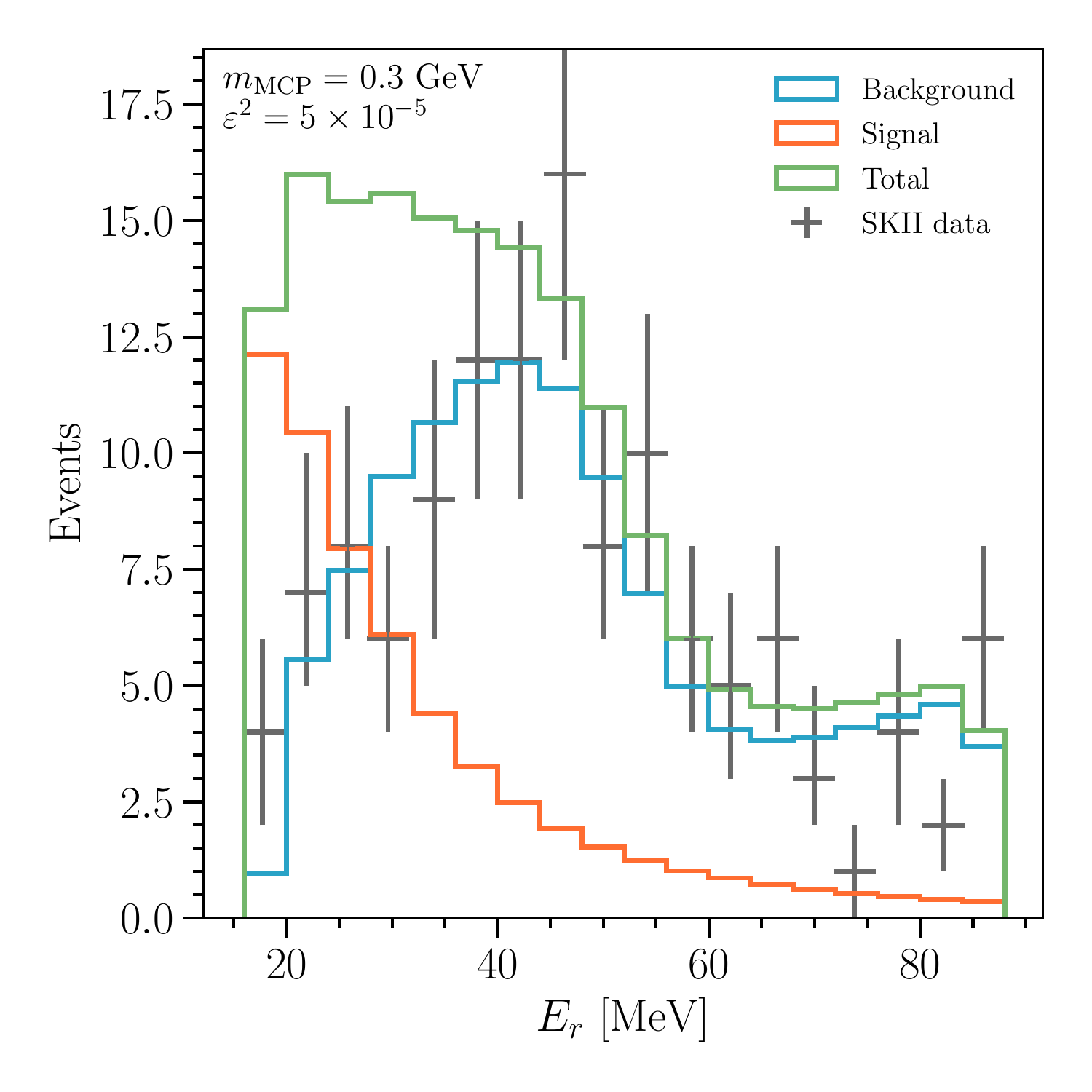}
    \includegraphics[width=0.45\linewidth]{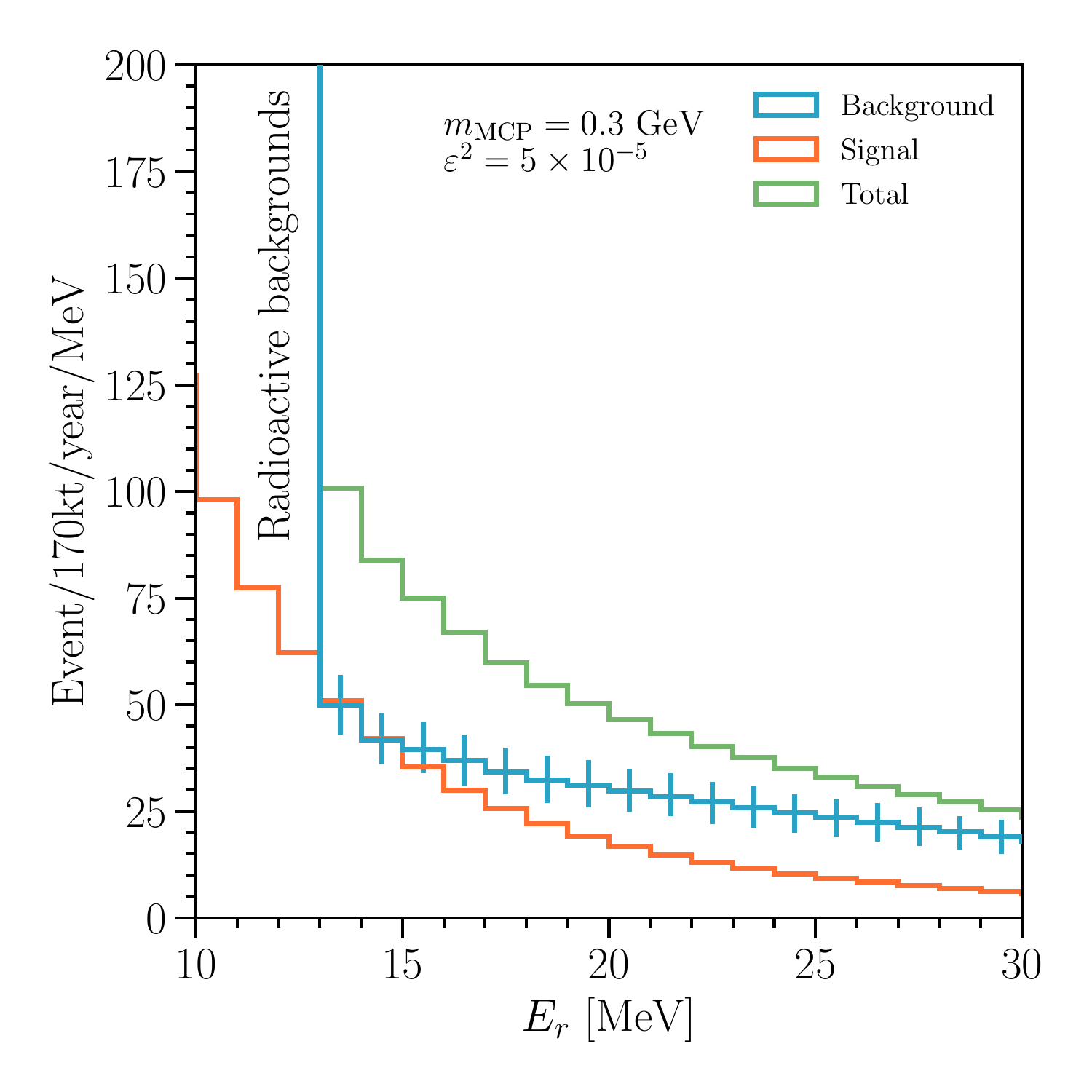}
    \caption{
    \textbf{\textit{Event distribution illustrations for Super-Kamiokande and JUNO.}}
    Left: Electron events measured in Super-Kamiokande are shown as crosses, total backgrounds are shown as a blue histogram, and a potential MCP hypothesis is shown in orange.
    Right: Expected distribution of MCP events in JUNO overlaid with backgrounds.
    }
    \label{fig:signatures}
\end{figure}

\subsubsection{Backgrounds\label{sec:backgrounds}}
Here we discuss the different sources of background events that contribute to such single-scattering millicharged particle searches in water Cherenkov or liquid scintillator detectors.
The expected background rates (summed over all contributions) in these two environments, along with a signal expectation, are shown in Fig.~\ref{fig:signatures} for Super-Kamiokande phase II (left) and JUNO (right). More details on the Super-Kamiokande event distributions are given in Appendix~\ref{app:SK}.

\paragraph{Penetrating muons:}
Our atmosphere is filled with muons generated from the decays of charged mesons that are dominated by pions at the lowest energies and kaons at higher energies~\cite{gaisser2016cosmic}.
These muons are highly boosted, and even though they have a short lifetime ($\sim 2 \times 10^{-6}\si\s$), they can easily reach the surface, propagate through the Earth and leave a signal in the detector.
The most efficient way to suppress this and other possible cosmogenic backgrounds is to have a sufficient amount of overburden over the location of the detector.
This background may be rejected by measuring the opening angle of the Cherenkov cone of the candidate electron in the event~\cite{Bays:2011si}.

\paragraph{Neutral-current neutrino events:}
Atmospheric neutrinos induce a source of background due to neutral-current interactions with nuclei in the medium.
In the case of liquid scintillator detectors, the background is induced by interactions with $^{12}$C whose de-excitation produces radiation.
In Cherenkov detectors, elastic, neutral-current scattering of neutrinos off the target nuclei can lead to small energy depositions with a similar spectrum to the millicharged particle signature.
However, this is a relatively small component of the background in Cherenkov detectors like Super-Kamiokande.

\paragraph{Charged-current neutrino events:}
Specifically in water or ice Cherenkov detectors (\textit{i.e.}, Super-Kamiokande, Hyper-Kamiokande, or IceCube), both $\nu_e/\overline{\nu}_e$ and $\nu_\mu/\overline{\nu}_\mu$ charged-current scattering events from the atmospheric neutrino flux may mimic our desired signal of a low-energy electron.
The former directly creates low-energy electrons indistinguishable from the signal as long as the hadronic particles produced in the interaction are below the Cherenkov threshold.
Despite the fact that they cannot be separated on an event-by-event basis, the recoil energy distribution of the electrons has a spectral shape different than the MCP signal, which allows them to be distinguished statistically.
Similarly, $\nu_\mu$ events can mimic the background if the outgoing $\mu^\pm$ is below the Cherenkov threshold, but the Michel $e^\pm$ coming from the $\mu^\pm$ decay are visible.
As in the previous case, the Michel electron energy distribution is distinct and can be disentangled statistically; see Ref.~\cite{Bays:2011si} for further details.

\paragraph{Radioactive and anthropogenic backgrounds:}
Radioactive backgrounds are produced by the materials in and around the detector.
These can produce electrons with energies similar to those struck by MCPs, \textit{e.g.}, in radioactive beta decays of nuclei, and depend on the experimental setup.
For example, the dominant radioactive backgrounds in JUNO, in the energy range relevant for this analysis, are $^{8}$B, $^{10}$C, $^{11}$C, and $^{11}$Be, which we reproduce from Ref.~\cite{An:2015jdp} and include as background in the right right of Fig.~\ref{fig:signatures}.
We note that this rate is significantly larger than currently allowed MCP signatures for $E_e \lesssim \SI{12}\MeV$ and will be the limiting factor for single-scattering searches in detectors like JUNO.
Searches for signals with an associated neutron in the final state, most prominently scattering of $\overline{\nu}_e$ from the diffuse supernova neutrino background, can reject these backgrounds by requiring coincidence with the signal of neutron capture.
Since MCP signatures do not produce an outgoing neutron, but only the electron recoil, this method cannot be used to reduce the background.
However, as we will discuss in Sec.~\ref{subsec:MultiHit}, multiple-hit signatures can be used to reduce radioactive backgrounds. The dominant anthropogenic backgrounds in the search of MCPs are produced by either nearby nuclear reactors or by accelerator facilities.
In JUNO, the antineutrinos produced by nuclear reactors are the source used to study neutrino oscillations and can be separated from the MCP signature by the presence of the coincident neutron capture mentioned above.
In Super-Kamiokande, accelerator neutrino events produced in the Tokai accelerator facility can be removed by searching for MCPs during off-beam times.

\subsubsection{Experimental setups considered}\label{subsubsec:ExpSetups}

In this section, we will briefly describe the experiments considered in this work and the constraints that can be established from the lack of significant MCP signal.
We will derive our constraints by performing a forward-folding binned-likelihood analysis, where the data and expectations are organized in equally sized bins of electron recoil energy.
To construct our test statistic, we compute, for each one of the experiments considered, the number of signal events expected in a given bin of electron recoil energy.
The number of signal events in reconstructed electron recoil energy is given by
\begin{equation}
\mu^s_{i} = T \epsilon(E_{i}) \int_{E_{i}-E_{b}/2}^{E_{i}+E_{b}/2} dE_{r} \int_{-1}^{+1} d\cos\theta \,  \frac{dN}{dE_{r}\, d\cos\theta},
\end{equation}
where the symbols inside the inner most integral are defined in Eq.~\eqref{eq:diff_rate}, $T$ is the time exposure, $E_{b}$ is the bin size, $E_i$ is the bin center, and $\epsilon(E_{i})$ is the average detection efficiency in a given bin.

To compute current experimental constraints, we construct a background-agnostic test statistic, comparing the observed data with that expected from an MCP with mass $m_\chi$ and mixing $\varepsilon$, ignoring any potential background contribution.
This procedure will always result in a relatively conservative upper limit on the parameter space of MCP and cannot allow for a potential preferred region.
We adopt this strategy for the current Super-Kamiokanade and XENON1T experimental results for different reasons, which we will detail in their respective paragraphs to follow.

The background-agnostic test statistic is calculated using the bin-by-bin likelihood function~\cite{Arguelles:2019ouk}
\begin{equation}\label{eq:BALikelihood}
\mathcal{L}_i =
\begin{cases}
P(d_i|\mu^s_i) & d_i < \mu^s_i \\ 1 & d_i \geq \mu^s_i
\end{cases},
\end{equation}
where $d_i$ and $\mu^s_i$ are the data and expected signal (given $m_\chi$ and $\varepsilon$) in bin $i$, respectively, and $P(d_i|\mu^s_i)$ is the Poissonian likelihood of observing $d_i$ given expectation $\mu^s_i$.
This form of the likelihood is both background-agnostic and one-sided, guaranteeing the setting of a constraint instead of a preferred region.
Our test statistic then is
\begin{equation}\label{eq:TS}
    \mathcal{TS} = -2\log{\left(\frac{\prod_{i} \mathcal{L}_i (m_\chi, \varepsilon)}{\prod_{i} \mathcal{L}_i (\varepsilon=0)}\right)},
\end{equation}
where the denominator is the signal-free likelihood function and will return $1$ when using the background-agnostic, one-sided likelihood of Eq.~\eqref{eq:BALikelihood}.

When deriving a constraint, we will assume that Wilks' theorem holds and set limits assuming we have two degrees of freedom, $m_\chi$ and $\varepsilon$.
In reality, the signal event distributions are all of the same shape (peaking at low electron recoil energy), where $\varepsilon$ scales with the rate and $m_\chi$ determines which mesons can contribute to the MCP flux, also impacting the overall normalization.
This relation implies that these two parameters could, in principle, be viewed as a single degree of freedom.
In light of this, our use of two degrees of freedom, combined with the above background-agnostic, one-sided likelihood function approach, should be viewed as conservative.

\paragraph{Super-K:} Our analysis uses the event selection in Ref.~\cite{Bays:2011si}, which was designed to search for diffuse supernova background neutrinos.
This event selection was previously discussed in Ref.~\cite{Plestid:2020kdm}, where they estimate event rates of MCPs produced in the atmosphere.
We use the data from SK-I, SK-II, and SK-III and perform a joint likelihood analysis combining the three phases.
For our analysis, we include the signal efficiencies provided in Ref.~\cite{Bays:2011si}.
The background expectations estimated by the collaboration can be extracted from the same reference.
We demonstrate the expected event rate from the collaboration's background as well as the observed data during SK-II operation in Fig.~\ref{fig:signatures} (left) as a function of electron recoil energy.
However, it must be stated that this background expectation does not include the potential signal from diffuse supernova neutrino background (DSNB) events, the signature of interest in Ref.~\cite{Bays:2011si}.
If we perform a likelihood analysis\footnote{Our likelihood analysis for Super-Kamiokande includes only statistical uncertainties. We have performed a version of our analysis incorporating uncorrelated bin-by-bin normalization uncertainties of 5\% (consistent with the systematic uncertainties discussed in Ref.~\cite{Bays:2011si}) and find no realizable differences in the constraints we obtain.} comparing the observed data with the expected background (with no DSNB contribution) plus the MCP signal, we obtain a moderate (${\sim}90\%$ CL) preference for the existence of MCP; however, this signal is degenerate with the DSNB one.
So, for this reason, we choose to adopt the conservative, background-agnostic approach discussed above to derive an upper limit on $\varepsilon^2$ as a function of $m_\chi$ at 90\% CL.

\paragraph{XENON1T:} Dark matter direct-detection experiments can measure electron recoil energies down to a few keV.
We use the recent XENON1T experiment~\cite{Aprile:2020tmw} result to search for signatures of millicharged particles produced in the Earth's atmosphere.
The background models presented in Ref.~\cite{Aprile:2020tmw} are notably insufficient to explain the observed rate of electron recoil events.
However, one potential background to explain the excess is the presence of an unconstrained tritium beta-decay background.
Because of this possibility, we choose to adopt the background-agnostic approach as we did for Super-Kamiokande.
If we calculate the likelihood using the given $B_0$ background model of Ref.~\cite{Aprile:2020tmw}, we observe a preferred region of parameter space at over 95\% CL. 
This preferred region of parameter space is nearly excluded completely by other constraints at $2\sigma$ confidence and will be robustly tested by JUNO.
See Appendix~\ref{app:XENON} for more details, including the preferred region of parameter space that we obtain.
When using Eq.~\eqref{eq:diff_rate}, we have assumed that all of the electrons in a detector are unbound -- this is not the case in XENON1T, especially when comparing the observed recoil energy distribution with the nuclear binding energies of interest.
In that context, our signal rates predicted in XENON1T will be overpredicted by a factor of several -- see, for instance, Refs.~\cite{Bloch:2020uzh,Harnik:2020ugb} for further discussion of this effect.
Given the scaling of our signal rate with $\varepsilon^4$ and the fact that XENON1T is not the most powerful experiment considered in this work, we disregard this effect in estimating our constraints.

\paragraph{JUNO:} Given that JUNO offers a future search for single-scattering events, we calculate our test statistic assuming that the collected data in each bin is consistent with the background expectation and perform a comparison with the expected signal-plus-background distributions.
See the right panel of Fig.~\ref{fig:signatures} for these backgrounds from Ref.~\cite{An:2015jdp} as well as a signal expectation.
We then calculate our bin-by-bin likelihood using
\begin{equation}
    \mathcal{L}_i = P(\mu^b_i|\mu^b_i + \mu^s_i),
\end{equation}
where $\mu^b_i$ ($\mu^s_i$) is the expected background (signal) in bin $i$. 
With this form of the binned likelihood, we then use the same expression for the test statistic $\mathcal{TS}$ in Eq.~\eqref{eq:TS} above and again, assume two degrees of freedom when projecting JUNO sensitivity at 90\% CL. 
For our analysis, we assume an exposure of $\SI{170}\kty$, which corresponds to approximately ten years of data collection.

\subsubsection{Single-scattering constraints and sensitivity projections}

\begin{figure}
\begin{center}
\includegraphics[width=0.8\columnwidth]{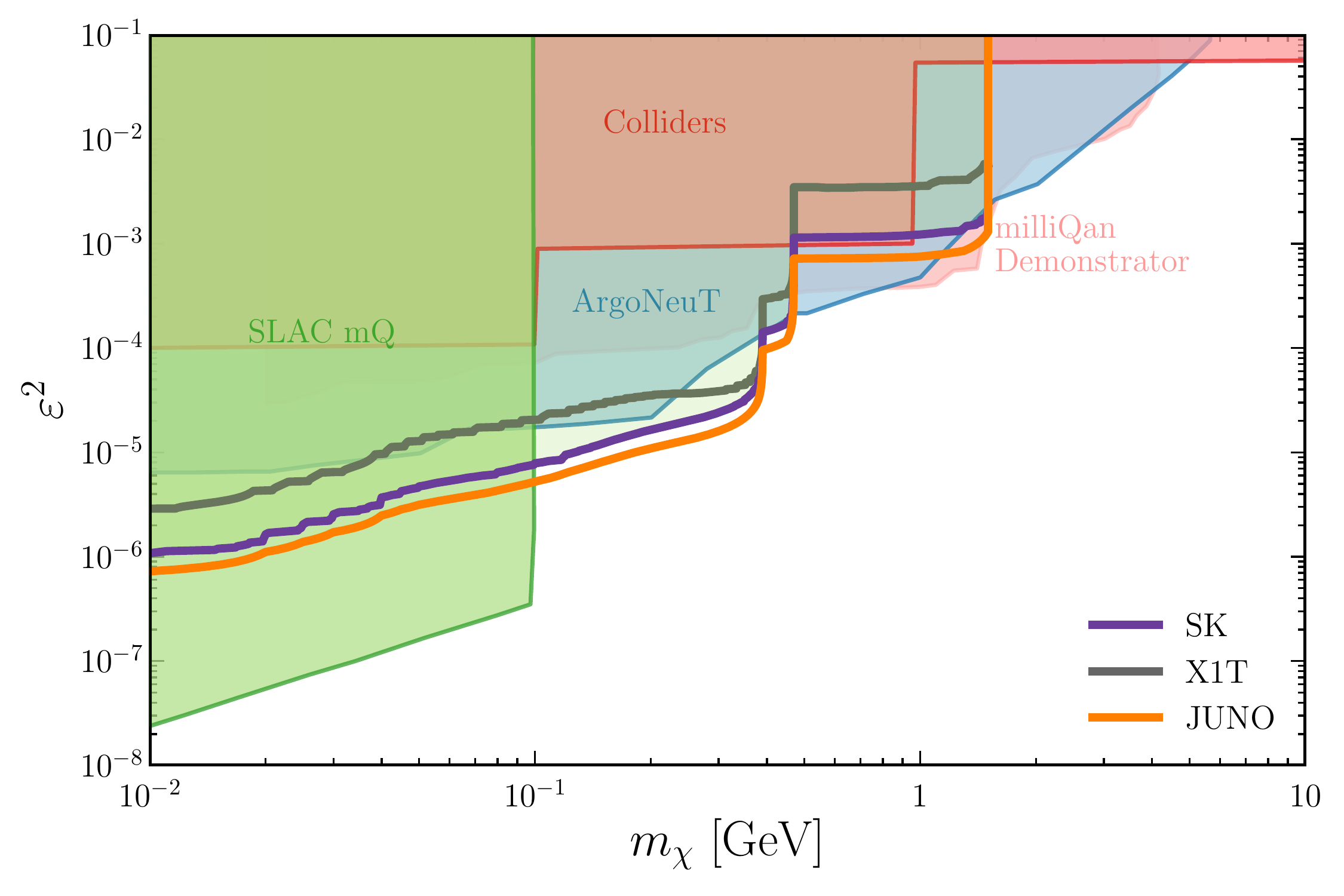} 
\caption{\textbf{\textit{Single-hit constraints on millicharged particles.}}
Constraints on millicharged particle parameter space from SLAC mQ~\cite{Prinz:2001qz}, ArgoNeuT~\cite{Acciarri:2019jly}, the milliQan Demonstrator~\cite{Ball:2021qrn}, and colliders~\cite{Davidson:2000hf}, compared with our 90\% CL single-hit constraints from Super-Kamiokande (purple), JUNO (orange), and XENON1T (grey).}
\label{fig:MCP_limits_single}
\end{center}
\end{figure}

Fig.~\ref{fig:MCP_limits_single} shows our new constraints at 90\% CL, compared with existing upper limits from the SLAC-mQ experiment~\cite{Prinz:2001qz}, ArgoNeuT~\cite{Acciarri:2019jly}, the milliQan Demonstrator~\cite{Ball:2021qrn}, and colliders~\cite{Davidson:2000hf}\footnote{We note here the different shape in this constraint relative to the one appearing in Ref.~\cite{Haas:2014dda}, appearing to be from the same source.
We believe the red region presented here represents the constraint from Ref.~\cite{Davidson:2000hf}.}. Additional constraints, potentially subject to large background uncertainties and systematics associated with hand-scanning, can also be found in Ref.~\cite{Marocco:2020dqu}.
Our Super-Kamiokande analysis, shown by the purple line and associated shaded region, yields results comparable results to those derived in Ref.~\cite{Plestid:2020kdm} for the bulk of parameter space.
The small discrepancies between these two results arise from the following differences. 
The analysis of Ref.~\cite{Plestid:2020kdm} set constraints assuming an exposure of $\SI{22.5}\kty$ at Super-Kamiokande, while here we have used the full 2853 day (or $7.81\times \SI{22.5}\kty$) exposure of SK I-III and the statistical analysis discussed above to set this constraint.
While the details of backgrounds are complicated, Hyper-Kamiokande should be able to improve on such constraints in the coming decades --- we direct the reader to Ref.~\cite{Plestid:2020kdm} for more on millicharged particle searches at Hyper-Kamiokande and Refs.~\cite{Abe:2018uyc,Moller:2018kpn} for a detailed discussion of searches for the DSNB at Hyper-Kamiokande and associated challenges.

As shown in Fig.~\ref{fig:MCP_limits_single}, the strongest current limit for $\SI{100}\MeV \lesssim m_\chi \lesssim \SI{500}\MeV$ comes from this Super-Kamiokande analysis.
However, a similar duration of a JUNO single-hit analysis (orange line) will be able to improve on this, due to JUNO's ability to reach very low electron recoil energy.
Although XENON1T (grey) can reach keV-scale electron recoils in this analysis, its small volume-exposure limits its capabilities relative to SK or JUNO.

%%%%%%%%%%%%%%%%%%%%%%%%%%%%%%%%%%%%%%%%%%%%%%%%%%%%%%%%%%%%%%%%%%%%%%%%%%%%%%%
%%%%%%%%%%%%%%%%%%%%%%%%%%%%%%%%%%%%%%%%%%%%%%%%%%%%%%%%%%%%%%%%%%%%%%%%%%%%%%%

\subsection{Multiple Scattering Searches}
\label{subsec:MultiHit}

Throughout Section~\ref{subsec:SingleHit}, we focused on scenarios where the atmospheric-produced MCP scatter inside a detector, providing enough energy to an electron to leave a distinct signature in the experiment.
When discussing the signal of single-scattering events, we pointed out that the rate of signal events grows with smaller electron recoil energy $E_r$.
However, in the case of both Super-Kamiokande and JUNO, the background do as well.
Especially for JUNO, the background rate (dominated by radioactive emissions) became prohibitive for $E_r \lesssim \SI{12}\MeV$, preventing a single-scattering analysis from reaching lower recoil energies.

One means for reducing (or even eliminating) these low-recoil-energy backgrounds is to require multiple scatterings within one small time window --- the radioactive backgrounds are each associated with some relatively large half-life and so the possibility of two such emissions in a ${\sim}10^{-6}$ s window is very rare.
Consequently, this type of analysis requires that a given MCP deposits energy at least twice as it traverses the detector, which will scale with even more powers of the millicharge $\varepsilon$ than our single-scattering analysis.
However, because this strategy allows us to search for even smaller recoil energies $E_r$ where the scattering cross section grows, we will see that large event rates are still possible.
This principle was exploited in Ref.~\cite{Harnik:2019zee} for beam-produced millicharged particle searches in ArgoNeuT/DUNE.

In the remainder of this subsection, we will utilize this same approach for our atmospheric searches, focusing on the JUNO experiment.
We will demonstrate that this approach can even outperform the single-scattering analysis presented in Section~\ref{subsec:SingleHit} due to the low-energy capabilities of the JUNO detector's liquid scintillator.

\subsubsection{Signal characteristics} 
We will determine the rate of multiple-scattering signal events in a given experimental setup by means of comparison with the single-scattering analyses discussed in Section~\ref{subsec:SingleHit}.
For simplicity, we will consider total event rates instead of the event distribution as a function of electron recoil energy $E_r$.
The important quantities to consider will be $T_H$, the minimum recoil energy for a hard-scattering event (\textit{i.e.}, those considered in the single-scattering analyses) and $T_S$, the minimum recoil energy capable of detecting events with multiple soft scatters in a small time window.

We perform the comparison between single- and multiple-scattering by computing the mean free path that a given MCP travels before either scattering in a ``hard'' manner (yielding an electron energetic enough to appear in single-scattering analyses) or in a ``soft'' manner (with electrons of too low an energy to be useful in single-hit searches, but energetic enough to be detected and used in a multiple-scattering search).

Assuming that a given MCP travels through a region with electron density $n_e$ and length $L_{\rm det.}$, we can approximate the probability that the MCP interacts using its hard-scattering mean free path $\lambda_{\rm H} = 1/(n_e \sigma_{\rm H})$,
\begin{equation}\label{eq:P1}
    P_1 \approx 1 - e^{-\frac{L_{\rm det.}}{\lambda_{\rm H}}},
\end{equation}
where $\sigma_{\rm H}$ is the hard-scattering cross section, i.e., the cross section where we require the outgoing electron to be energetic enough to be distinct from radioactive backgrounds.
In JUNO, this requirement is $E_r \gtrsim 12$ MeV. Ref.~\cite{Magill:2018tbb} approximates $\sigma_{\rm H}$ as
\begin{equation}\label{eq:HardScatter}
    \sigma_{\rm H} \approx \frac{2\pi \alpha_{\rm EM}^2 \varepsilon^2}{2m_e T_{\rm H}} = 2.6\times 10^{-25}\ \mathrm{cm}^2 \frac{\varepsilon^2}{\left(T_{\rm H}/1\ \mathrm{MeV}\right)}.
\end{equation}

The soft-scattering cross section and the corresponding mean free path between soft scatters are given by $\sigma_{\rm S}$ and $\lambda_{\rm S}$, respectively.
The cross section $\sigma_{\rm S}$ is identical to Eq.~\eqref{eq:HardScatter} with $T_{\rm S}$ replacing $T_{\rm H}$.
In principle, $T_{\rm S}$ can be significantly lower than $T_{\rm H}$, depending on the properties of the detector.
For JUNO, we estimate that $T_{\rm S} \approx 0.01-\SI{1}\MeV$ due to the $10^4$\,photons/MeV and $\mathcal{O}(10^{3})$\,photo-electrons/MeV of the detector's liquid scintillator~\cite{Fang:2019lej,Abusleme:2021zrw}.
The ratio of mean-free-paths $\lambda_{\rm H}/\lambda_{\rm S}$ then is proportional to $T_{\rm H}/T_{\rm S}$.

In addition to the single-scattering probability $P_1$ in Eq.~\eqref{eq:P1}, we can also consider the probability that an MCP scatters softly $n$ times inside the detector.
If we divide the total detector length into ``segments'' using $L_{\rm det.} = N_{\rm seg.} L_{\rm seg.}$, this probability is
\begin{equation}
    P_{(n)} = \frac{N_{\rm seg.}!}{n!(N_{\rm seg.}-n)!} \left( 1 - e^{-\frac{L_{\rm seg.}}{\lambda_{\rm S}}}\right)^n \left(e^{-\frac{L_{\rm seg.}}{\lambda_{\rm S}}}\right)^{N_{\rm seg.}-n}.
\end{equation}
Assuming we can take the $N_{\rm seg.} \to \infty$ (or $L_{\rm seg.} \to 0$) limit for an unsegmented detector, this probability approaches
\begin{equation}
    P_{(n)} = \frac{1}{n!}\left(\frac{L_{\rm det.}}{\lambda_{\rm S}}\right)^{n}e^{-\frac{L_{\rm det.}}{\lambda_{\rm S}}},
\end{equation}
The probability of two or more hits, $\sum_{n=2}^{\infty} P_{(n)}$ can be written as
\begin{equation}
    P_{n\geq 2} = 1 - e^{-\frac{L_{\rm det.}}{\lambda_{\rm S}}} \left( 1 + \frac{L_{\rm det.}}{\lambda_{\rm S}}\right).
\end{equation}

\begin{figure}
    \centering
    \includegraphics[width=0.8\linewidth]{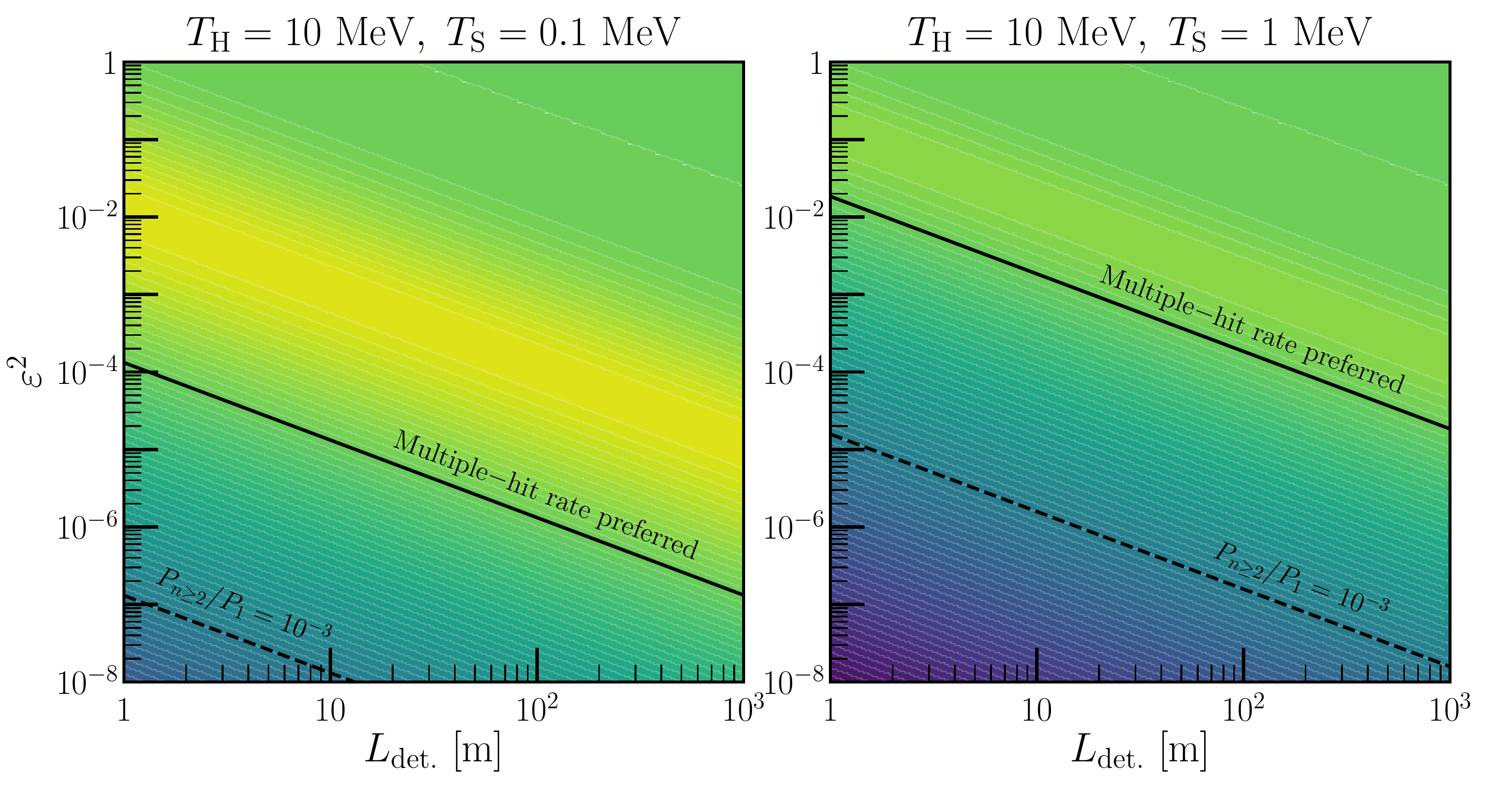}
    \caption{\textbf{\textit{Multiple-hit vs. single-hit search ratios.}}
    Ratio of multiple soft-scatter events to single hard-scatter events as a function of total detector length $L_{\rm det.}$ and $\varepsilon^2$ for two choices of $T_{\rm H}$ and $T_{\rm S}$, as labelled.
    Above the solid black lines, the double-hit rate exceeds the single-hit one.
    Above the dashed lines, the double-hit rate is at most three orders of magnitude smaller than the single-hit one.}
    \label{fig:SingleDouble}
\end{figure}

The ratio of multiple soft-scatter events to the single hard-scatter events is proportional to $P_{n\geq2}/P_1$.
This quantity depends on $\varepsilon^2$, the soft- and hard-scattering minimum recoil energies $T_{\rm S}$ and $T_{\rm H}$, and the total detector length $L_{\rm det.}$.
For two choices of $T_{\rm S}$ and $T_{\rm H}$, we present the ratio of these probabilities as a function of $L_{\rm det.}$ and $\varepsilon^2$ in Fig.~\ref{fig:SingleDouble}.
In each panel, above the solid black lines, the multiple-hit rate exceeds the single-hit one, implying that searches for these multiple-scattering events can be at least as powerful as the single-scattering searches.
If JUNO is capable of searching for $\SI{0.1}\MeV$ soft-scatters, this implies that these multiple scattering events can be favorable for $\varepsilon^2 \approx 10^{-5}$ (assuming path lengths on the order of $\SI{10}\m$).

One final feature of the multiple-scattering signature that is not very useful in single-scattering searches is directionality.
Because the single-scattering searches are seeking soft electrons, obtaining the direction of the incident millicharged particle is difficult.
With a multiple-scattering analysis, we can obtain the angular distribution of these events, which should match the flux prediction, including attenuation through Earth, as discussed in Section~\ref{sec:Propagation}.
This can be further used to statistically separate our signal from background events.

\subsubsection{Multiple-scattering backgrounds}
All of the backgrounds we discussed for the single-scattering analysis can contribute as backgrounds for the multiple-scattering searches.
However, they are only relevant if one or more of these stochastic backgrounds occurs within the time frame of an event for an experiment.
The large background rates present at low recoil energy for JUNO ($^{8}$B, $^{10}$C, $^{11}$C, and $^{11}$Be radioactive decays, specifically) will be suppressed by requiring multiple scatterings in this small window.
For JUNO, the emission timescale of the liquid scintillator is at most ${\sim}\SI{200}\ns$~\cite{An:2015jdp} --- if we can restrict the time window to be $\mathcal{O}(10^{-6} \mathrm{s})$, then these backgrounds are reduced to being $\mathcal{O}(1)$ for the ten years of data collection we assume for JUNO.

\subsubsection{Projected sensitivity with multiple scattering}
\begin{figure}
    \centering
    \includegraphics[width=0.8\linewidth]{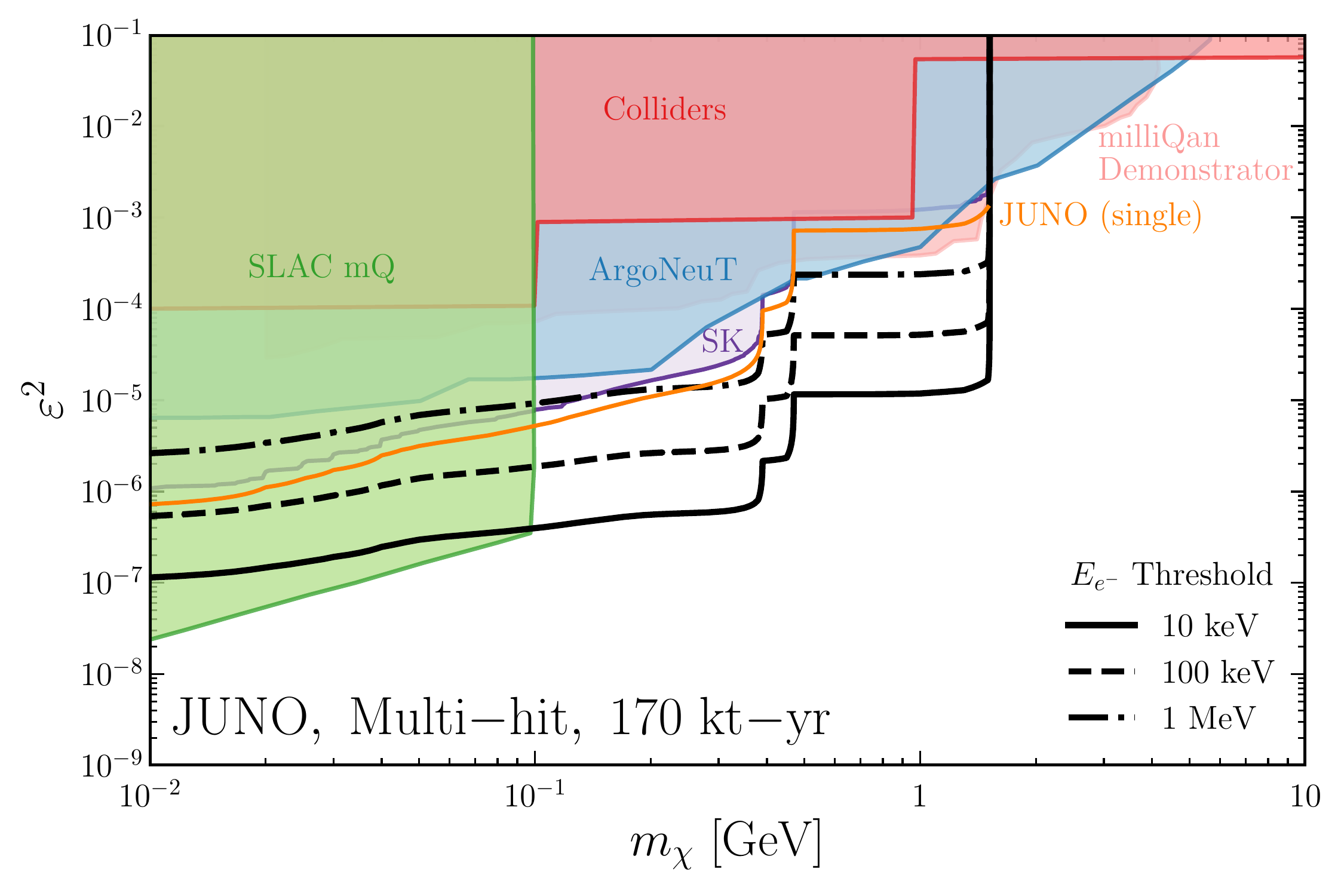}
    \caption{\textbf{\textit{Multiple-hit constraints on millicharged particles from JUNO.}} Black: Expected 90\% CL sensitivity to millicharged parameter space using the multiple hit strategy of JUNO and a 170 kt-yr exposure. Three different assumptions about the minimum observable electron recoil energy are assumed: $\SI{10}\keV$ (solid), $\SI{100}\keV$ (dashed), and $\SI{1}\MeV$ (dot-dashed). Existing constraints (including our Super-Kamiokande analysis in purple) are shown as filled in regions, and the single-hit analysis of JUNO from Fig.~\ref{fig:MCP_limits_single} is shown in orange.}
    \label{fig:MCPDoubleHit}
\end{figure}

Combining the above information, we project the sensitivity using this multiple-scattering search in JUNO here in Fig.~\ref{fig:MCPDoubleHit}.
In order to determine the ``detector length'' of JUNO that we discussed above, we simply average the path length of incoming MCP over the 35-m diameter spherical vessel.
The average path length through a sphere is $2D/3 = 23.3$ m for JUNO.

We assume that the backgrounds in JUNO are small enough that $10$ signal events are statistically significant in the 170 kt-yr exposure --- we do not use any spectral information about the energy of the electron events, as we expect that resolution at such sub-MeV energies will be difficult. We take three different assumptions about the minimum threshold energy of electrons that JUNO can detect --- $\SI{1}\MeV$ (dot-dashed), $\SI{100}\keV$ (dashed), and $\SI{10}\keV$ (solid).
If this 10-keV threshold is attainable (which could be possible given the $10^4$ photons/MeV of energy deposited in the liquid scintillator), we see that this multiple-hit strategy will far exceed single-scattering searches for atmospheric MCP.
Even with MeV thresholds, this is a complementary approach, especially at large $m_\chi$. The realistic $100$-keV threshold seems particularly promising as a target for JUNO.

\section{Millicharged Particles at Neutrino Telescopes}\label{sec:NeutrinoTelescopes}

The search for millicharged particles from natural sources is a statistically-limited problem, whose optimal detector would be a low-threshold, small-background, large-mass underground device.
In this work, we have so far studied kiloton-mass-scale low-threshold and small-background detectors, such as XENON1T, and tens of kilotons mass low-threshold detectors, such as JUNO, in this section we will study the sensitivity of megaton and gigaton detectors, known as neutrino telescopes.
These detectors use natural transparent media to detect Cherenkov light produced by neutrino interactions and are designed to detect faint high-energy astrophysical neutrino sources. 
However, their large detector volume allows them to also have unique capabilities to observe low-energy neutrinos produced in supernovae.

Since MCPs are long-lived and have small energy losses their signature in neutrino telescopes will be a long, faint track.
The main backgrounds for this search are coincident or dim atmospheric muons that can accidentally mimic the signature.
Searches for long-lived faint-tracks have been performed by the IceCube collaboration by looking for an isotropic flux of fractional charged particles with charges motivated by the quark charges~\cite{Verpoest,VanDriessche}.
This analysis yielded a sensitivity approximately ten times stronger than previous constraints from Kamiokande and MACRO, however the assumption of an isotropic flux of fractional charge particles is not consistent with the propagation of these particles through the Earth, as we discussed in Sec.~\ref{sec:Propagation}.
Despite this, Refs.~\cite{Verpoest,VanDriessche} motivate further study of the sensitivity to these signals.

The yield of Cherenkov photons produced per energy per unit path length of the distance travelled by a particle of charge $\varepsilon e$ is given by the Frank-Tamm equation~\cite{Zyla:2020zbs}
\begin{equation}
    \frac{d^2 N}{dEdX} = \frac{\alpha \varepsilon^2}{\hbar c} \sin^2\theta_c \approx 160 \varepsilon^2 \left(\frac{\sin^2\theta_c}{0.43}\right)\si\eV^{-1} \si\cm^{-1},
\end{equation}
where $\theta_c$ is the Cherenkov angle in the medium, which we have normalized to the Cherenkov angle in ice, $\theta_c \approx 41^{\circ}$~\cite{Aartsen:2013rt}.
The relevant Cherenkov photon wavelength for IceCube is around $\SI{400}\nm$, which results from the convolution of the IceCube PMT quantum efficiency and the glass-housing transmission probability~\cite{Abbasi:2008aa,Aartsen:2016nxy}.
At this wavelength the IceCube digital optical module (DOM) acceptance is approximately 0.15, and decreases by approximately half by reducing the wavelength to $\SI{350}\nm$ or increasing it to $\SI{550}\nm$.
Considering this, the yield of relevant Cherenkov photons is approximately $50\varepsilon^2\si\cm^{-1}$.
This implies that an IceCube-through-going MCP will produce, on average, $5 \times 10^6\times \varepsilon^2$ relevant Cherenkov photons. 
Next we need to estimate how many of these Cherenkov photons could reach an IceCube module, this implies taking into account the absorption of the Antartic ice~\cite{Aartsen:2013rt} and the single-photo-electron efficiency~\cite{IceCube:2020nwx}.
The number of photons from a single-point light source, which would be a given MCP energy loss, is reduced by a factor of $10^{-3}$ at a $\SI{10}\m$ distance and by $10^{-6}$ at a $\SI{100}\m$~\cite{Aartsen:2013vja}.
Thus, if an MCP passes at a $\SI{10}\m$ distance from an IceCube DOM the Cherenkov photon yield will be approximately $10\varepsilon^2$ photons.
This estimation implies that the relevant MCP parameter space that can produce enough Cherenkov photons in IceCube is above approximately $\SI{5}\GeV$, \textit{cf.} Fig.~\ref{fig:MCP_limits_single}. 
%However, the yield of Cherenkov photons is not very impressive.

In the above discussion, we have only taken into account the Cherenkov light, but there are other sources of light production from charged particles, such as ionization. 
To study the sensitivity of neutrino telescopes to millicharged particles, while taking into account these other losses, we have developed a dedicated Monte Carlo to estimate the trigger rate in detectors such as the IceCube Neutrino Observatory in the South Pole. This Monte Carlo can be found in our GitHub repository at \textcolor{BlueViolet}{\href{https://github.com/Harvard-Neutrino/HeavenlyMCP}{this https URL}}.

In order to estimate the trigger rate we need to know the precise locations of the energy depositions of the MCPs and their distance to the detection units.
Our Monte Carlo consists of the following stages:
\begin{enumerate}
\item We produce MCPs according to a power-law energy distribution and spread them uniformly across the surface of the Earth.
\item We propagate the MCPs from the Earth surface to a cylinder that contains the detector instrumented volume. 
To perform this we use \texttt{PROPOSAL}, which is a Monte Carlo package that simulates the energy losses of charged leptons in different media. 
The energy losses implemented in \texttt{PROPOSAL} are equivalent to the ones discussed in Sec.~\ref{sec:Propagation}, however, unlike our previous discussion which focused on the mean energy loss, \texttt{PROPOSAL} provides a detailed simulation of continuous energy losses and stochastic ones.
Thus, in this step, for each MCP particle in our Monte Carlo we obtain the location, type, and amount of energy loss along the particle trajectory.
\item We convert the energy losses produced in the detector vicinity to the number of Cherenkov photons produced in ice. 
In order to do this, we use the publicly available \texttt{PPC} photon propagation code, which uses the parameterizations given in~\cite{Koehne:2013gpa} to convert each type of loss (ionization, bremsstrahlung, photo-hadronic, and pair-production) into an ensemble of Cherenkov photons.
\item The in-ice photons are then propagated by \texttt{PPC} through the detector instrumented volume. 
Within this volume, we specify the detection units in \texttt{PPC} by providing the coordinate of each sensor.
With this information \texttt{PPC} returns the number of detection units hit by photons.
\end{enumerate}
Using this Monte Carlo approach one can estimate the trigger rate in a given detector. 
However, this is beyond the scope of this article and instead we provide this Monte Carlo and MCP fluxes computed in Sec.~\ref{sec:production} in order for experiments to estimate the MCP yield in their specific setup.

\begin{comment}
An example of one of the events produce by this Monte Carlo is shown in Fig.~\ref{}. 
Where we have used the IceCube geometry reported in~\cite{Aartsen:2016nxy} and the procedure above. 
In detectors, such as IceCube, the accidental single PMT noise rate is  approximately XXX~\cite{}.
Due to this large rate IceCube selects events that at least two hits in nearby detection units, a triggering criterium known as hard-local-coincidence~\cite{}.
In Fig.~\ref{}, we show the expected rate of MCP events as a function of the number of hit detector units.
\end{comment}

\section{Discussion and Conclusions}
\label{sec:Conclusions}

Whether particle charges are quantized and whether any charged particles exist beyond the Standard Model are two questions that have been asked for generations.
Searches for new particles with fractional charges work to address both of these questions, and significant progress has been made in these searches in recent years, particularly in the MeV to GeV mass range.

Focus in this mass range has been divided between two general categories --- searches for millicharged particles produced by collider and fixed-target experiments, and searches for millicharged particles naturally produced in the atmosphere.
We have focused on the latter approach.
In this work, we have revisited current constraints from the Super-Kamiokande neutrino experiment, qualitatively confirming the existing literature on searches of this type.
We have also analyzed existing data from the XENON1T dark matter direct-detection experiment and projected future capabilities of the JUNO reactor neutrino oscillation experiment in this parameter space, demonstrating paths for improvement in the next decade.

Going beyond this, we have combined a number of millicharged particle search strategies by proposing the search for multiple-scattering events in a liquid scintillator detector (specifically JUNO), allowing for sensitivity to significantly smaller millicharges than conventional, single-scattering searches.
This is because the multiple-scattering searches allow for analyses to probe even smaller energies where backgrounds dominate the conventional search.

We have focused predominantly on searches for these particles in tens-of-kiloton-scale detectors.
However, even larger detectors, such as the IceCube Neutrino Observatory (and forthcoming IceCube Upgrade) can offer an interesting, complementary means of searching for millicharged particles.
We demonstrated that detectors with longer path lengths (of traversing millicharged particles) are well-suited for multiple-scattering searches. 
IceCube is one such detector, and it can potentially perform a search for these ``faint track'' signatures in the coming years.
We have developed a Monte Carlo package to simulate the propagation and energy deposition of millicharged particles through a detector such as IceCube.

As we look forward to the next decade of searches for new, fractionally charged particles, it is imperative that we combine as many search strategies as possible.
This maximizes the chances of discovery, and, in the hopeful event of one discovery, a combined approach is our best way to interpret and understand such a momentous result.
Atmospheric searches, particularly those looking for multiple-scattering events, offer a powerful means to search for these particles, complementary to current and upcoming collider and fixed-target based searches.

\acknowledgments
We acknowledge useful discussions with Pilar Coloma, Pilar Hern{\'a}ndez, Ian Shoemaker, and Anatoli Fedynitch. We thank Ryan Plestid for useful discussions as well as comments on this manuscript.
We thank Jack Pairin for making the illustration shown in Fig.~\ref{fig:MCP_Artisitic} and Jean DeMerit for carefully reading our manuscript.
CAA is supported by the Faculty of Arts and Sciences of Harvard University, and the Alfred P. Sloan Foundation.
KJK is supported by Fermi Research Alliance, LLC, under contract DE-AC02-07CH11359 with the U.S. Department of Energy.
The work of V.M. is supported by CONICYT PFCHA/DOCTORADO BECAS CHILE/2018 - 72180000.

%%%%%% APPENDIX STARTS HERE

%\onecolumngrid
\appendix
\renewcommand{\theequation}{A\arabic{equation}}

\section{Details of Millicharged Particle Production}
\label{app:decay}

We consider production of millicharged particles via rare decays of neutral mesons $\pi^0$, $\eta$, $\rho$, $\omega$, $\phi$, and $J/\Psi$.
The first two (pseudoscalar mesons) can result in three-body decays $\mathfrak{m} \to \gamma \chi \overline{\chi}$, analogous to Dalitz decays.
The latter four (vector mesons) can decay via an off-shell photon $\mathfrak{m} \to \chi \overline{\chi}$.
Below we give the branching ratios of these processes as well as the energy spectrum of the daughter $\chi$ particles.

The $\omega$ meson has an additional decay channel that has a relatively large width -- $\omega \to \pi^0 e^+ e^-$, with a branching ratio of $7.7 \times 10^{-4}$ compared to $\mathrm{Br}(\omega \to e^+ e^-) = 7.36 \times 10^{-5}$~\cite{Zyla:2020zbs}.
The corresponding width of $\omega \to \pi^0 \chi \bar{\chi}$ is more complicated than the two- and three-body decay widths we present below.
In principle, there should be an additional $\chi$ flux nearly an order of magnitude larger than our estimates in the mass range $m_{\eta}/2 < m_\chi < 1/2(m_\omega - m_{\pi^0})$.
For simplicity, and due to the narrow mass range that this impacts, we choose to neglect this decay channel in our calculations.

\textbf{Two-Body Decays:} The branching ratio of the two-body decays for vector mesons into millicharged particle pairs can be expressed as
\begin{eqnarray}
\mathrm{Br}(\mathfrak{m} \rightarrow \chi\bar{\chi})= 2\varepsilon^{2}\,\mathrm{Br}(\mathfrak{m} \rightarrow e^{+}e^{-}) \,I^{(2)}\left(\frac{m_{\chi}^2}{m_{M}^2},\frac{m_{e}^2}{m_{M}^2}\right),
\end{eqnarray}
where $\mathrm{Br}(\mathfrak{m}\rightarrow e^{+} e^{-})$ is the experimentally-measured branching ratio of the meson into electron/positron pairs and $I^{(2)}(x,y)$ is a dimensionless quantity relating these two processes,
\begin{eqnarray}
I^{(2)}(x,y)=\frac{(1+2x)\sqrt{1-4x}}{(1+2y)\sqrt{1-4y}}.
\label{eq:I2}
\end{eqnarray}
The energy distribution of the millicharged particles in the parent meson rest frame is flat. 
The lab-frame distribution can be obtained by transforming between frames using the boost of the parent meson. The allowed energy of the millicharged particle in the lab frame $E_\chi$ can be determined by requiring
\begin{equation}
\frac{2E_\chi}{1 + \sqrt{\lambda(1,\frac{m_\chi^2}{m_{\mathfrak{m}}^2},\frac{m_\chi^2}{m_{\mathfrak{m}}^2})}} \leq E_{\mathfrak{m}} \leq \frac{2E_\chi}{1 - \sqrt{\lambda(1,\frac{m_\chi^2}{m_{\mathfrak{m}}^2},\frac{m_\chi^2}{m_{\mathfrak{m}}^2})}},
\end{equation}
where $\lambda(x,y,z)$ is the Källén function~\cite{Kallen:1964lxa}
\begin{equation}
\lambda(x,y,z)=x^{2}+y^{2}+z^{2}+2xy+2xz+2yz.
\end{equation}

\textbf{Three-Body Decays:} For the pseduoscalar meson ($\pi^0$ and $\eta$) decays, the process of interest is $\mathfrak{m} \to \gamma \chi \overline{\chi}$.
The branching ratio for this decay can be related to the branching ratio of $\mathfrak{m} \to \gamma\gamma$ using
\begin{eqnarray}
\mathrm{Br}\left(\mathfrak{m}\to \gamma\chi\overline{\chi}\right) = 2\varepsilon^2\alpha_{\rm EM} \mathrm{Br}\left(\mathfrak{m}\to \gamma\gamma\right) I^{(3)}\left(\frac{m_{\chi}^2}{m_{\mathfrak{m}}^2}\right),
\end{eqnarray}
where $I^{(3)}(x)$ is, similar to $I^{(2)}(x,y)$, a dimensionless function~\cite{Kelly:2018brz},
\begin{eqnarray}
I^{(3)}(x)=\frac{2}{3\pi} \int_{4x}^{1} dz \, \sqrt{1-\frac{4x}{z}} \frac{(1-z)^3}{z^2}(2x+z).
\label{eq:I3}
\end{eqnarray}
To obtain the lab-frame distribution of the millicharged particle energy, we use the quantity $z \equiv E_\chi/\gamma_\mathfrak{m}$, where $\gamma_\mathfrak{m}$ is the meson boost.
This distribution can be expressed as
\begin{equation}
\frac{1}{\Gamma} \frac{d \Gamma}{dz}= \frac{(z-m_{M})}{3 z^{3}\, Y_{\mathfrak{m}}(m_{\mathfrak{m}},m_{\chi})} F(z,m_{\mathfrak{m}},m_{\chi}),
\end{equation}
with
\begin{equation}
Y_{\mathfrak{m}}(m_{\mathfrak{m}},m_{\chi})= m_{\mathfrak{m}}^{7} -32m_{\mathfrak{m}}^{5}m_{\chi}^{2} -348m_{\mathfrak{m}}^{3}m_{\chi}^{4}\log(2m_{\chi}/m_{\mathfrak{m}}) + 512 m_{\mathfrak{m}}m_{\chi}^{6} - 256\frac{m_{\chi}^{8}}{m_{\mathfrak{m}}},
\end{equation}
and 
\begin{align}
F(z,m_{\mathfrak{m}},m_{\chi}) =\,& z (320m_{\mathfrak{m}}m_{\chi}^{6} + 144m_{\mathfrak{m}}^{3}m_{\chi}^{4} ) - z^{2}(m_{\chi}^{6} + 432 m_{\mathfrak{m}}^{2}m_{\chi}^{4} ) \nonumber\\
&+ z^{3}(108m_{\mathfrak{m}}^{3}m_{\chi}^{2} - 5m_{\mathfrak{m}}^{5}) - z^{4}(m_{\mathfrak{m}}^{2}m_{\chi}^{2} + m_{\mathfrak{m}}^{4}) \nonumber\\
&+4m_{\mathfrak{m}}^{3}z^{5} - 256 m_{\chi}^{6}m_{\mathfrak{m}}^{2}.
\end{align}
For a three-body decay we can determine the allowed range of $E_\chi$ using
\begin{equation}
\frac{m_{\mathfrak{m}} E_\chi}{E_{\rm max}+\sqrt{E_{\rm max}^2-m_{\chi}^2}} \leq E_{\mathfrak{m}} \leq \frac{m_{\mathfrak{m}} E_\chi}{E_{\rm max}-\sqrt{E_{\rm max}^2-m_{\chi}^2}},
\end{equation}
with $E_{\rm max}\equiv \frac{4m_{\chi}^2+m_{\mathfrak{m}}^2}{2m_{\mathfrak{m}}}$.

\section{Additional Details of the Super-Kamiokande Analysis}\label{app:SK}

\begin{figure}
\begin{center}
\includegraphics[width=0.32\columnwidth]{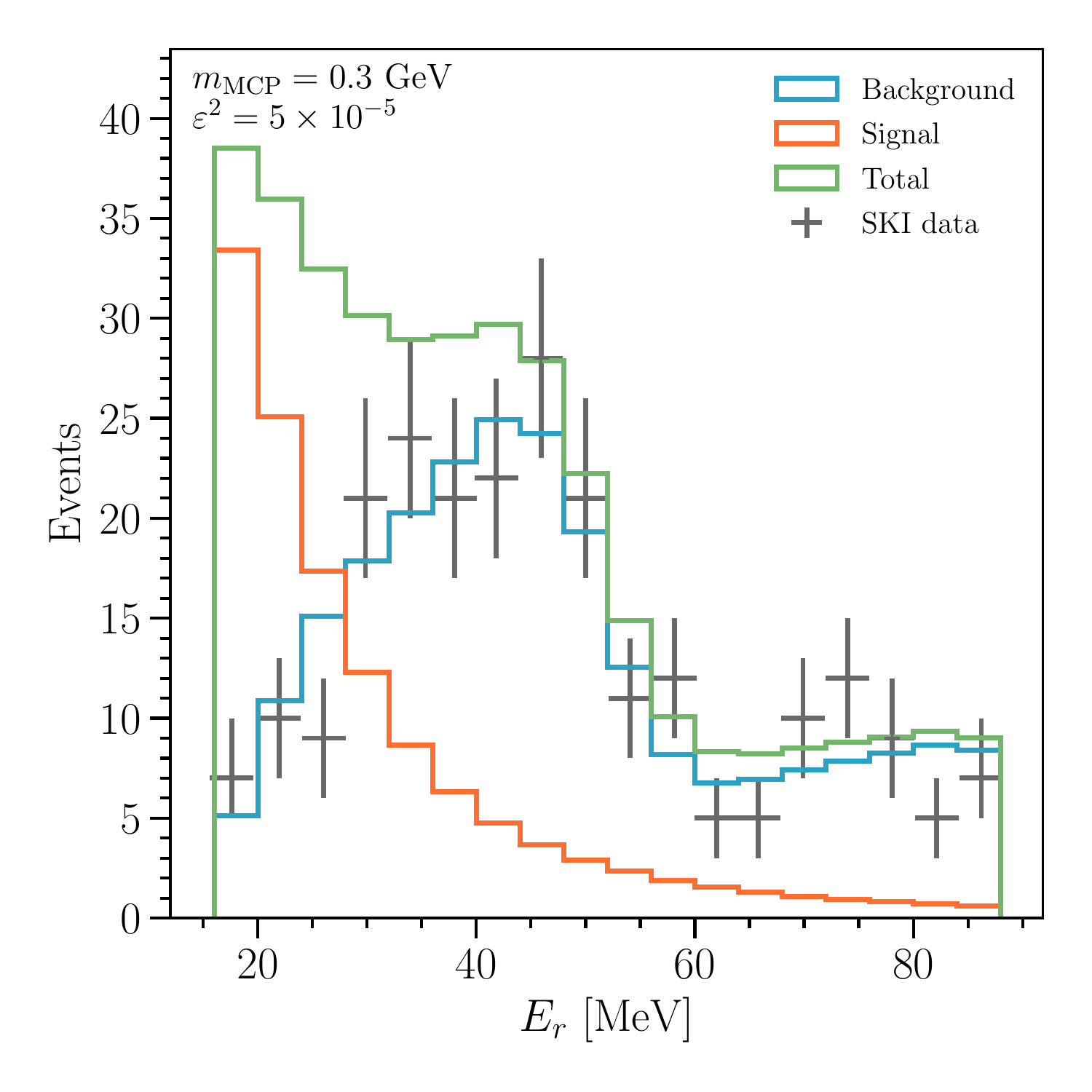} 
\includegraphics[width=0.32\columnwidth]{SKII_energy_distribution.pdf} 
\includegraphics[width=0.32\columnwidth]{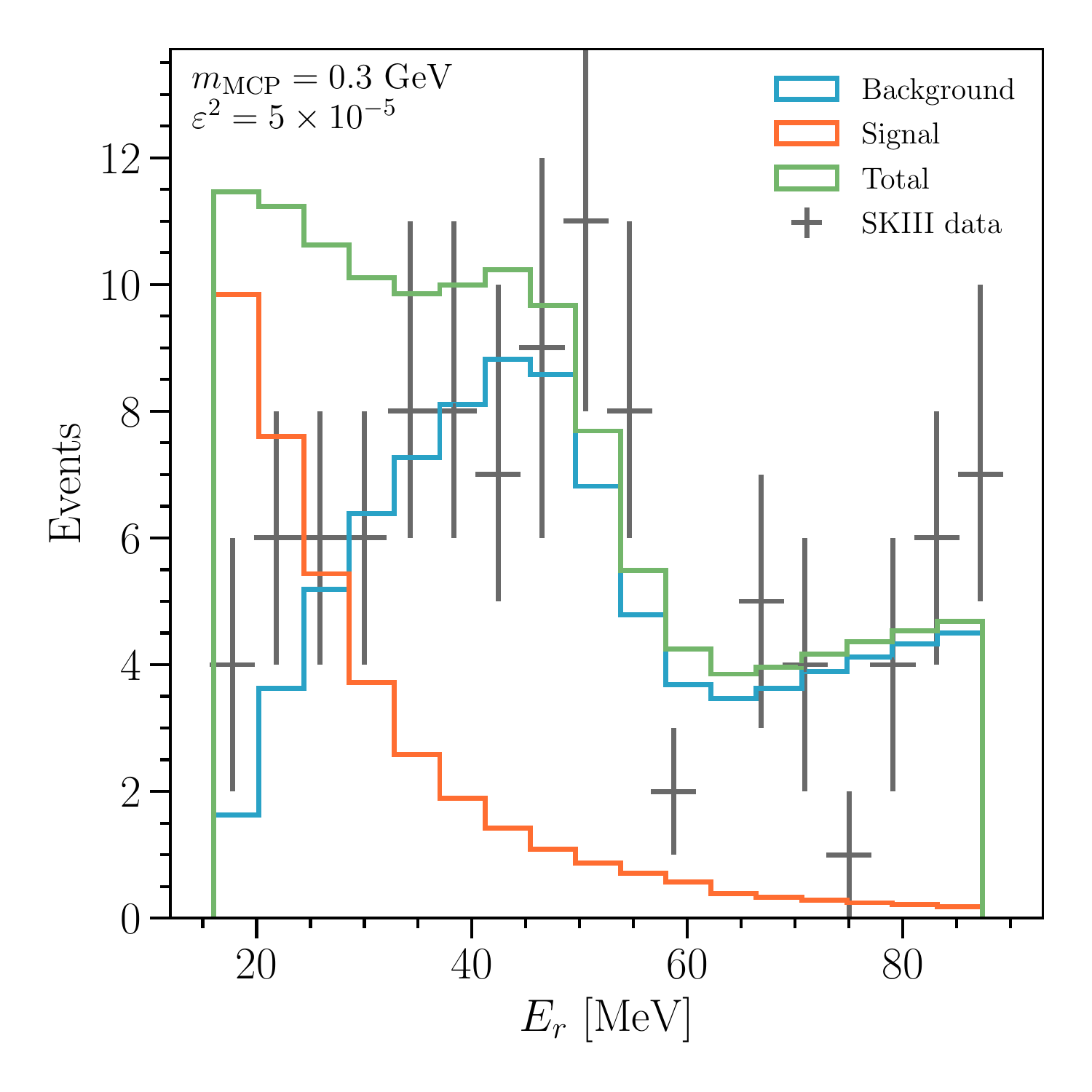} 
\caption{\textbf{\textit{Distributions for our millicharged particle search for the three stages of Super-Kamiokande.}}
Data is shown as crosses and expected background is shown as a blue histogram. 
An example millicharge particle distribution is shown, for the same parameters, as an orange histogram.
}
\label{fig:sk_suppl}
\end{center}
\end{figure}

When discussing expected signal and background event distributions in Super-Kamiokande (\textit{cf.} Fig.~\ref{fig:signatures} left), we showed the expected distributions with Super-Kamiokande II for simplicity.
For completeness in Fig.~\ref{fig:sk_suppl}, we provide the analogous distributions for all three stages of Super-Kamiokande data collection that enter our analysis.
Each panel displays the expected background events (blue), signal assuming $m_\chi = \SI{300}\MeV$ and $\varepsilon^2 = 5 \times 10^{-5}$ (orange), and their sum (green), compared against the observed data (black crosses).

In practice, we do not use the expected background distributions (blue) in our analysis.
This is because they do not include the possibility of any diffuse supernova neutrino background events, which would contribute at low recoil energy.
For this reason, we adopt the background-agnostic, one-sided likelihood approach discussed in Section~\ref{subsubsec:ExpSetups}.

%%%%%%%%%%%%%%%%%%%%%%%%%%%%%%%%%%%%%%%%%%%%%%%%%%%%%%%%%%%%%%%%%%%%%%%%%%%%%%%
%%%%%%%%%%%%%%%%%%%%%%%%%%%%%%%%%%%%%%%%%%%%%%%%%%%%%%%%%%%%%%%%%%%%%%%%%%%%%%%

\section{XENON1T Preferred Parameter Space}\label{app:XENON}

In Section~\ref{subsec:SingleHit} we discussed constraints on millicharged particle parameter space as a function of $m_\chi$ and $\varepsilon^2$ coming from single-hit searches, including Super-Kamiokande, XENON1T, and a future search in JUNO.
When considering current data from Super-Kamiokande and XENON1T, we took a background-agnostic approach with a one-sided likelihood function to derive a conservative upper limit on $\varepsilon^2$ as a function of $m_\chi$.
For XENON1T, this was done in part due to the much-discussed potential tritium background, absent in the nominal $B_0$ background model of Ref.~\cite{Aprile:2020tmw}.

In this appendix, we perform an alternate test of the XENON1T data where we assume that the $B_0$ model is robust and calculate the likelihood comparing $\mu^s_i$ (the expected number of signal events in bin $i$) plus $B_{0,i}$ (the expected background in bin $i$) with the observed data $d_i$.
This process, due to the excess electron-like events in the lowest energy bins, will yield a preferred region of millicharged parameter space instead of a constraint.

The potential signals proposed in Ref.~\cite{Aprile:2020tmw} include neutrino magnetic moments and solar axions, which improve the fit to data over the background-only explanation by $3.2-3.4\sigma$.
We find that the millicharged particle signature improves our test statistic over the background-only hypothesis by $10.43$ units. 
When we (conservatively) consider two degrees of freedom, this implies a preference of $2.8\sigma$ over the null hypothesis -- if we had considered the highly-correlated ($m_\chi$, $\varepsilon^2$) to account for a single degree of freedom, this would imply a preference of $3.2\sigma$, as preferable as the hypotheses presented by the XENON1T Collaboration.

%%%%%%%%%%%%%%%%%%%%%%%%%%%%%%%%%%%%%%%%%%%%%%%%%%%%%%%%%%%%%%%%%%%%%%%%%%%%%%%
\begin{figure}
\begin{center}
\includegraphics[width=0.8\columnwidth]{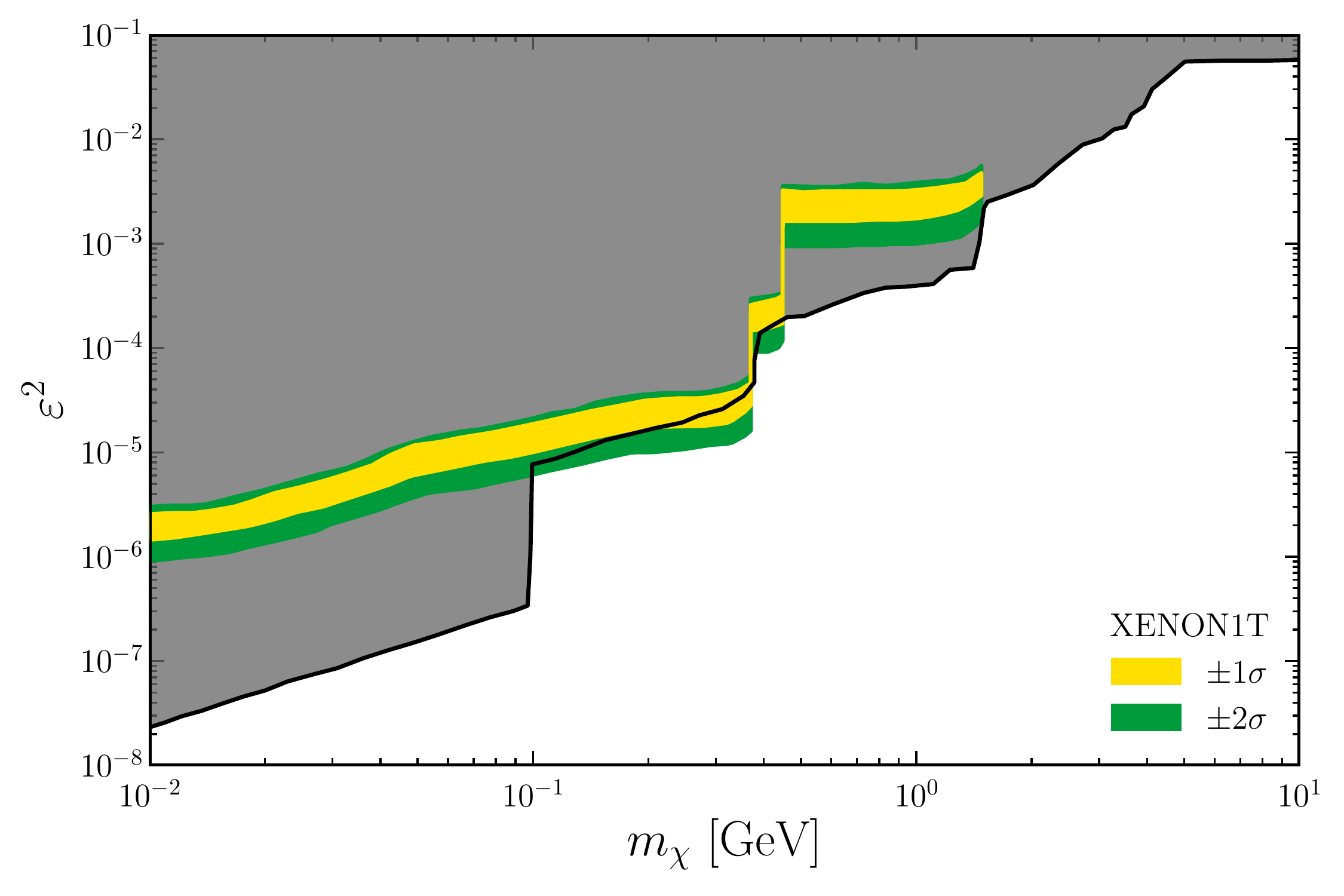} 
\caption{\textbf{\textit{Preferred parameter space from XENON1T.}} Yellow/green bands depict the $\pm 1\sigma$/$\pm 2\sigma$ preferred parameter space of our XENON1T analysis when we assume that the $B_0$ background of Ref.~\cite{Aprile:2020tmw} accurately represents all backgrounds. All existing constraints are shown in grey.}
\label{fig:MCP_Xenon_Pref}
\end{center}
\end{figure}
%%%%%%%%%%%%%%%%%%%%%%%%%%%%%%%%%%%%%%%%%%%%%%%%%%%%%%%%%%%%%%%%%%%%%%%%%%%%%%
Fig.~\ref{fig:MCP_Xenon_Pref} presents our $1\sigma$ (yellow) and $2\sigma$ (green) preferred region of parameter space from XENON1T, compared against all other existing limits (including our Super-Kamiokande analysis) in grey.
We see that the bulk of the $\pm2\sigma$ preferred parameter space is already excluded by one or more other constraint, however, some regions survive near $m_\chi \approx 350$ MeV.
These regions will be tested by JUNO's single-hit analysis and thoroughly explored by JUNO's double-hit analysis.

We note here, as in the main text, that our XENON1T analysis and signal-event prediction rate from Eq.~\eqref{eq:diff_rate} do not include the fact that the electrons in the XENON1T detector are bound~\cite{Bloch:2020uzh,Harnik:2020ugb}.
This is relevant because their binding energies are non-negligible compared to the recoil energy observed in the detector, and this implies that our signal rate predictions are optimistic.
For this reason, in addition to the tritium background discussion above, we caution the reader from inferring that the MCP solution to the XENON1T excess is a likely explanation.

\bibliography{mcp_atmo.bib}

%merlin.mbs apsrev4-1.bst 2010-07-25 4.21a (PWD, AO, DPC) hacked
%Control: key (0)
%Control: author (0) dotless jnrlst
%Control: editor formatted (1) identically to author
%Control: production of article title (0) allowed
%Control: page (1) range
%Control: year (0) verbatim
%Control: production of eprint (0) enabled
\begin{thebibliography}{84}%
\makeatletter
\providecommand \@ifxundefined [1]{%
 \@ifx{#1\undefined}
}%
\providecommand \@ifnum [1]{%
 \ifnum #1\expandafter \@firstoftwo
 \else \expandafter \@secondoftwo
 \fi
}%
\providecommand \@ifx [1]{%
 \ifx #1\expandafter \@firstoftwo
 \else \expandafter \@secondoftwo
 \fi
}%
\providecommand \natexlab [1]{#1}%
\providecommand \enquote  [1]{``#1''}%
\providecommand \bibnamefont  [1]{#1}%
\providecommand \bibfnamefont [1]{#1}%
\providecommand \citenamefont [1]{#1}%
\providecommand \href@noop [0]{\@secondoftwo}%
\providecommand \href [0]{\begingroup \@sanitize@url \@href}%
\providecommand \@href[1]{\@@startlink{#1}\@@href}%
\providecommand \@@href[1]{\endgroup#1\@@endlink}%
\providecommand \@sanitize@url [0]{\catcode `\\12\catcode `\$12\catcode
  `\&12\catcode `\#12\catcode `\^12\catcode `\_12\catcode `\%12\relax}%
\providecommand \@@startlink[1]{}%
\providecommand \@@endlink[0]{}%
\providecommand \url  [0]{\begingroup\@sanitize@url \@url }%
\providecommand \@url [1]{\endgroup\@href {#1}{\urlprefix }}%
\providecommand \urlprefix  [0]{URL }%
\providecommand \Eprint [0]{\href }%
\providecommand \doibase [0]{http://dx.doi.org/}%
\providecommand \selectlanguage [0]{\@gobble}%
\providecommand \bibinfo  [0]{\@secondoftwo}%
\providecommand \bibfield  [0]{\@secondoftwo}%
\providecommand \translation [1]{[#1]}%
\providecommand \BibitemOpen [0]{}%
\providecommand \bibitemStop [0]{}%
\providecommand \bibitemNoStop [0]{.\EOS\space}%
\providecommand \EOS [0]{\spacefactor3000\relax}%
\providecommand \BibitemShut  [1]{\csname bibitem#1\endcsname}%
\let\auto@bib@innerbib\@empty
%</preamble>
\bibitem [{\citenamefont {Pacini}(1912)}]{Pacini:2010hv}%
  \BibitemOpen
  \bibfield  {author} {\bibinfo {author} {\bibfnamefont {D.}~\bibnamefont
  {Pacini}},\ }\bibfield  {title} {\enquote {\bibinfo {title} {{Penetrating
  Radiation at the Surface of and in Water}},}\ }\href {\doibase
  10.1007/BF02957440} {\bibfield  {journal} {\bibinfo  {journal} {Nuovo Cim.}\
  }\textbf {\bibinfo {volume} {8}},\ \bibinfo {pages} {93--100} (\bibinfo
  {year} {1912})},\ \Eprint {http://arxiv.org/abs/1002.1810} {arXiv:1002.1810
  [physics.hist-ph]} \BibitemShut {NoStop}%
\bibitem [{\citenamefont {Hess}(2018)}]{Hess:2018twh}%
  \BibitemOpen
  \bibfield  {author} {\bibinfo {author} {\bibfnamefont {Victor}\ \bibnamefont
  {Hess}},\ }\bibfield  {title} {\enquote {\bibinfo {title} {{On the
  Observations of the Penetrating Radiation during Seven Balloon Flights}},}\
  }\href@noop {} {\  (\bibinfo {year} {2018})},\ \Eprint
  {http://arxiv.org/abs/1808.02927} {arXiv:1808.02927 [physics.hist-ph]}
  \BibitemShut {NoStop}%
\bibitem [{\citenamefont {De~Angelis}(2014)}]{DeAngelis:2014doa}%
  \BibitemOpen
  \bibfield  {author} {\bibinfo {author} {\bibfnamefont {Alessandro}\
  \bibnamefont {De~Angelis}},\ }\bibfield  {title} {\enquote {\bibinfo {title}
  {{Atmospheric ionization and cosmic rays: studies and measurements before
  1912}},}\ }\href {\doibase 10.1016/j.astropartphys.2013.05.010} {\bibfield
  {journal} {\bibinfo  {journal} {Astropart. Phys.}\ }\textbf {\bibinfo
  {volume} {53}},\ \bibinfo {pages} {19--26} (\bibinfo {year} {2014})},\
  \Eprint {http://arxiv.org/abs/1208.6527} {arXiv:1208.6527 [physics.hist-ph]}
  \BibitemShut {NoStop}%
\bibitem [{\citenamefont {Grupen}(2013)}]{Grupen:2013gja}%
  \BibitemOpen
  \bibfield  {author} {\bibinfo {author} {\bibfnamefont {Claus}\ \bibnamefont
  {Grupen}},\ }\bibfield  {title} {\enquote {\bibinfo {title} {{The History of
  Cosmic Ray Studies after Hess}},}\ }\href {\doibase
  10.1016/j.nuclphysbps.2013.05.003} {\bibfield  {journal} {\bibinfo  {journal}
  {Nucl. Phys. B Proc. Suppl.}\ }\textbf {\bibinfo {volume} {239-240}},\
  \bibinfo {pages} {19--25} (\bibinfo {year} {2013})}\BibitemShut {NoStop}%
\bibitem [{\citenamefont {Neddermeyer}\ and\ \citenamefont
  {Anderson}(1937)}]{Neddermeyer:1937md}%
  \BibitemOpen
  \bibfield  {author} {\bibinfo {author} {\bibfnamefont {S.H.}\ \bibnamefont
  {Neddermeyer}}\ and\ \bibinfo {author} {\bibfnamefont {C.D.}\ \bibnamefont
  {Anderson}},\ }\bibfield  {title} {\enquote {\bibinfo {title} {{Note on the
  Nature of Cosmic Ray Particles}},}\ }\href {\doibase 10.1103/PhysRev.51.884}
  {\bibfield  {journal} {\bibinfo  {journal} {Phys. Rev.}\ }\textbf {\bibinfo
  {volume} {51}},\ \bibinfo {pages} {884--886} (\bibinfo {year}
  {1937})}\BibitemShut {NoStop}%
\bibitem [{\citenamefont {Lattes}\ \emph {et~al.}(1947)\citenamefont {Lattes},
  \citenamefont {Muirhead}, \citenamefont {Occhialini},\ and\ \citenamefont
  {Powell}}]{Lattes:1947mw}%
  \BibitemOpen
  \bibfield  {author} {\bibinfo {author} {\bibfnamefont {C.M.G.}\ \bibnamefont
  {Lattes}}, \bibinfo {author} {\bibfnamefont {H.}~\bibnamefont {Muirhead}},
  \bibinfo {author} {\bibfnamefont {G.P.S.}\ \bibnamefont {Occhialini}}, \ and\
  \bibinfo {author} {\bibfnamefont {C.F.}\ \bibnamefont {Powell}},\ }\bibfield
  {title} {\enquote {\bibinfo {title} {{PROCESSES INVOLVING CHARGED MESONS}},}\
  }\href {\doibase 10.1038/159694a0} {\bibfield  {journal} {\bibinfo  {journal}
  {Nature}\ }\textbf {\bibinfo {volume} {159}},\ \bibinfo {pages} {694--697}
  (\bibinfo {year} {1947})}\BibitemShut {NoStop}%
\bibitem [{\citenamefont {Dobroliubov}\ and\ \citenamefont
  {Ignatiev}(1990)}]{Dobroliubov:1989mr}%
  \BibitemOpen
  \bibfield  {author} {\bibinfo {author} {\bibfnamefont {M.I.}\ \bibnamefont
  {Dobroliubov}}\ and\ \bibinfo {author} {\bibfnamefont {A.Yu.}\ \bibnamefont
  {Ignatiev}},\ }\bibfield  {title} {\enquote {\bibinfo {title} {{MILLICHARGED
  PARTICLES}},}\ }\href {\doibase 10.1103/PhysRevLett.65.679} {\bibfield
  {journal} {\bibinfo  {journal} {Phys. Rev. Lett.}\ }\textbf {\bibinfo
  {volume} {65}},\ \bibinfo {pages} {679--682} (\bibinfo {year}
  {1990})}\BibitemShut {NoStop}%
\bibitem [{\citenamefont {Gell-Mann}(1964)}]{GellMann:1964nj}%
  \BibitemOpen
  \bibfield  {author} {\bibinfo {author} {\bibfnamefont {Murray}\ \bibnamefont
  {Gell-Mann}},\ }\bibfield  {title} {\enquote {\bibinfo {title} {{A Schematic
  Model of Baryons and Mesons}},}\ }\href {\doibase
  10.1016/S0031-9163(64)92001-3} {\bibfield  {journal} {\bibinfo  {journal}
  {Phys. Lett.}\ }\textbf {\bibinfo {volume} {8}},\ \bibinfo {pages} {214--215}
  (\bibinfo {year} {1964})}\BibitemShut {NoStop}%
\bibitem [{\citenamefont {Unnikrishnan}\ and\ \citenamefont
  {Gillies}(2004)}]{Unnikrishnan_2004}%
  \BibitemOpen
  \bibfield  {author} {\bibinfo {author} {\bibfnamefont {C~S}\ \bibnamefont
  {Unnikrishnan}}\ and\ \bibinfo {author} {\bibfnamefont {G~T}\ \bibnamefont
  {Gillies}},\ }\bibfield  {title} {\enquote {\bibinfo {title} {The electrical
  neutrality of atoms and of bulk matter},}\ }\href {\doibase
  10.1088/0026-1394/41/5/s03} {\bibfield  {journal} {\bibinfo  {journal}
  {Metrologia}\ }\textbf {\bibinfo {volume} {41}},\ \bibinfo {pages}
  {S125--S135} (\bibinfo {year} {2004})}\BibitemShut {NoStop}%
\bibitem [{\citenamefont {Nishijima}(1996)}]{Nishijima:1996ji}%
  \BibitemOpen
  \bibfield  {author} {\bibinfo {author} {\bibfnamefont {K.}~\bibnamefont
  {Nishijima}},\ }\bibfield  {title} {\enquote {\bibinfo {title} {{BRS
  invariance, asymptotic freedom and color confinement. (A review)}},}\ }\href
  {\doibase 10.1007/BF01692238} {\bibfield  {journal} {\bibinfo  {journal}
  {Czech. J. Phys.}\ }\textbf {\bibinfo {volume} {46}},\ \bibinfo {pages}
  {1--124} (\bibinfo {year} {1996})}\BibitemShut {NoStop}%
\bibitem [{\citenamefont {Jones}(1977)}]{RevModPhys.49.717}%
  \BibitemOpen
  \bibfield  {author} {\bibinfo {author} {\bibfnamefont {Lawrence~W.}\
  \bibnamefont {Jones}},\ }\bibfield  {title} {\enquote {\bibinfo {title} {A
  review of quark search experiments},}\ }\href {\doibase
  10.1103/RevModPhys.49.717} {\bibfield  {journal} {\bibinfo  {journal} {Rev.
  Mod. Phys.}\ }\textbf {\bibinfo {volume} {49}},\ \bibinfo {pages} {717--752}
  (\bibinfo {year} {1977})}\BibitemShut {NoStop}%
\bibitem [{\citenamefont {Lyons}(1985)}]{LYONS1985225}%
  \BibitemOpen
  \bibfield  {author} {\bibinfo {author} {\bibfnamefont {L.}~\bibnamefont
  {Lyons}},\ }\bibfield  {title} {\enquote {\bibinfo {title} {Quark search
  experiments at accelerators and in cosmic rays},}\ }\href {\doibase
  https://doi.org/10.1016/0370-1573(85)90011-0} {\bibfield  {journal} {\bibinfo
   {journal} {Physics Reports}\ }\textbf {\bibinfo {volume} {129}},\ \bibinfo
  {pages} {225 -- 284} (\bibinfo {year} {1985})}\BibitemShut {NoStop}%
\bibitem [{\citenamefont {Perl}\ \emph {et~al.}(2009)\citenamefont {Perl},
  \citenamefont {Lee},\ and\ \citenamefont
  {Loomba}}]{doi:10.1146/annurev-nucl-121908-122035}%
  \BibitemOpen
  \bibfield  {author} {\bibinfo {author} {\bibfnamefont {Martin~L.}\
  \bibnamefont {Perl}}, \bibinfo {author} {\bibfnamefont {Eric~R.}\
  \bibnamefont {Lee}}, \ and\ \bibinfo {author} {\bibfnamefont {Dinesh}\
  \bibnamefont {Loomba}},\ }\bibfield  {title} {\enquote {\bibinfo {title}
  {Searches for fractionally charged particles},}\ }\href {\doibase
  10.1146/annurev-nucl-121908-122035} {\bibfield  {journal} {\bibinfo
  {journal} {Annual Review of Nuclear and Particle Science}\ }\textbf {\bibinfo
  {volume} {59}},\ \bibinfo {pages} {47--65} (\bibinfo {year} {2009})},\
  \Eprint
  {http://arxiv.org/abs/https://doi.org/10.1146/annurev-nucl-121908-122035}
  {https://doi.org/10.1146/annurev-nucl-121908-122035} \BibitemShut {NoStop}%
\bibitem [{\citenamefont {Marinelli}\ and\ \citenamefont
  {Morpurgo}(1984)}]{Marinelli:1983nd}%
  \BibitemOpen
  \bibfield  {author} {\bibinfo {author} {\bibfnamefont {M.}~\bibnamefont
  {Marinelli}}\ and\ \bibinfo {author} {\bibfnamefont {Giacomo}\ \bibnamefont
  {Morpurgo}},\ }\bibfield  {title} {\enquote {\bibinfo {title} {{The Electric
  Neutrality of Matter: A Summary}},}\ }\href {\doibase
  10.1016/0370-2693(84)91752-0} {\bibfield  {journal} {\bibinfo  {journal}
  {Phys. Lett. B}\ }\textbf {\bibinfo {volume} {137}},\ \bibinfo {pages}
  {439--442} (\bibinfo {year} {1984})}\BibitemShut {NoStop}%
\bibitem [{\citenamefont {Hillas}\ and\ \citenamefont
  {Cranshaw}(1959)}]{osti_4211117}%
  \BibitemOpen
  \bibfield  {author} {\bibinfo {author} {\bibfnamefont {A~M}\ \bibnamefont
  {Hillas}}\ and\ \bibinfo {author} {\bibfnamefont {T~E}\ \bibnamefont
  {Cranshaw}},\ }\bibfield  {title} {\enquote {\bibinfo {title} {A comparison
  of the charges of the electron, proton and neutron},}\ }\href {\doibase
  10.1038/184892a0} {\bibfield  {journal} {\bibinfo  {journal} {Nature}\
  }\textbf {\bibinfo {volume} {Vol: 184, Suppl. 12,}} (\bibinfo {year}
  {1959}),\ 10.1038/184892a0}\BibitemShut {NoStop}%
\bibitem [{\citenamefont {Lyttleton}\ and\ \citenamefont
  {Bondi}(1959)}]{Lyttleton:1959zz}%
  \BibitemOpen
  \bibfield  {author} {\bibinfo {author} {\bibfnamefont {R.A.}\ \bibnamefont
  {Lyttleton}}\ and\ \bibinfo {author} {\bibfnamefont {H.}~\bibnamefont
  {Bondi}},\ }\bibfield  {title} {\enquote {\bibinfo {title} {{On the physical
  consequences of a general excess of charge}},}\ }\href {\doibase
  10.1098/rspa.1959.0155} {\bibfield  {journal} {\bibinfo  {journal} {Proc.
  Roy. Soc. Lond. A}\ }\textbf {\bibinfo {volume} {252}},\ \bibinfo {pages}
  {313--333} (\bibinfo {year} {1959})}\BibitemShut {NoStop}%
\bibitem [{\citenamefont {Zyla}\ \emph {et~al.}(2020)\citenamefont {Zyla} \emph
  {et~al.}}]{Zyla:2020zbs}%
  \BibitemOpen
  \bibfield  {author} {\bibinfo {author} {\bibfnamefont {P.~A.}\ \bibnamefont
  {Zyla}} \emph {et~al.} (\bibinfo {collaboration} {Particle Data Group}),\
  }\bibfield  {title} {\enquote {\bibinfo {title} {{Review of Particle
  Physics}},}\ }\href {\doibase 10.1093/ptep/ptaa104} {\bibfield  {journal}
  {\bibinfo  {journal} {PTEP}\ }\textbf {\bibinfo {volume} {2020}},\ \bibinfo
  {pages} {083C01} (\bibinfo {year} {2020})}\BibitemShut {NoStop}%
\bibitem [{\citenamefont {Holdom}(1986)}]{Holdom:1985ag}%
  \BibitemOpen
  \bibfield  {author} {\bibinfo {author} {\bibfnamefont {Bob}\ \bibnamefont
  {Holdom}},\ }\bibfield  {title} {\enquote {\bibinfo {title} {{Two U(1)'s and
  Epsilon Charge Shifts}},}\ }\href {\doibase 10.1016/0370-2693(86)91377-8}
  {\bibfield  {journal} {\bibinfo  {journal} {Phys. Lett. B}\ }\textbf
  {\bibinfo {volume} {166}},\ \bibinfo {pages} {196--198} (\bibinfo {year}
  {1986})}\BibitemShut {NoStop}%
\bibitem [{\citenamefont {FOOT}(2004)}]{FOOT_2004}%
  \BibitemOpen
  \bibfield  {author} {\bibinfo {author} {\bibfnamefont {R.}~\bibnamefont
  {FOOT}},\ }\bibfield  {title} {\enquote {\bibinfo {title} {Mirror matter-type
  dark matter},}\ }\href {\doibase 10.1142/s0218271804006449} {\bibfield
  {journal} {\bibinfo  {journal} {International Journal of Modern Physics D}\
  }\textbf {\bibinfo {volume} {13}},\ \bibinfo {pages} {2161–2192} (\bibinfo
  {year} {2004})}\BibitemShut {NoStop}%
\bibitem [{\citenamefont {Magill}\ \emph {et~al.}(2019)\citenamefont {Magill},
  \citenamefont {Plestid}, \citenamefont {Pospelov},\ and\ \citenamefont
  {Tsai}}]{Magill:2018tbb}%
  \BibitemOpen
  \bibfield  {author} {\bibinfo {author} {\bibfnamefont {Gabriel}\ \bibnamefont
  {Magill}}, \bibinfo {author} {\bibfnamefont {Ryan}\ \bibnamefont {Plestid}},
  \bibinfo {author} {\bibfnamefont {Maxim}\ \bibnamefont {Pospelov}}, \ and\
  \bibinfo {author} {\bibfnamefont {Yu-Dai}\ \bibnamefont {Tsai}},\ }\bibfield
  {title} {\enquote {\bibinfo {title} {{Millicharged particles in neutrino
  experiments}},}\ }\href {\doibase 10.1103/PhysRevLett.122.071801} {\bibfield
  {journal} {\bibinfo  {journal} {Phys. Rev. Lett.}\ }\textbf {\bibinfo
  {volume} {122}},\ \bibinfo {pages} {071801} (\bibinfo {year} {2019})},\
  \Eprint {http://arxiv.org/abs/1806.03310} {arXiv:1806.03310 [hep-ph]}
  \BibitemShut {NoStop}%
\bibitem [{\citenamefont {Harnik}\ \emph {et~al.}(2019)\citenamefont {Harnik},
  \citenamefont {Liu},\ and\ \citenamefont {Palamara}}]{Harnik:2019zee}%
  \BibitemOpen
  \bibfield  {author} {\bibinfo {author} {\bibfnamefont {Roni}\ \bibnamefont
  {Harnik}}, \bibinfo {author} {\bibfnamefont {Zhen}\ \bibnamefont {Liu}}, \
  and\ \bibinfo {author} {\bibfnamefont {Ornella}\ \bibnamefont {Palamara}},\
  }\bibfield  {title} {\enquote {\bibinfo {title} {{Millicharged Particles in
  Liquid Argon Neutrino Experiments}},}\ }\href {\doibase
  10.1007/JHEP07(2019)170} {\bibfield  {journal} {\bibinfo  {journal} {JHEP}\
  }\textbf {\bibinfo {volume} {07}},\ \bibinfo {pages} {170} (\bibinfo {year}
  {2019})},\ \Eprint {http://arxiv.org/abs/1902.03246} {arXiv:1902.03246
  [hep-ph]} \BibitemShut {NoStop}%
\bibitem [{\citenamefont {Haas}\ \emph {et~al.}(2015)\citenamefont {Haas},
  \citenamefont {Hill}, \citenamefont {Izaguirre},\ and\ \citenamefont
  {Yavin}}]{Haas:2014dda}%
  \BibitemOpen
  \bibfield  {author} {\bibinfo {author} {\bibfnamefont {Andrew}\ \bibnamefont
  {Haas}}, \bibinfo {author} {\bibfnamefont {Christopher~S.}\ \bibnamefont
  {Hill}}, \bibinfo {author} {\bibfnamefont {Eder}\ \bibnamefont {Izaguirre}},
  \ and\ \bibinfo {author} {\bibfnamefont {Itay}\ \bibnamefont {Yavin}},\
  }\bibfield  {title} {\enquote {\bibinfo {title} {{Looking for milli-charged
  particles with a new experiment at the LHC}},}\ }\href {\doibase
  10.1016/j.physletb.2015.04.062} {\bibfield  {journal} {\bibinfo  {journal}
  {Phys. Lett. B}\ }\textbf {\bibinfo {volume} {746}},\ \bibinfo {pages}
  {117--120} (\bibinfo {year} {2015})},\ \Eprint
  {http://arxiv.org/abs/1410.6816} {arXiv:1410.6816 [hep-ph]} \BibitemShut
  {NoStop}%
\bibitem [{\citenamefont {Yoo}(2019)}]{Yoo:2018lhk}%
  \BibitemOpen
  \bibfield  {author} {\bibinfo {author} {\bibfnamefont {Jae~Hyeok}\
  \bibnamefont {Yoo}} (\bibinfo {collaboration} {milliQan}),\ }\bibfield
  {title} {\enquote {\bibinfo {title} {{The milliQan Experiment: Search for
  milli-charged Particles at the LHC}},}\ }\href {\doibase 10.22323/1.340.0520}
  {\bibfield  {journal} {\bibinfo  {journal} {PoS}\ }\textbf {\bibinfo {volume}
  {ICHEP2018}},\ \bibinfo {pages} {520} (\bibinfo {year} {2019})},\ \Eprint
  {http://arxiv.org/abs/1810.06733} {arXiv:1810.06733 [physics.ins-det]}
  \BibitemShut {NoStop}%
\bibitem [{\citenamefont {Kelly}\ and\ \citenamefont
  {Tsai}(2019)}]{Kelly:2018brz}%
  \BibitemOpen
  \bibfield  {author} {\bibinfo {author} {\bibfnamefont {Kevin~J.}\
  \bibnamefont {Kelly}}\ and\ \bibinfo {author} {\bibfnamefont {Yu-Dai}\
  \bibnamefont {Tsai}},\ }\bibfield  {title} {\enquote {\bibinfo {title}
  {{Proton fixed-target scintillation experiment to search for millicharged
  dark matter}},}\ }\href {\doibase 10.1103/PhysRevD.100.015043} {\bibfield
  {journal} {\bibinfo  {journal} {Phys. Rev. D}\ }\textbf {\bibinfo {volume}
  {100}},\ \bibinfo {pages} {015043} (\bibinfo {year} {2019})},\ \Eprint
  {http://arxiv.org/abs/1812.03998} {arXiv:1812.03998 [hep-ph]} \BibitemShut
  {NoStop}%
\bibitem [{\citenamefont {Foroughi-Abari}\ \emph {et~al.}(2020)\citenamefont
  {Foroughi-Abari}, \citenamefont {Kling},\ and\ \citenamefont
  {Tsai}}]{Foroughi-Abari:2020qar}%
  \BibitemOpen
  \bibfield  {author} {\bibinfo {author} {\bibfnamefont {Saeid}\ \bibnamefont
  {Foroughi-Abari}}, \bibinfo {author} {\bibfnamefont {Felix}\ \bibnamefont
  {Kling}}, \ and\ \bibinfo {author} {\bibfnamefont {Yu-Dai}\ \bibnamefont
  {Tsai}},\ }\bibfield  {title} {\enquote {\bibinfo {title} {{FORMOSA: Looking
  Forward to Millicharged Dark Sectors}},}\ }\href@noop {} {\  (\bibinfo {year}
  {2020})},\ \Eprint {http://arxiv.org/abs/2010.07941} {arXiv:2010.07941
  [hep-ph]} \BibitemShut {NoStop}%
\bibitem [{\citenamefont {Kim}\ \emph {et~al.}(2021)\citenamefont {Kim},
  \citenamefont {Hwang},\ and\ \citenamefont {Yoo}}]{Kim:2021eix}%
  \BibitemOpen
  \bibfield  {author} {\bibinfo {author} {\bibfnamefont {Jeong~Hwa}\
  \bibnamefont {Kim}}, \bibinfo {author} {\bibfnamefont {In~Sung}\ \bibnamefont
  {Hwang}}, \ and\ \bibinfo {author} {\bibfnamefont {Jae~Hyeok}\ \bibnamefont
  {Yoo}},\ }\bibfield  {title} {\enquote {\bibinfo {title} {{Search for
  sub-millicharged particles at J-PARC}},}\ }\href@noop {} {\  (\bibinfo {year}
  {2021})},\ \Eprint {http://arxiv.org/abs/2102.11493} {arXiv:2102.11493
  [hep-ex]} \BibitemShut {NoStop}%
\bibitem [{\citenamefont {Gorbunov}\ \emph {et~al.}(2021)\citenamefont
  {Gorbunov}, \citenamefont {Krasnov}, \citenamefont {Kudenko},\ and\
  \citenamefont {Suvorov}}]{Gorbunov:2021jog}%
  \BibitemOpen
  \bibfield  {author} {\bibinfo {author} {\bibfnamefont {Dmitry}\ \bibnamefont
  {Gorbunov}}, \bibinfo {author} {\bibfnamefont {Igor}\ \bibnamefont
  {Krasnov}}, \bibinfo {author} {\bibfnamefont {Yury}\ \bibnamefont {Kudenko}},
  \ and\ \bibinfo {author} {\bibfnamefont {Sergey}\ \bibnamefont {Suvorov}},\
  }\bibfield  {title} {\enquote {\bibinfo {title} {{Double-Hit Signature of
  Millicharged Particles in 3D segmented neutrino detector}},}\ }\href@noop {}
  {\  (\bibinfo {year} {2021})},\ \Eprint {http://arxiv.org/abs/2103.11814}
  {arXiv:2103.11814 [hep-ph]} \BibitemShut {NoStop}%
\bibitem [{\citenamefont {Marocco}\ and\ \citenamefont
  {Sarkar}(2021)}]{Marocco:2020dqu}%
  \BibitemOpen
  \bibfield  {author} {\bibinfo {author} {\bibfnamefont {Giacomo}\ \bibnamefont
  {Marocco}}\ and\ \bibinfo {author} {\bibfnamefont {Subir}\ \bibnamefont
  {Sarkar}},\ }\bibfield  {title} {\enquote {\bibinfo {title} {{Blast from the
  past: Constraints on the dark sector from the BEBC WA66 beam dump
  experiment}},}\ }\href {\doibase 10.21468/SciPostPhys.10.2.043} {\bibfield
  {journal} {\bibinfo  {journal} {SciPost Phys.}\ }\textbf {\bibinfo {volume}
  {10}},\ \bibinfo {pages} {043} (\bibinfo {year} {2021})},\ \Eprint
  {http://arxiv.org/abs/2011.08153} {arXiv:2011.08153 [hep-ph]} \BibitemShut
  {NoStop}%
\bibitem [{\citenamefont {Acciarri}\ \emph {et~al.}(2020)\citenamefont
  {Acciarri} \emph {et~al.}}]{Acciarri:2019jly}%
  \BibitemOpen
  \bibfield  {author} {\bibinfo {author} {\bibfnamefont {R.}~\bibnamefont
  {Acciarri}} \emph {et~al.} (\bibinfo {collaboration} {ArgoNeuT}),\ }\bibfield
   {title} {\enquote {\bibinfo {title} {{Improved Limits on Millicharged
  Particles Using the ArgoNeuT Experiment at Fermilab}},}\ }\href {\doibase
  10.1103/PhysRevLett.124.131801} {\bibfield  {journal} {\bibinfo  {journal}
  {Phys. Rev. Lett.}\ }\textbf {\bibinfo {volume} {124}},\ \bibinfo {pages}
  {131801} (\bibinfo {year} {2020})},\ \Eprint
  {http://arxiv.org/abs/1911.07996} {arXiv:1911.07996 [hep-ex]} \BibitemShut
  {NoStop}%
\bibitem [{\citenamefont {Plestid}\ \emph {et~al.}(2020)\citenamefont
  {Plestid}, \citenamefont {Takhistov}, \citenamefont {Tsai}, \citenamefont
  {Bringmann}, \citenamefont {Kusenko},\ and\ \citenamefont
  {Pospelov}}]{Plestid:2020kdm}%
  \BibitemOpen
  \bibfield  {author} {\bibinfo {author} {\bibfnamefont {Ryan}\ \bibnamefont
  {Plestid}}, \bibinfo {author} {\bibfnamefont {Volodymyr}\ \bibnamefont
  {Takhistov}}, \bibinfo {author} {\bibfnamefont {Yu-Dai}\ \bibnamefont
  {Tsai}}, \bibinfo {author} {\bibfnamefont {Torsten}\ \bibnamefont
  {Bringmann}}, \bibinfo {author} {\bibfnamefont {Alexander}\ \bibnamefont
  {Kusenko}}, \ and\ \bibinfo {author} {\bibfnamefont {Maxim}\ \bibnamefont
  {Pospelov}},\ }\bibfield  {title} {\enquote {\bibinfo {title} {{New
  Constraints on Millicharged Particles from Cosmic-ray Production}},}\
  }\href@noop {} {\  (\bibinfo {year} {2020})},\ \Eprint
  {http://arxiv.org/abs/2002.11732} {arXiv:2002.11732 [hep-ph]} \BibitemShut
  {NoStop}%
\bibitem [{\citenamefont {Harnik}\ \emph {et~al.}(2020)\citenamefont {Harnik},
  \citenamefont {Plestid}, \citenamefont {Pospelov},\ and\ \citenamefont
  {Ramani}}]{Harnik:2020ugb}%
  \BibitemOpen
  \bibfield  {author} {\bibinfo {author} {\bibfnamefont {Roni}\ \bibnamefont
  {Harnik}}, \bibinfo {author} {\bibfnamefont {Ryan}\ \bibnamefont {Plestid}},
  \bibinfo {author} {\bibfnamefont {Maxim}\ \bibnamefont {Pospelov}}, \ and\
  \bibinfo {author} {\bibfnamefont {Harikrishnan}\ \bibnamefont {Ramani}},\
  }\bibfield  {title} {\enquote {\bibinfo {title} {{Millicharged Cosmic Rays
  and Low Recoil Detectors}},}\ }\href@noop {} {\  (\bibinfo {year} {2020})},\
  \Eprint {http://arxiv.org/abs/2010.11190} {arXiv:2010.11190 [hep-ph]}
  \BibitemShut {NoStop}%
\bibitem [{\citenamefont {Pospelov}\ and\ \citenamefont
  {Ramani}(2020)}]{Pospelov:2020ktu}%
  \BibitemOpen
  \bibfield  {author} {\bibinfo {author} {\bibfnamefont {Maxim}\ \bibnamefont
  {Pospelov}}\ and\ \bibinfo {author} {\bibfnamefont {Harikrishnan}\
  \bibnamefont {Ramani}},\ }\bibfield  {title} {\enquote {\bibinfo {title}
  {{Earth-bound Milli-charge Relics}},}\ }\href@noop {} {\  (\bibinfo {year}
  {2020})},\ \Eprint {http://arxiv.org/abs/2012.03957} {arXiv:2012.03957
  [hep-ph]} \BibitemShut {NoStop}%
\bibitem [{\citenamefont {An}\ \emph {et~al.}(2016)\citenamefont {An} \emph
  {et~al.}}]{An:2015jdp}%
  \BibitemOpen
  \bibfield  {author} {\bibinfo {author} {\bibfnamefont {Fengpeng}\
  \bibnamefont {An}} \emph {et~al.} (\bibinfo {collaboration} {JUNO}),\
  }\bibfield  {title} {\enquote {\bibinfo {title} {{Neutrino Physics with
  JUNO}},}\ }\href {\doibase 10.1088/0954-3899/43/3/030401} {\bibfield
  {journal} {\bibinfo  {journal} {J. Phys. G}\ }\textbf {\bibinfo {volume}
  {43}},\ \bibinfo {pages} {030401} (\bibinfo {year} {2016})},\ \Eprint
  {http://arxiv.org/abs/1507.05613} {arXiv:1507.05613 [physics.ins-det]}
  \BibitemShut {NoStop}%
\bibitem [{\citenamefont {Abusleme}\ \emph {et~al.}(2021)\citenamefont
  {Abusleme} \emph {et~al.}}]{Abusleme:2021zrw}%
  \BibitemOpen
  \bibfield  {author} {\bibinfo {author} {\bibfnamefont {Angel}\ \bibnamefont
  {Abusleme}} \emph {et~al.} (\bibinfo {collaboration} {JUNO}),\ }\bibfield
  {title} {\enquote {\bibinfo {title} {{JUNO Physics and Detector}},}\
  }\href@noop {} {\  (\bibinfo {year} {2021})},\ \Eprint
  {http://arxiv.org/abs/2104.02565} {arXiv:2104.02565 [hep-ex]} \BibitemShut
  {NoStop}%
\bibitem [{\citenamefont {Bowman}\ \emph {et~al.}(2018)\citenamefont {Bowman},
  \citenamefont {Rogers}, \citenamefont {Monsalve}, \citenamefont {Mozdzen},\
  and\ \citenamefont {Mahesh}}]{Bowman:2018yin}%
  \BibitemOpen
  \bibfield  {author} {\bibinfo {author} {\bibfnamefont {Judd~D.}\ \bibnamefont
  {Bowman}}, \bibinfo {author} {\bibfnamefont {Alan E.~E.}\ \bibnamefont
  {Rogers}}, \bibinfo {author} {\bibfnamefont {Raul~A.}\ \bibnamefont
  {Monsalve}}, \bibinfo {author} {\bibfnamefont {Thomas~J.}\ \bibnamefont
  {Mozdzen}}, \ and\ \bibinfo {author} {\bibfnamefont {Nivedita}\ \bibnamefont
  {Mahesh}},\ }\bibfield  {title} {\enquote {\bibinfo {title} {{An absorption
  profile centred at 78 megahertz in the sky-averaged spectrum}},}\ }\href
  {\doibase 10.1038/nature25792} {\bibfield  {journal} {\bibinfo  {journal}
  {Nature}\ }\textbf {\bibinfo {volume} {555}},\ \bibinfo {pages} {67--70}
  (\bibinfo {year} {2018})},\ \Eprint {http://arxiv.org/abs/1810.05912}
  {arXiv:1810.05912 [astro-ph.CO]} \BibitemShut {NoStop}%
\bibitem [{\citenamefont {Berlin}\ \emph {et~al.}(2018)\citenamefont {Berlin},
  \citenamefont {Hooper}, \citenamefont {Krnjaic},\ and\ \citenamefont
  {McDermott}}]{Berlin:2018sjs}%
  \BibitemOpen
  \bibfield  {author} {\bibinfo {author} {\bibfnamefont {Asher}\ \bibnamefont
  {Berlin}}, \bibinfo {author} {\bibfnamefont {Dan}\ \bibnamefont {Hooper}},
  \bibinfo {author} {\bibfnamefont {Gordan}\ \bibnamefont {Krnjaic}}, \ and\
  \bibinfo {author} {\bibfnamefont {Samuel~D.}\ \bibnamefont {McDermott}},\
  }\bibfield  {title} {\enquote {\bibinfo {title} {{Severely Constraining Dark
  Matter Interpretations of the 21-cm Anomaly}},}\ }\href {\doibase
  10.1103/PhysRevLett.121.011102} {\bibfield  {journal} {\bibinfo  {journal}
  {Phys. Rev. Lett.}\ }\textbf {\bibinfo {volume} {121}},\ \bibinfo {pages}
  {011102} (\bibinfo {year} {2018})},\ \Eprint
  {http://arxiv.org/abs/1803.02804} {arXiv:1803.02804 [hep-ph]} \BibitemShut
  {NoStop}%
\bibitem [{\citenamefont {Kovetz}\ \emph {et~al.}(2018)\citenamefont {Kovetz},
  \citenamefont {Poulin}, \citenamefont {Gluscevic}, \citenamefont {Boddy},
  \citenamefont {Barkana},\ and\ \citenamefont
  {Kamionkowski}}]{Kovetz:2018zan}%
  \BibitemOpen
  \bibfield  {author} {\bibinfo {author} {\bibfnamefont {Ely~D.}\ \bibnamefont
  {Kovetz}}, \bibinfo {author} {\bibfnamefont {Vivian}\ \bibnamefont {Poulin}},
  \bibinfo {author} {\bibfnamefont {Vera}\ \bibnamefont {Gluscevic}}, \bibinfo
  {author} {\bibfnamefont {Kimberly~K.}\ \bibnamefont {Boddy}}, \bibinfo
  {author} {\bibfnamefont {Rennan}\ \bibnamefont {Barkana}}, \ and\ \bibinfo
  {author} {\bibfnamefont {Marc}\ \bibnamefont {Kamionkowski}},\ }\bibfield
  {title} {\enquote {\bibinfo {title} {{Tighter limits on dark matter
  explanations of the anomalous EDGES 21 cm signal}},}\ }\href {\doibase
  10.1103/PhysRevD.98.103529} {\bibfield  {journal} {\bibinfo  {journal} {Phys.
  Rev. D}\ }\textbf {\bibinfo {volume} {98}},\ \bibinfo {pages} {103529}
  (\bibinfo {year} {2018})},\ \Eprint {http://arxiv.org/abs/1807.11482}
  {arXiv:1807.11482 [astro-ph.CO]} \BibitemShut {NoStop}%
\bibitem [{\citenamefont {Creque-Sarbinowski}\ \emph
  {et~al.}(2019)\citenamefont {Creque-Sarbinowski}, \citenamefont {Ji},
  \citenamefont {Kovetz},\ and\ \citenamefont
  {Kamionkowski}}]{Creque-Sarbinowski:2019mcm}%
  \BibitemOpen
  \bibfield  {author} {\bibinfo {author} {\bibfnamefont {Cyril}\ \bibnamefont
  {Creque-Sarbinowski}}, \bibinfo {author} {\bibfnamefont {Lingyuan}\
  \bibnamefont {Ji}}, \bibinfo {author} {\bibfnamefont {Ely~D.}\ \bibnamefont
  {Kovetz}}, \ and\ \bibinfo {author} {\bibfnamefont {Marc}\ \bibnamefont
  {Kamionkowski}},\ }\bibfield  {title} {\enquote {\bibinfo {title} {{Direct
  millicharged dark matter cannot explain the EDGES signal}},}\ }\href
  {\doibase 10.1103/PhysRevD.100.023528} {\bibfield  {journal} {\bibinfo
  {journal} {Phys. Rev. D}\ }\textbf {\bibinfo {volume} {100}},\ \bibinfo
  {pages} {023528} (\bibinfo {year} {2019})},\ \Eprint
  {http://arxiv.org/abs/1903.09154} {arXiv:1903.09154 [astro-ph.CO]}
  \BibitemShut {NoStop}%
\bibitem [{\citenamefont {Boehm}\ \emph {et~al.}(2013)\citenamefont {Boehm},
  \citenamefont {Dolan},\ and\ \citenamefont {McCabe}}]{Boehm:2013jpa}%
  \BibitemOpen
  \bibfield  {author} {\bibinfo {author} {\bibfnamefont {C\'eline}\
  \bibnamefont {Boehm}}, \bibinfo {author} {\bibfnamefont {Matthew~J.}\
  \bibnamefont {Dolan}}, \ and\ \bibinfo {author} {\bibfnamefont {Christopher}\
  \bibnamefont {McCabe}},\ }\bibfield  {title} {\enquote {\bibinfo {title} {{A
  Lower Bound on the Mass of Cold Thermal Dark Matter from Planck}},}\ }\href
  {\doibase 10.1088/1475-7516/2013/08/041} {\bibfield  {journal} {\bibinfo
  {journal} {JCAP}\ }\textbf {\bibinfo {volume} {08}},\ \bibinfo {pages} {041}
  (\bibinfo {year} {2013})},\ \Eprint {http://arxiv.org/abs/1303.6270}
  {arXiv:1303.6270 [hep-ph]} \BibitemShut {NoStop}%
\bibitem [{\citenamefont {Vogel}\ and\ \citenamefont
  {Redondo}(2014)}]{Vogel:2013raa}%
  \BibitemOpen
  \bibfield  {author} {\bibinfo {author} {\bibfnamefont {Hendrik}\ \bibnamefont
  {Vogel}}\ and\ \bibinfo {author} {\bibfnamefont {Javier}\ \bibnamefont
  {Redondo}},\ }\bibfield  {title} {\enquote {\bibinfo {title} {{Dark Radiation
  constraints on minicharged particles in models with a hidden photon}},}\
  }\href {\doibase 10.1088/1475-7516/2014/02/029} {\bibfield  {journal}
  {\bibinfo  {journal} {JCAP}\ }\textbf {\bibinfo {volume} {02}},\ \bibinfo
  {pages} {029} (\bibinfo {year} {2014})},\ \Eprint
  {http://arxiv.org/abs/1311.2600} {arXiv:1311.2600 [hep-ph]} \BibitemShut
  {NoStop}%
\bibitem [{\citenamefont {Chang}\ \emph {et~al.}(2018)\citenamefont {Chang},
  \citenamefont {Essig},\ and\ \citenamefont {McDermott}}]{Chang:2018rso}%
  \BibitemOpen
  \bibfield  {author} {\bibinfo {author} {\bibfnamefont {Jae~Hyeok}\
  \bibnamefont {Chang}}, \bibinfo {author} {\bibfnamefont {Rouven}\
  \bibnamefont {Essig}}, \ and\ \bibinfo {author} {\bibfnamefont {Samuel~D.}\
  \bibnamefont {McDermott}},\ }\bibfield  {title} {\enquote {\bibinfo {title}
  {{Supernova 1987A Constraints on Sub-GeV Dark Sectors, Millicharged
  Particles, the QCD Axion, and an Axion-like Particle}},}\ }\href {\doibase
  10.1007/JHEP09(2018)051} {\bibfield  {journal} {\bibinfo  {journal} {JHEP}\
  }\textbf {\bibinfo {volume} {09}},\ \bibinfo {pages} {051} (\bibinfo {year}
  {2018})},\ \Eprint {http://arxiv.org/abs/1803.00993} {arXiv:1803.00993
  [hep-ph]} \BibitemShut {NoStop}%
\bibitem [{\citenamefont {Emken}\ \emph {et~al.}(2019)\citenamefont {Emken},
  \citenamefont {Essig}, \citenamefont {Kouvaris},\ and\ \citenamefont
  {Sholapurkar}}]{Emken:2019tni}%
  \BibitemOpen
  \bibfield  {author} {\bibinfo {author} {\bibfnamefont {Timon}\ \bibnamefont
  {Emken}}, \bibinfo {author} {\bibfnamefont {Rouven}\ \bibnamefont {Essig}},
  \bibinfo {author} {\bibfnamefont {Chris}\ \bibnamefont {Kouvaris}}, \ and\
  \bibinfo {author} {\bibfnamefont {Mukul}\ \bibnamefont {Sholapurkar}},\
  }\bibfield  {title} {\enquote {\bibinfo {title} {{Direct Detection of
  Strongly Interacting Sub-GeV Dark Matter via Electron Recoils}},}\ }\href
  {\doibase 10.1088/1475-7516/2019/09/070} {\bibfield  {journal} {\bibinfo
  {journal} {JCAP}\ }\textbf {\bibinfo {volume} {09}},\ \bibinfo {pages} {070}
  (\bibinfo {year} {2019})},\ \Eprint {http://arxiv.org/abs/1905.06348}
  {arXiv:1905.06348 [hep-ph]} \BibitemShut {NoStop}%
\bibitem [{\citenamefont {Carney}\ \emph {et~al.}(2021)\citenamefont {Carney},
  \citenamefont {H\"affner}, \citenamefont {Moore},\ and\ \citenamefont
  {Taylor}}]{Carney:2021irt}%
  \BibitemOpen
  \bibfield  {author} {\bibinfo {author} {\bibfnamefont {Daniel}\ \bibnamefont
  {Carney}}, \bibinfo {author} {\bibfnamefont {Hartmut}\ \bibnamefont
  {H\"affner}}, \bibinfo {author} {\bibfnamefont {David~C.}\ \bibnamefont
  {Moore}}, \ and\ \bibinfo {author} {\bibfnamefont {Jacob~M.}\ \bibnamefont
  {Taylor}},\ }\bibfield  {title} {\enquote {\bibinfo {title} {{Trapped
  electrons and ions as particle detectors}},}\ }\href@noop {} {\  (\bibinfo
  {year} {2021})},\ \Eprint {http://arxiv.org/abs/2104.05737} {arXiv:2104.05737
  [quant-ph]} \BibitemShut {NoStop}%
\bibitem [{\citenamefont {Bird}\ \emph {et~al.}(1995)\citenamefont {Bird} \emph
  {et~al.}}]{Bird:1994uy}%
  \BibitemOpen
  \bibfield  {author} {\bibinfo {author} {\bibfnamefont {D.~J.}\ \bibnamefont
  {Bird}} \emph {et~al.},\ }\bibfield  {title} {\enquote {\bibinfo {title}
  {{Detection of a cosmic ray with measured energy well beyond the expected
  spectral cutoff due to cosmic microwave radiation}},}\ }\href {\doibase
  10.1086/175344} {\bibfield  {journal} {\bibinfo  {journal} {Astrophys. J.}\
  }\textbf {\bibinfo {volume} {441}},\ \bibinfo {pages} {144--150} (\bibinfo
  {year} {1995})},\ \Eprint {http://arxiv.org/abs/astro-ph/9410067}
  {arXiv:astro-ph/9410067} \BibitemShut {NoStop}%
\bibitem [{\citenamefont {Gondolo}\ \emph {et~al.}(1996)\citenamefont
  {Gondolo}, \citenamefont {Ingelman},\ and\ \citenamefont
  {Thunman}}]{Gondolo:1995fq}%
  \BibitemOpen
  \bibfield  {author} {\bibinfo {author} {\bibfnamefont {P.}~\bibnamefont
  {Gondolo}}, \bibinfo {author} {\bibfnamefont {G.}~\bibnamefont {Ingelman}}, \
  and\ \bibinfo {author} {\bibfnamefont {M.}~\bibnamefont {Thunman}},\
  }\bibfield  {title} {\enquote {\bibinfo {title} {{Charm production and
  high-energy atmospheric muon and neutrino fluxes}},}\ }\href {\doibase
  10.1016/0927-6505(96)00033-3} {\bibfield  {journal} {\bibinfo  {journal}
  {Astropart. Phys.}\ }\textbf {\bibinfo {volume} {5}},\ \bibinfo {pages}
  {309--332} (\bibinfo {year} {1996})},\ \Eprint
  {http://arxiv.org/abs/hep-ph/9505417} {arXiv:hep-ph/9505417} \BibitemShut
  {NoStop}%
\bibitem [{\citenamefont {Fedynitch}\ \emph {et~al.}(2015)\citenamefont
  {Fedynitch}, \citenamefont {Engel}, \citenamefont {Gaisser}, \citenamefont
  {Riehn},\ and\ \citenamefont {Stanev}}]{Fedynitch:2015zma}%
  \BibitemOpen
  \bibfield  {author} {\bibinfo {author} {\bibfnamefont {Anatoli}\ \bibnamefont
  {Fedynitch}}, \bibinfo {author} {\bibfnamefont {Ralph}\ \bibnamefont
  {Engel}}, \bibinfo {author} {\bibfnamefont {Thomas~K.}\ \bibnamefont
  {Gaisser}}, \bibinfo {author} {\bibfnamefont {Felix}\ \bibnamefont {Riehn}},
  \ and\ \bibinfo {author} {\bibfnamefont {Todor}\ \bibnamefont {Stanev}},\
  }\bibfield  {title} {\enquote {\bibinfo {title} {{Calculation of conventional
  and prompt lepton fluxes at very high energy}},}\ }\bibfield  {booktitle}
  {\emph {\bibinfo {booktitle} {{Proceedings, 18th International Symposium on
  Very High Energy Cosmic Ray Interactions (ISVHECRI 2014): Geneva,
  Switzerland, August 18-22, 2014}}},\ }\href {\doibase
  10.1051/epjconf/20159908001} {\bibfield  {journal} {\bibinfo  {journal} {EPJ
  Web Conf.}\ }\textbf {\bibinfo {volume} {99}},\ \bibinfo {pages} {08001}
  (\bibinfo {year} {2015})},\ \Eprint {http://arxiv.org/abs/1503.00544}
  {arXiv:1503.00544 [hep-ph]} \BibitemShut {NoStop}%
%%CITATION = ARXIV:1503.00544;%%
\bibitem [{\citenamefont {Fedynitch}\ \emph {et~al.}(2012)\citenamefont
  {Fedynitch}, \citenamefont {Becker~Tjus},\ and\ \citenamefont
  {Desiati}}]{Fedynitch:2012fs}%
  \BibitemOpen
  \bibfield  {author} {\bibinfo {author} {\bibfnamefont {Anatoli}\ \bibnamefont
  {Fedynitch}}, \bibinfo {author} {\bibfnamefont {Julia}\ \bibnamefont
  {Becker~Tjus}}, \ and\ \bibinfo {author} {\bibfnamefont {Paolo}\ \bibnamefont
  {Desiati}},\ }\bibfield  {title} {\enquote {\bibinfo {title} {{Influence of
  hadronic interaction models and the cosmic ray spectrum on the high energy
  atmospheric muon and neutrino flux}},}\ }\href {\doibase
  10.1103/PhysRevD.86.114024} {\bibfield  {journal} {\bibinfo  {journal} {Phys.
  Rev.}\ }\textbf {\bibinfo {volume} {D86}},\ \bibinfo {pages} {114024}
  (\bibinfo {year} {2012})},\ \Eprint {http://arxiv.org/abs/1206.6710}
  {arXiv:1206.6710 [astro-ph.HE]} \BibitemShut {NoStop}%
%%CITATION = ARXIV:1206.6710;%%
\bibitem [{\citenamefont {Fedynitch}\ \emph {et~al.}(2018)\citenamefont
  {Fedynitch}, \citenamefont {Riehn}, \citenamefont {Engel}, \citenamefont
  {Gaisser},\ and\ \citenamefont {Stanev}}]{Fedynitch:2018cbl}%
  \BibitemOpen
  \bibfield  {author} {\bibinfo {author} {\bibfnamefont {Anatoli}\ \bibnamefont
  {Fedynitch}}, \bibinfo {author} {\bibfnamefont {Felix}\ \bibnamefont
  {Riehn}}, \bibinfo {author} {\bibfnamefont {Ralph}\ \bibnamefont {Engel}},
  \bibinfo {author} {\bibfnamefont {Thomas~K.}\ \bibnamefont {Gaisser}}, \ and\
  \bibinfo {author} {\bibfnamefont {Todor}\ \bibnamefont {Stanev}},\ }\bibfield
   {title} {\enquote {\bibinfo {title} {{The hadronic interaction model
  Sibyll-2.3c and inclusive lepton fluxes}},}\ }\href@noop {} {\  (\bibinfo
  {year} {2018})},\ \Eprint {http://arxiv.org/abs/1806.04140} {arXiv:1806.04140
  [hep-ph]} \BibitemShut {NoStop}%
%%CITATION = ARXIV:1806.04140;%%
\bibitem [{\citenamefont {Gaisser}(2012)}]{Gaisser:2011cc}%
  \BibitemOpen
  \bibfield  {author} {\bibinfo {author} {\bibfnamefont {Thomas~K.}\
  \bibnamefont {Gaisser}},\ }\bibfield  {title} {\enquote {\bibinfo {title}
  {{Spectrum of cosmic-ray nucleons, kaon production, and the atmospheric muon
  charge ratio}},}\ }\href {\doibase 10.1016/j.astropartphys.2012.02.010}
  {\bibfield  {journal} {\bibinfo  {journal} {Astropart. Phys.}\ }\textbf
  {\bibinfo {volume} {35}},\ \bibinfo {pages} {801--806} (\bibinfo {year}
  {2012})},\ \Eprint {http://arxiv.org/abs/1111.6675} {arXiv:1111.6675
  [astro-ph.HE]} \BibitemShut {NoStop}%
%%CITATION = ARXIV:1111.6675;%%
\bibitem [{\citenamefont {Picone}\ \emph {et~al.}(2002)\citenamefont {Picone},
  \citenamefont {Hedin}, \citenamefont {Drob},\ and\ \citenamefont
  {Aikin}}]{Picone:2002go}%
  \BibitemOpen
  \bibfield  {author} {\bibinfo {author} {\bibfnamefont {J.~M.}\ \bibnamefont
  {Picone}}, \bibinfo {author} {\bibfnamefont {A.~E.}\ \bibnamefont {Hedin}},
  \bibinfo {author} {\bibfnamefont {D.~P.}\ \bibnamefont {Drob}}, \ and\
  \bibinfo {author} {\bibfnamefont {A.~C.}\ \bibnamefont {Aikin}},\ }\bibfield
  {title} {\enquote {\bibinfo {title} {Nrlmsise-00 empirical model of the
  atmosphere: Statistical comparisons and scientific issues},}\ }\href
  {\doibase 10.1029/2002JA009430} {\bibfield  {journal} {\bibinfo  {journal}
  {Journal of Geophysical Research: Space Physics}\ }\textbf {\bibinfo {volume}
  {107}},\ \bibinfo {pages} {SIA 15--1--SIA 15--16} (\bibinfo {year}
  {2002})}\BibitemShut {NoStop}%
\bibitem [{\citenamefont {Kachelrie\ss{}}\ and\ \citenamefont
  {Tjemsland}(2021)}]{Kachelriess:2021lpm}%
  \BibitemOpen
  \bibfield  {author} {\bibinfo {author} {\bibfnamefont {M.}~\bibnamefont
  {Kachelrie\ss{}}}\ and\ \bibinfo {author} {\bibfnamefont {J.}~\bibnamefont
  {Tjemsland}},\ }\bibfield  {title} {\enquote {\bibinfo {title} {{Meson
  production in air showers and the search for light exotic particles}},}\
  }\href@noop {} {\  (\bibinfo {year} {2021})},\ \Eprint
  {http://arxiv.org/abs/2104.06811} {arXiv:2104.06811 [hep-ph]} \BibitemShut
  {NoStop}%
\bibitem [{\citenamefont {Gaisser}(2005)}]{Gaisser:2005tu}%
  \BibitemOpen
  \bibfield  {author} {\bibinfo {author} {\bibfnamefont {Thomas~K.}\
  \bibnamefont {Gaisser}},\ }\bibfield  {title} {\enquote {\bibinfo {title}
  {{Outstanding problems in particle astrophysics}},}\ }in\ \href@noop {}
  {\emph {\bibinfo {booktitle} {{International School of Cosmic Ray
  Astrophysics: 14th Course: Neutrinos and Explosive Events in the Universe: A
  NATO Advanced Study Institute}}}}\ (\bibinfo {year} {2005})\ \Eprint
  {http://arxiv.org/abs/astro-ph/0501195} {arXiv:astro-ph/0501195} \BibitemShut
  {NoStop}%
\bibitem [{\citenamefont {Gaisser}\ and\ \citenamefont
  {Stanev}(2006)}]{Gaisser:2006sf}%
  \BibitemOpen
  \bibfield  {author} {\bibinfo {author} {\bibfnamefont {Thomas~K.}\
  \bibnamefont {Gaisser}}\ and\ \bibinfo {author} {\bibfnamefont {Todor}\
  \bibnamefont {Stanev}},\ }\bibfield  {title} {\enquote {\bibinfo {title}
  {{High-energy cosmic rays}},}\ }\href {\doibase
  10.1016/j.nuclphysa.2005.01.024} {\bibfield  {journal} {\bibinfo  {journal}
  {Nucl. Phys. A}\ }\textbf {\bibinfo {volume} {777}},\ \bibinfo {pages}
  {98--110} (\bibinfo {year} {2006})},\ \Eprint
  {http://arxiv.org/abs/astro-ph/0510321} {arXiv:astro-ph/0510321} \BibitemShut
  {NoStop}%
\bibitem [{\citenamefont {Fedynitch}\ \emph {et~al.}(2019)\citenamefont
  {Fedynitch}, \citenamefont {Riehn}, \citenamefont {Engel}, \citenamefont
  {Gaisser},\ and\ \citenamefont {Stanev}}]{Fedynitch_2019}%
  \BibitemOpen
  \bibfield  {author} {\bibinfo {author} {\bibfnamefont {Anatoli}\ \bibnamefont
  {Fedynitch}}, \bibinfo {author} {\bibfnamefont {Felix}\ \bibnamefont
  {Riehn}}, \bibinfo {author} {\bibfnamefont {Ralph}\ \bibnamefont {Engel}},
  \bibinfo {author} {\bibfnamefont {Thomas~K.}\ \bibnamefont {Gaisser}}, \ and\
  \bibinfo {author} {\bibfnamefont {Todor}\ \bibnamefont {Stanev}},\ }\bibfield
   {title} {\enquote {\bibinfo {title} {Hadronic interaction model sibyll 2.3c
  and inclusive lepton fluxes},}\ }\href {\doibase 10.1103/physrevd.100.103018}
  {\bibfield  {journal} {\bibinfo  {journal} {Physical Review D}\ }\textbf
  {\bibinfo {volume} {100}} (\bibinfo {year} {2019}),\
  10.1103/physrevd.100.103018}\BibitemShut {NoStop}%
\bibitem [{\citenamefont {Aartsen}\ \emph {et~al.}(2016)\citenamefont
  {Aartsen}, \citenamefont {Abraham}, \citenamefont {Ackermann}, \citenamefont
  {Adams}, \citenamefont {Aguilar}, \citenamefont {Ahlers}, \citenamefont
  {Ahrens}, \citenamefont {Altmann}, \citenamefont {Anderson}, \citenamefont
  {Archinger},\ and\ \citenamefont {et~al.}}]{Aartsen_2016}%
  \BibitemOpen
  \bibfield  {author} {\bibinfo {author} {\bibfnamefont {M.G.}\ \bibnamefont
  {Aartsen}}, \bibinfo {author} {\bibfnamefont {K.}~\bibnamefont {Abraham}},
  \bibinfo {author} {\bibfnamefont {M.}~\bibnamefont {Ackermann}}, \bibinfo
  {author} {\bibfnamefont {J.}~\bibnamefont {Adams}}, \bibinfo {author}
  {\bibfnamefont {J.A.}\ \bibnamefont {Aguilar}}, \bibinfo {author}
  {\bibfnamefont {M.}~\bibnamefont {Ahlers}}, \bibinfo {author} {\bibfnamefont
  {M.}~\bibnamefont {Ahrens}}, \bibinfo {author} {\bibfnamefont
  {D.}~\bibnamefont {Altmann}}, \bibinfo {author} {\bibfnamefont
  {T.}~\bibnamefont {Anderson}}, \bibinfo {author} {\bibfnamefont
  {M.}~\bibnamefont {Archinger}}, \ and\ \bibinfo {author} {\bibnamefont
  {et~al.}},\ }\bibfield  {title} {\enquote {\bibinfo {title} {Characterization
  of the atmospheric muon flux in icecube},}\ }\href {\doibase
  10.1016/j.astropartphys.2016.01.006} {\bibfield  {journal} {\bibinfo
  {journal} {Astroparticle Physics}\ }\textbf {\bibinfo {volume} {78}},\
  \bibinfo {pages} {1–27} (\bibinfo {year} {2016})}\BibitemShut {NoStop}%
\bibitem [{\citenamefont {Gaisser}\ \emph {et~al.}(2013)\citenamefont
  {Gaisser}, \citenamefont {Stanev},\ and\ \citenamefont
  {Tilav}}]{gaisser2013cosmic}%
  \BibitemOpen
  \bibfield  {author} {\bibinfo {author} {\bibfnamefont {Thomas~K.}\
  \bibnamefont {Gaisser}}, \bibinfo {author} {\bibfnamefont {Todor}\
  \bibnamefont {Stanev}}, \ and\ \bibinfo {author} {\bibfnamefont {Serap}\
  \bibnamefont {Tilav}},\ }\href@noop {} {\enquote {\bibinfo {title} {Cosmic
  ray energy spectrum from measurements of air showers},}\ } (\bibinfo {year}
  {2013}),\ \Eprint {http://arxiv.org/abs/1303.3565} {arXiv:1303.3565
  [astro-ph.HE]} \BibitemShut {NoStop}%
\bibitem [{\citenamefont {Hörandel}(2003)}]{H_randel_2003}%
  \BibitemOpen
  \bibfield  {author} {\bibinfo {author} {\bibfnamefont {Jörg~R.}\
  \bibnamefont {Hörandel}},\ }\bibfield  {title} {\enquote {\bibinfo {title}
  {On the knee in the energy spectrum of cosmic rays},}\ }\href {\doibase
  10.1016/s0927-6505(02)00198-6} {\bibfield  {journal} {\bibinfo  {journal}
  {Astroparticle Physics}\ }\textbf {\bibinfo {volume} {19}},\ \bibinfo {pages}
  {193–220} (\bibinfo {year} {2003})}\BibitemShut {NoStop}%
\bibitem [{\citenamefont {Ostapchenko}(2011)}]{Ostapchenko_2011}%
  \BibitemOpen
  \bibfield  {author} {\bibinfo {author} {\bibfnamefont {S.}~\bibnamefont
  {Ostapchenko}},\ }\bibfield  {title} {\enquote {\bibinfo {title} {Monte carlo
  treatment of hadronic interactions in enhanced pomeron scheme: Qgsjet-ii
  model},}\ }\href {\doibase 10.1103/physrevd.83.014018} {\bibfield  {journal}
  {\bibinfo  {journal} {Physical Review D}\ }\textbf {\bibinfo {volume} {83}}
  (\bibinfo {year} {2011}),\ 10.1103/physrevd.83.014018}\BibitemShut {NoStop}%
\bibitem [{\citenamefont {Roesler}\ \emph {et~al.}(2001)\citenamefont
  {Roesler}, \citenamefont {Engel},\ and\ \citenamefont
  {Ranft}}]{Roesler_2001}%
  \BibitemOpen
  \bibfield  {author} {\bibinfo {author} {\bibfnamefont {S.}~\bibnamefont
  {Roesler}}, \bibinfo {author} {\bibfnamefont {R.}~\bibnamefont {Engel}}, \
  and\ \bibinfo {author} {\bibfnamefont {J.}~\bibnamefont {Ranft}},\ }\bibfield
   {title} {\enquote {\bibinfo {title} {The monte carlo event generator
  dpmjet-iii},}\ }\href {\doibase 10.1007/978-3-642-18211-2_166} {\bibfield
  {journal} {\bibinfo  {journal} {Advanced Monte Carlo for Radiation Physics,
  Particle Transport Simulation and Applications}\ ,\ \bibinfo {pages}
  {1033–1038}} (\bibinfo {year} {2001})}\BibitemShut {NoStop}%
\bibitem [{\citenamefont {Pierog}\ and\ \citenamefont
  {Werner}(2009)}]{Pierog_2009}%
  \BibitemOpen
  \bibfield  {author} {\bibinfo {author} {\bibfnamefont {T.}~\bibnamefont
  {Pierog}}\ and\ \bibinfo {author} {\bibfnamefont {K.}~\bibnamefont
  {Werner}},\ }\bibfield  {title} {\enquote {\bibinfo {title} {Epos model and
  ultra high energy cosmic rays},}\ }\href {\doibase
  10.1016/j.nuclphysbps.2009.09.017} {\bibfield  {journal} {\bibinfo  {journal}
  {Nuclear Physics B - Proceedings Supplements}\ }\textbf {\bibinfo {volume}
  {196}},\ \bibinfo {pages} {102–105} (\bibinfo {year} {2009})}\BibitemShut
  {NoStop}%
\bibitem [{\citenamefont {Aguilar-Benitez}\ \emph {et~al.}(1991)\citenamefont
  {Aguilar-Benitez} \emph {et~al.}}]{AguilarBenitez:1991yy}%
  \BibitemOpen
  \bibfield  {author} {\bibinfo {author} {\bibfnamefont {M.}~\bibnamefont
  {Aguilar-Benitez}} \emph {et~al.},\ }\bibfield  {title} {\enquote {\bibinfo
  {title} {{Inclusive particle production in 400-GeV/c p p interactions}},}\
  }\href {\doibase 10.1007/BF01551452} {\bibfield  {journal} {\bibinfo
  {journal} {Z. Phys. C}\ }\textbf {\bibinfo {volume} {50}},\ \bibinfo {pages}
  {405--426} (\bibinfo {year} {1991})}\BibitemShut {NoStop}%
\bibitem [{\citenamefont {Gribov}(1967)}]{Gribov:1968fc}%
  \BibitemOpen
  \bibfield  {author} {\bibinfo {author} {\bibfnamefont {V.~N.}\ \bibnamefont
  {Gribov}},\ }\bibfield  {title} {\enquote {\bibinfo {title} {{A REGGEON
  DIAGRAM TECHNIQUE}},}\ }\href@noop {} {\bibfield  {journal} {\bibinfo
  {journal} {Zh. Eksp. Teor. Fiz.}\ }\textbf {\bibinfo {volume} {53}},\
  \bibinfo {pages} {654--672} (\bibinfo {year} {1967})}\BibitemShut {NoStop}%
\bibitem [{\citenamefont {Prinz}(2001)}]{Prinz:2001qz}%
  \BibitemOpen
  \bibfield  {author} {\bibinfo {author} {\bibfnamefont {Alyssa~Ann}\
  \bibnamefont {Prinz}},\ }\emph {\bibinfo {title} {{The Search for
  millicharged particles at SLAC}}},\ \href@noop {} {\bibinfo {type} {Other
  thesis}} (\bibinfo {year} {2001})\BibitemShut {NoStop}%
\bibitem [{\citenamefont {Koehne}\ \emph {et~al.}(2013)\citenamefont {Koehne},
  \citenamefont {Frantzen}, \citenamefont {Schmitz}, \citenamefont {Fuchs},
  \citenamefont {Rhode}, \citenamefont {Chirkin},\ and\ \citenamefont
  {Becker~Tjus}}]{Koehne:2013gpa}%
  \BibitemOpen
  \bibfield  {author} {\bibinfo {author} {\bibfnamefont {J.~H.}\ \bibnamefont
  {Koehne}}, \bibinfo {author} {\bibfnamefont {K.}~\bibnamefont {Frantzen}},
  \bibinfo {author} {\bibfnamefont {M.}~\bibnamefont {Schmitz}}, \bibinfo
  {author} {\bibfnamefont {T.}~\bibnamefont {Fuchs}}, \bibinfo {author}
  {\bibfnamefont {W.}~\bibnamefont {Rhode}}, \bibinfo {author} {\bibfnamefont
  {D.}~\bibnamefont {Chirkin}}, \ and\ \bibinfo {author} {\bibfnamefont
  {J.}~\bibnamefont {Becker~Tjus}},\ }\bibfield  {title} {\enquote {\bibinfo
  {title} {{PROPOSAL: A tool for propagation of charged leptons}},}\ }\href
  {\doibase 10.1016/j.cpc.2013.04.001} {\bibfield  {journal} {\bibinfo
  {journal} {Comput. Phys. Commun.}\ }\textbf {\bibinfo {volume} {184}},\
  \bibinfo {pages} {2070--2090} (\bibinfo {year} {2013})}\BibitemShut {NoStop}%
\bibitem [{\citenamefont {Davidson}\ \emph {et~al.}(2000)\citenamefont
  {Davidson}, \citenamefont {Hannestad},\ and\ \citenamefont
  {Raffelt}}]{Davidson:2000hf}%
  \BibitemOpen
  \bibfield  {author} {\bibinfo {author} {\bibfnamefont {Sacha}\ \bibnamefont
  {Davidson}}, \bibinfo {author} {\bibfnamefont {Steen}\ \bibnamefont
  {Hannestad}}, \ and\ \bibinfo {author} {\bibfnamefont {Georg}\ \bibnamefont
  {Raffelt}},\ }\bibfield  {title} {\enquote {\bibinfo {title} {{Updated bounds
  on millicharged particles}},}\ }\href {\doibase
  10.1088/1126-6708/2000/05/003} {\bibfield  {journal} {\bibinfo  {journal}
  {JHEP}\ }\textbf {\bibinfo {volume} {05}},\ \bibinfo {pages} {003} (\bibinfo
  {year} {2000})},\ \Eprint {http://arxiv.org/abs/hep-ph/0001179}
  {arXiv:hep-ph/0001179} \BibitemShut {NoStop}%
\bibitem [{\citenamefont {Jaeckel}\ \emph {et~al.}(2013)\citenamefont
  {Jaeckel}, \citenamefont {Jankowiak},\ and\ \citenamefont
  {Spannowsky}}]{Jaeckel:2012yz}%
  \BibitemOpen
  \bibfield  {author} {\bibinfo {author} {\bibfnamefont {Joerg}\ \bibnamefont
  {Jaeckel}}, \bibinfo {author} {\bibfnamefont {Martin}\ \bibnamefont
  {Jankowiak}}, \ and\ \bibinfo {author} {\bibfnamefont {Michael}\ \bibnamefont
  {Spannowsky}},\ }\bibfield  {title} {\enquote {\bibinfo {title} {{LHC probes
  the hidden sector}},}\ }\href {\doibase 10.1016/j.dark.2013.06.001}
  {\bibfield  {journal} {\bibinfo  {journal} {Phys. Dark Univ.}\ }\textbf
  {\bibinfo {volume} {2}},\ \bibinfo {pages} {111--117} (\bibinfo {year}
  {2013})},\ \Eprint {http://arxiv.org/abs/1212.3620} {arXiv:1212.3620
  [hep-ph]} \BibitemShut {NoStop}%
\bibitem [{\citenamefont {Ball}\ \emph {et~al.}(2020)\citenamefont {Ball} \emph
  {et~al.}}]{Ball:2020dnx}%
  \BibitemOpen
  \bibfield  {author} {\bibinfo {author} {\bibfnamefont {A.}~\bibnamefont
  {Ball}} \emph {et~al.},\ }\bibfield  {title} {\enquote {\bibinfo {title}
  {{Search for millicharged particles in proton-proton collisions at $\sqrt{s}
  = 13$ TeV}},}\ }\href {\doibase 10.1103/PhysRevD.102.032002} {\bibfield
  {journal} {\bibinfo  {journal} {Phys. Rev. D}\ }\textbf {\bibinfo {volume}
  {102}},\ \bibinfo {pages} {032002} (\bibinfo {year} {2020})},\ \Eprint
  {http://arxiv.org/abs/2005.06518} {arXiv:2005.06518 [hep-ex]} \BibitemShut
  {NoStop}%
\bibitem [{\citenamefont {Ball}\ \emph {et~al.}(2021)\citenamefont {Ball} \emph
  {et~al.}}]{Ball:2021qrn}%
  \BibitemOpen
  \bibfield  {author} {\bibinfo {author} {\bibfnamefont {A.}~\bibnamefont
  {Ball}} \emph {et~al.} (\bibinfo {collaboration} {milliQan}),\ }\bibfield
  {title} {\enquote {\bibinfo {title} {{Sensitivity to millicharged particles
  in future proton-proton collisions at the LHC}},}\ }\href@noop {} {\
  (\bibinfo {year} {2021})},\ \Eprint {http://arxiv.org/abs/2104.07151}
  {arXiv:2104.07151 [hep-ex]} \BibitemShut {NoStop}%
\bibitem [{\citenamefont {Abe}\ \emph {et~al.}(2018)\citenamefont {Abe} \emph
  {et~al.}}]{Abe:2018uyc}%
  \BibitemOpen
  \bibfield  {author} {\bibinfo {author} {\bibfnamefont {K.}~\bibnamefont
  {Abe}} \emph {et~al.} (\bibinfo {collaboration} {Hyper-Kamiokande}),\
  }\bibfield  {title} {\enquote {\bibinfo {title} {{Hyper-Kamiokande Design
  Report}},}\ }\href@noop {} {\  (\bibinfo {year} {2018})},\ \Eprint
  {http://arxiv.org/abs/1805.04163} {arXiv:1805.04163 [physics.ins-det]}
  \BibitemShut {NoStop}%
\bibitem [{\citenamefont {Gaisser}\ \emph {et~al.}(2016)\citenamefont
  {Gaisser}, \citenamefont {Engel},\ and\ \citenamefont
  {Resconi}}]{gaisser2016cosmic}%
  \BibitemOpen
  \bibfield  {author} {\bibinfo {author} {\bibfnamefont {T.K.}\ \bibnamefont
  {Gaisser}}, \bibinfo {author} {\bibfnamefont {R.}~\bibnamefont {Engel}}, \
  and\ \bibinfo {author} {\bibfnamefont {E.}~\bibnamefont {Resconi}},\ }\href
  {https://books.google.com/books?id=nHFNDAAAQBAJ} {\emph {\bibinfo {title}
  {Cosmic Rays and Particle Physics}}}\ (\bibinfo  {publisher} {Cambridge
  University Press},\ \bibinfo {year} {2016})\BibitemShut {NoStop}%
\bibitem [{\citenamefont {Bays}\ \emph {et~al.}(2012)\citenamefont {Bays} \emph
  {et~al.}}]{Bays:2011si}%
  \BibitemOpen
  \bibfield  {author} {\bibinfo {author} {\bibfnamefont {K.}~\bibnamefont
  {Bays}} \emph {et~al.} (\bibinfo {collaboration} {Super-Kamiokande}),\
  }\bibfield  {title} {\enquote {\bibinfo {title} {{Supernova Relic Neutrino
  Search at Super-Kamiokande}},}\ }\href {\doibase 10.1103/PhysRevD.85.052007}
  {\bibfield  {journal} {\bibinfo  {journal} {Phys. Rev. D}\ }\textbf {\bibinfo
  {volume} {85}},\ \bibinfo {pages} {052007} (\bibinfo {year} {2012})},\
  \Eprint {http://arxiv.org/abs/1111.5031} {arXiv:1111.5031 [hep-ex]}
  \BibitemShut {NoStop}%
\bibitem [{\citenamefont {Arg\"uelles}\ \emph {et~al.}(2019)\citenamefont
  {Arg\"uelles}, \citenamefont {Diaz}, \citenamefont {Kheirandish},
  \citenamefont {Olivares-Del-Campo}, \citenamefont {Safa},\ and\ \citenamefont
  {Vincent}}]{Arguelles:2019ouk}%
  \BibitemOpen
  \bibfield  {author} {\bibinfo {author} {\bibfnamefont {Carlos~A.}\
  \bibnamefont {Arg\"uelles}}, \bibinfo {author} {\bibfnamefont {Alejandro}\
  \bibnamefont {Diaz}}, \bibinfo {author} {\bibfnamefont {Ali}\ \bibnamefont
  {Kheirandish}}, \bibinfo {author} {\bibfnamefont {Andr\'es}\ \bibnamefont
  {Olivares-Del-Campo}}, \bibinfo {author} {\bibfnamefont {Ibrahim}\
  \bibnamefont {Safa}}, \ and\ \bibinfo {author} {\bibfnamefont {Aaron~C.}\
  \bibnamefont {Vincent}},\ }\bibfield  {title} {\enquote {\bibinfo {title}
  {{Dark Matter Annihilation to Neutrinos}},}\ }\href@noop {} {\  (\bibinfo
  {year} {2019})},\ \Eprint {http://arxiv.org/abs/1912.09486} {arXiv:1912.09486
  [hep-ph]} \BibitemShut {NoStop}%
\bibitem [{\citenamefont {Aprile}\ \emph {et~al.}(2020)\citenamefont {Aprile}
  \emph {et~al.}}]{Aprile:2020tmw}%
  \BibitemOpen
  \bibfield  {author} {\bibinfo {author} {\bibfnamefont {E.}~\bibnamefont
  {Aprile}} \emph {et~al.} (\bibinfo {collaboration} {XENON}),\ }\bibfield
  {title} {\enquote {\bibinfo {title} {{Excess electronic recoil events in
  XENON1T}},}\ }\href {\doibase 10.1103/PhysRevD.102.072004} {\bibfield
  {journal} {\bibinfo  {journal} {Phys. Rev. D}\ }\textbf {\bibinfo {volume}
  {102}},\ \bibinfo {pages} {072004} (\bibinfo {year} {2020})},\ \Eprint
  {http://arxiv.org/abs/2006.09721} {arXiv:2006.09721 [hep-ex]} \BibitemShut
  {NoStop}%
\bibitem [{\citenamefont {Bloch}\ \emph {et~al.}(2021)\citenamefont {Bloch},
  \citenamefont {Caputo}, \citenamefont {Essig}, \citenamefont {Redigolo},
  \citenamefont {Sholapurkar},\ and\ \citenamefont {Volansky}}]{Bloch:2020uzh}%
  \BibitemOpen
  \bibfield  {author} {\bibinfo {author} {\bibfnamefont {Itay~M.}\ \bibnamefont
  {Bloch}}, \bibinfo {author} {\bibfnamefont {Andrea}\ \bibnamefont {Caputo}},
  \bibinfo {author} {\bibfnamefont {Rouven}\ \bibnamefont {Essig}}, \bibinfo
  {author} {\bibfnamefont {Diego}\ \bibnamefont {Redigolo}}, \bibinfo {author}
  {\bibfnamefont {Mukul}\ \bibnamefont {Sholapurkar}}, \ and\ \bibinfo {author}
  {\bibfnamefont {Tomer}\ \bibnamefont {Volansky}},\ }\bibfield  {title}
  {\enquote {\bibinfo {title} {{Exploring new physics with O(keV) electron
  recoils in direct detection experiments}},}\ }\href {\doibase
  10.1007/JHEP01(2021)178} {\bibfield  {journal} {\bibinfo  {journal} {JHEP}\
  }\textbf {\bibinfo {volume} {01}},\ \bibinfo {pages} {178} (\bibinfo {year}
  {2021})},\ \Eprint {http://arxiv.org/abs/2006.14521} {arXiv:2006.14521
  [hep-ph]} \BibitemShut {NoStop}%
\bibitem [{\citenamefont {M\o{}ller}\ \emph {et~al.}(2018)\citenamefont
  {M\o{}ller}, \citenamefont {Suliga}, \citenamefont {Tamborra},\ and\
  \citenamefont {Denton}}]{Moller:2018kpn}%
  \BibitemOpen
  \bibfield  {author} {\bibinfo {author} {\bibfnamefont {Klaes}\ \bibnamefont
  {M\o{}ller}}, \bibinfo {author} {\bibfnamefont {Anna~M.}\ \bibnamefont
  {Suliga}}, \bibinfo {author} {\bibfnamefont {Irene}\ \bibnamefont
  {Tamborra}}, \ and\ \bibinfo {author} {\bibfnamefont {Peter~B.}\ \bibnamefont
  {Denton}},\ }\bibfield  {title} {\enquote {\bibinfo {title} {{Measuring the
  supernova unknowns at the next-generation neutrino telescopes through the
  diffuse neutrino background}},}\ }\href {\doibase
  10.1088/1475-7516/2018/05/066} {\bibfield  {journal} {\bibinfo  {journal}
  {JCAP}\ }\textbf {\bibinfo {volume} {05}},\ \bibinfo {pages} {066} (\bibinfo
  {year} {2018})},\ \Eprint {http://arxiv.org/abs/1804.03157} {arXiv:1804.03157
  [astro-ph.HE]} \BibitemShut {NoStop}%
\bibitem [{\citenamefont {Fang}\ \emph {et~al.}(2020)\citenamefont {Fang},
  \citenamefont {Zhang}, \citenamefont {Gong}, \citenamefont {Cao},
  \citenamefont {Lin}, \citenamefont {Yang},\ and\ \citenamefont
  {Li}}]{Fang:2019lej}%
  \BibitemOpen
  \bibfield  {author} {\bibinfo {author} {\bibfnamefont {X.}~\bibnamefont
  {Fang}}, \bibinfo {author} {\bibfnamefont {Y.}~\bibnamefont {Zhang}},
  \bibinfo {author} {\bibfnamefont {G.~H.}\ \bibnamefont {Gong}}, \bibinfo
  {author} {\bibfnamefont {G.~F.}\ \bibnamefont {Cao}}, \bibinfo {author}
  {\bibfnamefont {T.}~\bibnamefont {Lin}}, \bibinfo {author} {\bibfnamefont
  {C.~W.}\ \bibnamefont {Yang}}, \ and\ \bibinfo {author} {\bibfnamefont
  {W.~D.}\ \bibnamefont {Li}},\ }\bibfield  {title} {\enquote {\bibinfo {title}
  {{Capability of detecting low energy events in JUNO Central Detector}},}\
  }\href {\doibase 10.1088/1748-0221/15/03/P03020} {\bibfield  {journal}
  {\bibinfo  {journal} {JINST}\ }\textbf {\bibinfo {volume} {15}},\ \bibinfo
  {pages} {P03020} (\bibinfo {year} {2020})},\ \Eprint
  {http://arxiv.org/abs/1912.01864} {arXiv:1912.01864 [physics.ins-det]}
  \BibitemShut {NoStop}%
\bibitem [{\citenamefont {Verpoest}(2018)}]{Verpoest}%
  \BibitemOpen
  \bibfield  {author} {\bibinfo {author} {\bibfnamefont {Stef}\ \bibnamefont
  {Verpoest}},\ }\emph {\bibinfo {title} {Search for particles with fractional
  charges in IceCube based on anomalous energy loss}},\ \href@noop {} {Master's
  thesis},\ \bibinfo  {school} {Ghent University} (\bibinfo {year}
  {2018})\BibitemShut {NoStop}%
\bibitem [{\citenamefont {Van~Driessche}(2019)}]{VanDriessche}%
  \BibitemOpen
  \bibfield  {author} {\bibinfo {author} {\bibfnamefont {Ward}\ \bibnamefont
  {Van~Driessche}},\ }\emph {\bibinfo {title} {Search for particles with
  anomalous charge in the IceCube detector}},\ \href@noop {} {Ph.D. thesis},\
  \bibinfo  {school} {Ghent University} (\bibinfo {year} {2019})\BibitemShut
  {NoStop}%
\bibitem [{\citenamefont {Aartsen}\ \emph {et~al.}(2013)\citenamefont {Aartsen}
  \emph {et~al.}}]{Aartsen:2013rt}%
  \BibitemOpen
  \bibfield  {author} {\bibinfo {author} {\bibfnamefont {M.G.}\ \bibnamefont
  {Aartsen}} \emph {et~al.} (\bibinfo {collaboration} {IceCube}),\ }\bibfield
  {title} {\enquote {\bibinfo {title} {{Measurement of South Pole ice
  transparency with the IceCube LED calibration system}},}\ }\href {\doibase
  10.1016/j.nima.2013.01.054} {\bibfield  {journal} {\bibinfo  {journal} {Nucl.
  Instrum. Meth. A}\ }\textbf {\bibinfo {volume} {711}},\ \bibinfo {pages}
  {73--89} (\bibinfo {year} {2013})},\ \Eprint {http://arxiv.org/abs/1301.5361}
  {arXiv:1301.5361 [astro-ph.IM]} \BibitemShut {NoStop}%
\bibitem [{\citenamefont {Abbasi}\ \emph {et~al.}(2009)\citenamefont {Abbasi}
  \emph {et~al.}}]{Abbasi:2008aa}%
  \BibitemOpen
  \bibfield  {author} {\bibinfo {author} {\bibfnamefont {R.}~\bibnamefont
  {Abbasi}} \emph {et~al.} (\bibinfo {collaboration} {IceCube}),\ }\bibfield
  {title} {\enquote {\bibinfo {title} {{The IceCube Data Acquisition System:
  Signal Capture, Digitization, and Timestamping}},}\ }\href {\doibase
  10.1016/j.nima.2009.01.001} {\bibfield  {journal} {\bibinfo  {journal} {Nucl.
  Instrum. Meth. A}\ }\textbf {\bibinfo {volume} {601}},\ \bibinfo {pages}
  {294--316} (\bibinfo {year} {2009})},\ \Eprint
  {http://arxiv.org/abs/0810.4930} {arXiv:0810.4930 [physics.ins-det]}
  \BibitemShut {NoStop}%
\bibitem [{\citenamefont {Aartsen}\ \emph {et~al.}(2017)\citenamefont {Aartsen}
  \emph {et~al.}}]{Aartsen:2016nxy}%
  \BibitemOpen
  \bibfield  {author} {\bibinfo {author} {\bibfnamefont {M.~G.}\ \bibnamefont
  {Aartsen}} \emph {et~al.} (\bibinfo {collaboration} {IceCube}),\ }\bibfield
  {title} {\enquote {\bibinfo {title} {{The IceCube Neutrino Observatory:
  Instrumentation and Online Systems}},}\ }\href {\doibase
  10.1088/1748-0221/12/03/P03012} {\bibfield  {journal} {\bibinfo  {journal}
  {JINST}\ }\textbf {\bibinfo {volume} {12}},\ \bibinfo {pages} {P03012}
  (\bibinfo {year} {2017})},\ \Eprint {http://arxiv.org/abs/1612.05093}
  {arXiv:1612.05093 [astro-ph.IM]} \BibitemShut {NoStop}%
\bibitem [{\citenamefont {Aartsen}\ \emph {et~al.}(2020)\citenamefont {Aartsen}
  \emph {et~al.}}]{IceCube:2020nwx}%
  \BibitemOpen
  \bibfield  {author} {\bibinfo {author} {\bibfnamefont {M.~G.}\ \bibnamefont
  {Aartsen}} \emph {et~al.} (\bibinfo {collaboration} {IceCube}),\ }\bibfield
  {title} {\enquote {\bibinfo {title} {{In-situ calibration of the
  single-photoelectron charge response of the IceCube photomultiplier
  tubes}},}\ }\href {\doibase 10.1088/1748-0221/15/06/P06032} {\bibfield
  {journal} {\bibinfo  {journal} {JINST}\ }\textbf {\bibinfo {volume} {15}},\
  \bibinfo {pages} {06} (\bibinfo {year} {2020})},\ \Eprint
  {http://arxiv.org/abs/2002.00997} {arXiv:2002.00997 [physics.ins-det]}
  \BibitemShut {NoStop}%
\bibitem [{\citenamefont {Aartsen}\ \emph {et~al.}(2014)\citenamefont {Aartsen}
  \emph {et~al.}}]{Aartsen:2013vja}%
  \BibitemOpen
  \bibfield  {author} {\bibinfo {author} {\bibfnamefont {M.~G.}\ \bibnamefont
  {Aartsen}} \emph {et~al.} (\bibinfo {collaboration} {IceCube}),\ }\bibfield
  {title} {\enquote {\bibinfo {title} {{Energy Reconstruction Methods in the
  IceCube Neutrino Telescope}},}\ }\href {\doibase
  10.1088/1748-0221/9/03/P03009} {\bibfield  {journal} {\bibinfo  {journal}
  {JINST}\ }\textbf {\bibinfo {volume} {9}},\ \bibinfo {pages} {P03009}
  (\bibinfo {year} {2014})},\ \Eprint {http://arxiv.org/abs/1311.4767}
  {arXiv:1311.4767 [physics.ins-det]} \BibitemShut {NoStop}%
\bibitem [{\citenamefont {K\"all\'en}(1964)}]{Kallen:1964lxa}%
  \BibitemOpen
  \bibfield  {author} {\bibinfo {author} {\bibfnamefont {Gunnar}\ \bibnamefont
  {K\"all\'en}},\ }\href@noop {} {\emph {\bibinfo {title} {{Elementary particle
  physics}}}}\ (\bibinfo  {publisher} {Addison-Wesley},\ \bibinfo {address}
  {Reading, MA},\ \bibinfo {year} {1964})\BibitemShut {NoStop}%
\end{thebibliography}%

\end{document}